\documentclass[prb,showpacs,twocolumn,aps,superscriptaddress,a4paper]{revtex4-1}
\usepackage{graphicx}
\usepackage{latexsym}
\usepackage{amsfonts}
\usepackage{amssymb}
\usepackage{rotating}
\usepackage{bbm}
\usepackage{cancel}
\usepackage{color}

\newcommand{\be}{\begin{equation}}
\newcommand{\ee}{\end{equation}}
\newcommand{\bea}{\begin{eqnarray}}
\newcommand{\eea}{\end{eqnarray}}

\def\a{\alpha}
\def\b{\beta}
\def\g{\gamma}
\def\G{\Gamma}
\def\d{\delta}
\def\D{\Delta}
\def\e{\epsilon}
\def\ve{\varepsilon}
\def\h{\eta}
\def\th{\theta}
\def\k{\kappa}
\def\l{\lambda}
\def\L{\Lambda}
\def\m{\mu}
\def\n{\nu}
\def\c{\xi}

\def\p{\pi}
\def\P{\Pi}
\def\r{\rho}
\def\s{\sigma}
\def\S{\Sigma}
\def\t{\tau}
\def\f{\phi}
\def\vf{\varphi}
\def\F{\Phi}
\def\x{\chi}
\def\w{\omega}
\def\W{\Omega}
\def\q{\psi}
\def\Q{\Psi}
\def\z{\zeta}

\def\bgG{\mbox{\boldmath $\Gamma$}}

\def\bgD{\mbox{\boldmath $\Delta$}}
\def\bge{\mbox{\boldmath $\epsilon$}}

\def\bgL{\mbox{\boldmath $\Lambda$}}

\def\bgr{\mbox{\boldmath $\rho$}}

\def\bgS{\mbox{\boldmath $\Sigma$}}

\def\blr{{\mathbf r}}

\def\blA{{\mathbf A}}

\def\blG{{\mathbf G}}
\def\blH{{\mathbf H}}

\def\callA{\mbox{$\mathcal{A}$}}

\def\callG{\mbox{$\mathcal{G}$}}

\def\callS{\mbox{$\mathcal{S}$}}
\def\callT{\mbox{$\mathcal{T}$}}

\def\bcallG{\mbox{\boldmath $\mathcal{G}$}}
\def\bcallH{\mbox{\boldmath $\mathcal{H}$}}

\def\ua{\uparrow}
\def\da{\downarrow}
\def\ra{\rightarrow}

\def\de{\partial}
\def\iif{\infty}
\def\bra{\langle}
\def\ket{\rangle}
\def\grad{\mbox{\boldmath $\nabla$}}
\def\Tr{{\rm Tr}}
\def\Re{{\rm Re}}
\def\Im{{\rm Im}}

\def\1op{\hat{\mathbbm{1}}}
\def\nn{\nonumber}

\DeclareGraphicsExtensions{.eps,.pdf,.gif,.jpg,.png}

\renewcommand{\vr} {{\bf r}}
\newcommand{\Id}[1] {\int \! \! {\rm d}^3 #1}
\newcommand{\ID}[1] {\int \!  \frac{{\rm d} #1}{2 \pi}}
\renewcommand{\Re}{ {\rm Re} \, }
\renewcommand{\Im}{ {\rm Im} \, }
\newcommand{\tr}[1] { {\rm Tr} \left\{ #1 \right\} }

\begin{document}  

\def\gC{\mbox{\boldmath $C$}}
\def\gZ{\mbox{\boldmath $Z$}}
\def\gR{\mbox{\boldmath $R$}}
\def\gN{\mbox{\boldmath $N$}}
\def\gG{\mbox{\boldmath $G$}}
\def\green{\mbox{\boldmath ${\cal G}$}}
\def\grn{\mbox{${\cal G}$}}
\def\gH{\mbox{\boldmath $H$}}
\def\bA{{\bf A}}
\def\bJ{{\bf J}}
\def\bG{{\bf G}}
\def\bF{{\bf F}}
\def\bH{\mbox{\boldmath $H$}}
\def\bQ{\mbox{\boldmath $Q$}}
\def\bgS{\mbox{\boldmath $\Sigma$}}
\def\bT{\mbox{\boldmath $T$}}
\def\bU{\mbox{\boldmath $U$}}
\def\bV{\mbox{\boldmath $V$}}
\def\bgG{\mbox{\boldmath $\Gamma$}}
\def\bgL{\mbox{\boldmath $\Lambda$}}
\def\ubG{\underline{{\bf G}}}
\def\ubH{\underline{{\bf H}}}
\def\ubQ{\underline{{\bf Q}}}
\def\ubS{\underline{{\bf S}}}
\def\ubg{\underline{{\bf g}}}
\def\ubq{\underline{{\bf q}}}
\def\ubp{\underline{{\bf p}}}
\def\ubgS{\underline{{\bf \Sigma}}}
\def\bge{\mbox{\boldmath $\epsilon$}}
\def\bgD{{\bf \Delta}}

\def\bDelta{\mbox{\boldmath $\Delta$}}
\def\bcalE{\mbox{\boldmath ${\cal E}$}}
\def\bcalF{\mbox{\boldmath ${\cal F}$}}
\def\bcalG{\mbox{\boldmath $G$}}
\def\ubcalG{\mbox{\underline{\boldmath $G$}}}
\def\callG{\mathcal{G}}
\def\callA{\mathcal{A}}
\def\callT{\mathcal{T}}
\def\ubcalA{\mbox{\underline{\boldmath $A$}}}
\def\ubcalB{\mbox{\underline{\boldmath $B$}}}
\def\ubcalC{\mbox{\underline{\boldmath $C$}}}
\def\ubcalg{\mbox{\underline{\boldmath $g$}}}
\def\ubcalH{\mbox{\underline{\boldmath $H$}}}
\def\bcalK{\mbox{\boldmath $K$}}
\def\ubcalK{\mbox{\underline{\boldmath $K$}}}
\def\bcalV{\mbox{\boldmath ${\cal V}$}}
\def\ubcalV{\mbox{\underline{\boldmath $V$}}}
\def\bcalU{\mbox{\boldmath ${\cal U}$}}
\def\ubcalz{\mbox{\underline{\boldmath $z$}}}
\def\bcalQ{\mbox{\boldmath ${\cal Q}$}}
\def\ubcalQ{\mbox{\underline{\boldmath $Q$}}}
\def\ubcalP{\mbox{\underline{\boldmath $P$}}}
\def\bSS{\mbox{\boldmath $S$}}
\def\ubcalS{\mbox{\underline{\boldmath $S$}}}
\def\bff{\mbox{\boldmath $f$}}
\def\bg{\mbox{\boldmath $g$}}
\def\bh{\mbox{\boldmath $h$}}
\def\bk{\mbox{\boldmath $k$}}
\def\bq{\mbox{\boldmath $q$}}
\def\bp{\mbox{\boldmath $p$}}
\def\br{\mbox{\boldmath $r$}}
\def\bt{\mbox{\boldmath $t$}}
\def\ubh{\mbox{\underline{\boldmath $h$}}}
\def\ubt{\mbox{\underline{\boldmath $t$}}}
\def\ubk{\mbox{\underline{\boldmath $k$}}}
\def\ua{\uparrow}
\def\da{\downarrow}
\def\a{\alpha}
\def\b{\beta}
\def\g{\gamma}
\def\G{\Gamma}
\def\d{\delta}
\def\D{\Delta}
\def\e{\epsilon}
\def\ve{\varepsilon}
\def\z{\zeta}
\def\h{\eta}
\def\th{\theta}
\def\vth{\vartheta}
\def\k{\kappa}
\def\l{\lambda}
\def\L{\Lambda}
\def\m{\mu}
\def\n{\nu}
\def\x{\xi}
\def\X{\Xi}
\def\p{\pi}
\def\P{\Pi}
\def\r{\rho}
\def\bgr{\mbox{\boldmath $\rho$}}
\def\s{\sigma}
\def\us{\mbox{\underline{\boldmath $\sigma$}}}
\def\ubgm{\mbox{\underline{\boldmath $\mu$}}}
\def\S{\Sigma}
\def\ubcgS{\mbox{\underline{\boldmath $\Sigma$}}}
\def\t{\tau}
\def\f{\phi}
\def\vf{\varphi}
\def\F{\Phi}
\def\c{\chi}
\def\k{\kappa}
\def\w{\omega}
\def\W{\Omega}
\def\Q{\Psi}
\def\q{\psi}
\def\de{\partial}
\def\inf{\infty}
\def\ra{\rightarrow}
\def\bra{\langle}
\def\ket{\rangle}
\def\bbra{\langle\langle}
\def\kket{\rangle\rangle}
\def\grad{\mbox{\boldmath $\nabla$}}
\def\no{\bf 1}
\def\ze{\bf 0}
\def\uno{\underline{\bf 1}}
\def\zero{\underline{\bf 0}}

\def\dr{{\rm d}}
\def\bj{{\bf j}}
\def\br{{\bf r}}
\def\bz{\bar{z}}
\def\bart{\bar{t}}

\title{Transport through correlated systems with density functional theory}

\author{S. Kurth}
\affiliation{Nano-Bio Spectroscopy Group and ETSF, 
Dpto. de F\'{i}sica de Materiales,
Universidad del Pa\'{i}s Vasco UPV/EHU, Av. Tolosa 72, 
E-20018 San Sebasti\'{a}n, Spain}
\affiliation{IKERBASQUE, Basque Foundation for Science, Maria Diaz de Haro 3, 
E-48013 Bilbao, Spain}

\author{G. Stefanucci}
\affiliation{Dipartimento di Fisica, Universit\`{a} di Roma Tor Vergata,
Via della Ricerca Scientifica 1, 00133 Rome, Italy; European Theoretical 
Spectroscopy Facility (ETSF)}
\affiliation{INFN, Sezione di Roma Tor Vergata, Via della Ricerca 
Scientifica 1, 00133 Rome, Italy}

\begin{abstract}
We present recent advances in Density Functional Theory (DFT) for applications
to the field of quantum transport, with particular emphasis on transport
through strongly correlated systems. We review the foundations of the
popular Landauer-B\"uttiker(LB)+DFT approach. This formalism, 
when using approximations to the exchange-correlation (xc) potential with 
steps at integer occupation, correctly captures the Kondo plateau in the zero 
bias conductance at zero temperature but completely fails to capture the 
transition to the Coulomb blockade (CB) regime as temperature increases. 
Both of these 
effects are hallmarks of strong electronic correlations. To overcome the 
limitations of LB+DFT the quantum transport problem is treated from a 
time-dependent (TD) perspective using TDDFT, an exact framework to deal with 
nonequilibrium situations. The steady-state limit of TDDFT shows that in 
addition to an xc potential in the junction, there also exists an
xc correction to the applied bias. Open shell molecules in the 
CB regime provide the most striking examples of the importance of
the xc bias correction. Using the Anderson model as guidance we estimate
these corrections for a class of systems in the limit of zero bias. For the
general case we put forward a steady-state DFT which is based on the
one-to-one correspondence between the pair of basic variables steady density
on and steady current across the junction and the pair local potential on and
bias across the junction. Like TDDFT, this framework also leads to both an
xc potential in the junction and an xc correction to the bias. Unlike in
TDDFT, these potentials are independent of history. We highlight the universal
features of both xc potential and xc bias corrections for junctions in the
CB regime. We also provide an accurate parametrization of both xc potentials
for the Anderson model at arbitrary temperatures and interaction strengths 
thus providing a unified DFT description for both Kondo and CB 
regimes and the transition between them.
\end{abstract}


\maketitle

\section{Introduction}

Density Functional Theory (DFT) is probably the most popular method for an 
ab-initio description of atoms, molecules, and solids both in 
\cite{DreizlerGross:90} and out of (thermal) equilibrium 
\cite{Ullrich:12,Maitra:16}. 
The main reason for this popularity is its relative 
numerical simplicity which arises due to the mapping of the interacting 
many-electron problem onto an effective non-interacting one. 
This simplicity has given rise to a plethora of applications 
of DFT to the description of more and more complex systems in 
atomistic detail. However, there are also physical situations where 
DFT has been less successful, one of those being the description of 
strongly correlated systems. Since the fundamental theorems of DFT, both in 
thermal equilibrium and in its time-dependent out-of-equilibrium incarnation, 
give the framework a sound theoretical foundation, the failures of DFT 
have to be attributed to the inadequacy of known approximations to the 
famous exchange-correlation (xc) functional.

In this Topical Review we are concerned with a DFT treatment of the 
particular physical situation of quantum transport, i.e., the description of 
electronic transport through a nanoscale region contacted by metallic leads 
and driven out of equilibrium by application of an external bias. 
We will discuss different DFT frameworks for the 
description of quantum transport with a particular focus on 
recent advances in DFT 
approximations to deal with strongly correlated nanojunctions. 
It turns out that in all these frameworks, if one wants to capture 
strong correlations within DFT, the corresponding xc functionals need to have 
step features at integer occupation. These steps are intimately related to the 
famous derivative discontinuity of the exact xc potential.

The first application of DFT to electronic 
transport goes back to a seminal paper of Lang \cite{Lang:95} where, following 
ideas of Landauer \cite{Landauer:57} and B\"uttiker \cite{Buettiker:86}, 
transport in the steady state is treated as a scattering problem of effectively 
non-interacting electrons. The resulting formalism, known as 
Landauer-B\"uttiker plus DFT (LB+DFT), or its equivalent formulation 
in terms of non-equilibrium Green functions 
(NEGF)~\cite{TGW.2001,TGW2.2001,BMOTS.2002} are by far the most 
widely used DFT schemes for 
transport. In Sec.~\ref{lb_dft} we will present this formalism and critically 
discuss its merits and shortcomings. 
We will show that the LB+DFT framework, at zero temperature and 
in the limit of small bias, is capable of correctly capturing features 
related to the Kondo effect, a hallmark of strong electronic correlation. On 
the other hand, at finite temperatures LB+DFT fails to correctly describe 
Coulomb blockade, another ubiquitous correlation effect.

The Landauer-B\"uttiker formalism is only concerned with steady-state 
electronic transport under application of a DC bias voltage. Alternatively 
one may view transport as an explictly time-dependent problem where a system 
contacted to metallic leads is initially in thermal equilibrium. At a given 
time $t_0$, the system is driven out of equilibrium by application of a bias 
and one follows its time evolution~\cite{C.1980,sa-1.2004}. In a DFT 
framework, this situations can naturally be described with Time-Dependent 
DFT (TDDFT) \cite{rg.1984,Ullrich:12,sa-2.2004,ksarg.2005}. Of course, 
application of TDDFT to transport is ideally suited to treat systems moving 
under the influence of explicitly time-dependent fields (e.g., AC bias). It 
turns out, however, that even for the steady state which develops in the 
long-time limit after application of a DC bias, TDDFT in principle leads to 
corrections to the LB+DFT 
formalism~\cite{sa-2.2004,ewk.2004,kbe.2006,skgr.2007}. The TDDFT approach 
to transport will be discussed in Sec.~\ref{tddft}.
The TDDFT corrections to LB+DFT are shown to be crucial in a 
correct description of Coulomb blockade in the zero-bias conductance. However, 
the price to be paid now is a lack of Kondo features in this framework.

Finally, in Sec.~\ref{idft} we review yet another, very recent DFT approach to 
steady-state transport called i-DFT \cite{StefanucciKurth:15}. Compared to 
the LB+DFT approach, the novelty is that the corresponding Kohn-Sham (KS) 
system is characterized not only in terms of the exchange-correlation (xc) 
potential in the nanojunction but also in terms of an xc 
contribution to the bias. 
For this framework we first construct approximations which are 
capable of correctly describing Coulomb blockade not only in the zero-bias 
limit but also at finite bias. Again, these approximations miss Kondo physics 
which, however, can be included using rather simple arguments. The 
i-DFT framework thus allows for a unified description of both Kondo effect and 
Coulomb blockade in finite-bias electronic transport as well as a 
transition between these regimes as temperature increases.

\section{Quantum Transport with Density Functional Theory}

\subsection{The standard approach: Landauer-B\"uttiker Formalism}
\label{lb_dft}

In the typical setup of transport, a mesoscopic or nanoscopic central region 
$C$ is connected to two (or more) metallic leads and one is interested in the 
current flowing through $C$ upon application of a bias between any two leads. 

\begin{figure}[t]
\centerline{
\includegraphics[width=0.5\textwidth]{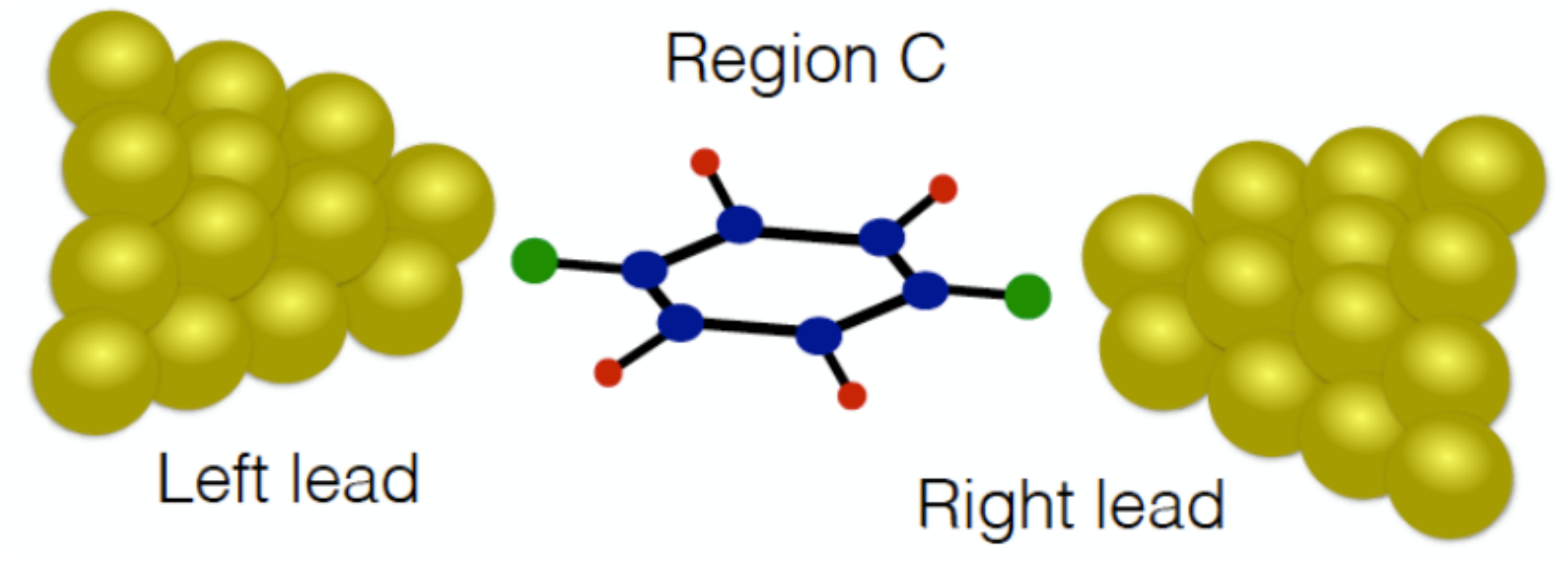}}
\caption{Schematic illustration of the quantum transport setup.}
\label{setup}
\end{figure}

For simplicity, here we only consider the case of the central region $C$ 
connected to left ($L$) and right ($R$) leads, see 
Fig.~\ref{setup} for an illustration of the setup.  
A DFT description of such a system in thermal equilibrium at 
temperature $T=1/\b$ and chemical potential $\m$
requires the self-consistent solution of the Kohn-Sham (KS) equations (we use 
atomic units unless otherwise stated)
\bea
\left[ -\frac{\nabla^2}{2} + v_0(\vr) + v_{\rm H}[n](\vr) + v_{\rm xc}[n](\vr)
\right] \q_k(\vr) = \varepsilon_k \q_k(\vr) \nn\\
\label{ks_eq}
\eea
where $v_{0}$ is the external potential generated by the positively 
charged nuclei and
\be
n(\vr) = \sum_k f(\varepsilon_k) | \q_k(\vr)|^2 
\label{dens}
\ee
is the electronic density, $f(\e)=1/(e^{\b(\e-\m)}+1)$ being the 
Fermi function. In Eq.~(\ref{ks_eq}) there appear two 
functionals of the density. These are the Hartree potential
\be
v_{\rm H}[n](\vr) = \Id{r'} w(\vr-\vr') n(\vr'), 
\label{hartree}
\ee
with $w(\vr - \vr')$ the electron-electron interaction,  and
the exchange-correlation (xc) potential
\be
v_{\rm xc}[n](\vr) = \frac{\delta E_{\rm xc}[n]}{\delta n(\vr)}.
\label{v_xc}
\ee
The xc potential is the functional 
derivative of the xc (free) energy functional $E_{\rm xc}[n]$.

A first technical difficulty when trying to solve the KS equations in the 
geometry of (semi-infinite) left and right leads connected to a central region 
arises from the fact that such a system is neither finite nor periodic. Thus 
standard techniques for electronic structure calculations of either finite 
or periodic systems need to be adapted to the problem. 

One convenient way of dealing with the transport geometry is to use the 
language of Green's functions. To begin with, we introduce a localized 
single-particle basis $|j q \rangle$ 
where $j$ denotes an atomic site and $q$ labels the different basis functions 
localized at this site. For simplicity, we assume that the single-particle 
basis is orthonormal and complete, i.e., $\langle j q|j' q' \rangle = 
\delta_{j,j'} \delta_{q,q'}$ and $\sum_{j,q} |jq\rangle \langle jq| = \mathbbm{1}$.  
In this basis, the KS Hamiltonian $\bcallH_s$  can be written in 
$3\times3$ block form as 
\be
\bcallH_s = \left[ 
\begin{array}{ccc}
{\mathbf H}_{LL} & {\mathbf H}_{LC} & 0 \\
{\mathbf H}_{CL} & {\mathbf H}_{CC} & {\mathbf H}_{CR} \\
0 & {\mathbf H}_{RC} & {\mathbf H}_{RR} 
\end{array} \right] 
\label{blockh}
\ee
where ${\mathbf H}_{\alpha \alpha'}$ collects all matrix elements connecting 
regions $\alpha$ and $\alpha'$ ($\alpha, \alpha' \in \{L,C,R\})$. 
Hereafter we use boldface letters to denote matrices in the 
one-electron basis.
Note that in Eq.~(\ref{blockh}) we have assumed that all matrix elements connecting the left and right 
leads vanish.  The KS single-particle orbitals $\q_k$ can be expanded 
in the localized basis as 
\be
\q_k(\vr) = \sum_{jq} c_{k,jq} \langle \vr| iq \rangle
\ee
which allows to write the density according to 
\be
n(\vr) = \sum_{jq} \sum_{j'q'} \bgr^{\rm eq}_{jq, j'q'} \langle jq | \vr 
\rangle \langle \vr| j'q' \rangle.
\label{ks_dens}
\ee 
In Eq.~(\ref{ks_dens}) $\bgr^{\rm eq}$ is the equilibrium one-particle density matrix 
whose matrix elements are given by 
\be
\bgr^{\rm eq}_{jq, j'q'} = 2
\sum_k f(\varepsilon_k) c_{k,jq}^* c_{k,j'q'}  \;,
\label{dens_matrix}
\ee
where the factor of $2$ comes from spin.

The retarded Green's function $\blG(\omega)$ at 
energy $\omega$ is defined through 
\be
\left( (\omega+i\eta) \mathbbm{1} - \bcallH_s\right) \bcallG
(\omega) = \mathbbm{1}
\label{gf_ret}
\ee
with the infinitesimal $\eta \to 0^+$. We write 
the Green's function in the same block 
structure as the Hamiltonian
\be
\bcallG(\omega) = \left[ 
\begin{array}{ccc}
\blG_{LL}(\omega) & \blG_{LC}(\omega) & \blG_{LR}(\omega) \\
\blG_{CL}(\omega) & \blG_{CC}(\omega) & \blG_{CR}(\omega) \\
\blG_{RL}(\omega) & \blG_{RC}(\omega) & \blG_{RR}(\omega) 
\end{array} \right] \;. 
\ee
Using Eq.~(\ref{gf_ret}) we can easily solve for $\blG_{CC}$ and 
find 
\be
\blG_{CC}(\omega) = \frac{1}{(\omega+i\eta) \mathbbm{1}_C - 
{\mathbf H}_{CC} 
- {\mathbf \Sigma}_{L}^{\rm eq}(\omega) - {\mathbf \Sigma}_{R}^{\rm eq}(\omega)},
\label{gcceq}
\ee
where the (retarded) embedding self energy for lead $\alpha$ is 
defined as 
\be
{\mathbf \Sigma}_{\alpha}^{\rm eq}(\omega) = {\mathbf H}_{C\alpha} 
\frac{1}{ (\omega+i\eta) \mathbbm{1}_{\alpha} - 
{\mathbf H}_{\alpha \alpha}} {\mathbf H}_{\alpha C} .
\ee
Hereafter we will omit the subscript $CC$ from all matrices with both 
indices in the central region. 
Knowledge of $\blG$ allows us to obtain the central block 
$\bgr^{\rm eq}$ of the density matrix according to
\be
\bgr^{\rm eq} = 2\sum_{\alpha=L,R} \ID{\omega} f(\omega) 
\blG(\omega) {\mathbf \Gamma}^{\rm eq}_{\alpha}(\omega) 
\blG^{\dagger}(\omega),
\label{dens_matrix_sc}
\ee
where the broadening matrix
\be
{\mathbf \Gamma}_{\alpha}^{\rm eq}(\omega) = i \left( {\mathbf \Sigma}_{\alpha}^{\rm eq}(\omega) 
- {\mathbf \Sigma}_{\alpha}^{{\rm eq},\dagger}(\omega) \right).
\label{broadmatx}
\ee
It is worth noting that $\blG$ depends, 
through the Hartree-xc potential $v_{\rm Hxc}\equiv v_{\rm H}+v_{\rm xc}$, on the density and thus, via 
Eq.~(\ref{ks_dens}), on the density matrix. In principle, $v_{\rm Hxc}$ depends 
on the density both in the leads and the central region and therefore 
Eq.~(\ref{dens_matrix_sc}) is not a closed equation for 
$\bgr^{\rm eq}$. However, if any local or semilocal approximation 
such as LDA or GGA is employed and, at the same time, the embedding self 
energies ${\mathbf \Sigma}_{\alpha}$ for the leads are known, then 
Eq.~(\ref{dens_matrix_sc}) indeed becomes a self-consistent equation for 
$\bgr^{\rm eq}$.

So far, we have been discussing the situation in thermal equilibrium. In 
transport, however, one is interested in the scenario where the system is 
driven out of equilibrium by application of a bias. Most commonly, one is 
concerned with the steady state current of the system under application of a 
DC bias. In the picture of Landauer \cite{Landauer:57} and B\"uttiker 
\cite{Buettiker:86}, this steady current may be viewed as the result of lead 
electrons scattering off the potential of the central region $C$. In a seminal 
paper \cite{Lang:95}, Lang proposed to use the KS potential of 
DFT in the central region as the scattering potential. The idea is to 
calculate the scattering states deep in the left and right leads and filling 
them up to the chemical potentials $\mu_{\alpha}=\mu-V_{\alpha}$ ($\alpha =L,R$) 
shifted by the bias $V_{\alpha}$ in lead $\alpha$. One can now proceed by 
directly calculating the scattering states via, e.g., the Lippmann-Schwinger 
equation \cite{Lang:95,BMOTS.2002}. Equivalently, one may use the 
non-equilibrium Green's function (NEGF) formalism \cite{svl-book} to calculate 
the steady-state density matrix or Green's function of the central region. 
The central block of the steady-state density matrix $\bgr$ 
can then be obtained by the following equation
\be
\bgr = 2\sum_{\alpha=L,R} \ID{\omega} f(\omega-V_{\alpha}) 
\blG(\omega) {\mathbf \Gamma}_{\alpha}(\omega) 
\blG^{\dagger}(\omega) \;
\label{dens_matrix_cc_noneq}
\ee
which is structurally very similar to Eq.~(\ref{dens_matrix_sc}), the 
only difference being that $\blG$ is calculated with 
$\bgS_{\a}(\w)= \bgS_{\a}^{\rm eq}(\w-V_{\a})$,  the 
broadening matrix $\bgG_{\a}(\w)=\bgG_{\a}^{\rm eq}(\w-V_{\a})$ 
and the Fermi function for lead $\alpha$ 
is shifted by $V_{\alpha}$. With the density 
matrix one can then calculate the electronic density as in 
Eq.~(\ref{ks_dens}) using $\bgr$ in place of $\bgr^{\rm 
eq}$. After some elementary algebra one finds
\be
n(\blr)=2\sum_{\a=L,R}\ID{\omega}f(\w-V_{\a})A_{\a,s}(\blr,\w)
\label{neqdensity}
\ee
where
\be
A_{\a,s}(\blr,\w)=\sum_{jq,j'q'}[\blA_{\a,s}(\w)]_{jq,j'q'}
\langle jq | \vr 
\rangle \langle \vr| j'q' \rangle,
\ee
and 
\be
\blA_{\a,s}(\w)\equiv 
\blG(\omega) {\mathbf \Gamma}_{\alpha}(\omega) 
\blG^{\dagger}(\omega).
\label{partialAKS}
\ee
is the partial KS spectral function. Again, through the 
dependence of the Green's function on the Hxc potential and thus on the 
density, this defines a self-consistency problem. From the local 
density we can also calculate the total number of electrons in region 
$C$ as 
\bea
N&=&\int_{C} {\rm d}^{3}r\, n(\blr)
\nn\\&=&
2\sum_{\alpha=L,R} \ID{\omega} f(\omega-V_{\alpha}) 
\Tr[\blA_{\a,s}(\w)]
\label{neqN}
\eea
where we have taken into account that the states $jq$ form a complete 
set in region $C$ and the trace is over the single-particle basis in 
region $C$ only.

The steady current can be calculated using the self-consistent 
Green's function via the celebrated Landauer-B\"uttiker (LB) formula
\bea
I_{s} &=& 2 \ID{\w} \left( f(\w-V_L) - f(\w-V_R) \right) \nn \\
&& \times \Tr \left[ 
\blG(\w) \bgG_L(\w) \blG^{\dagger}(\w) \bgG_R(\w) \right] 
\label{lb_current}
\eea
The subscript $s$ in $I_{s}$ highlights the fact that 
Eq.~(\ref{lb_current}) gives the steady current of the KS 
system. There exist no formal proof that $I_{s}$ is the same as the 
steady current $I$ of the interacting system. In fact, in this review 
we will present relevant cases for which $I_{s}\neq I$.
From Eq.~(\ref{lb_current}) we can also calculate the 
KS zero-bias conductance 
\bea
G_{s}&=&\lim_{V_{L}-V_{R}\to 0}\frac{I_{s}}{V_{L}-V_{R}} \nn\\
&=& - 2 \ID{\w} f'(\w) \Tr \left[ 
\blG(\w) \bgG_L(\w) \blG^{\dagger}(\w) \bgG_R(\w) \right] 
\label{lb_conductance}
\eea
where all quantities in the trace are evaluated at $V_{\a}=0$, i.e., 
at equilibrium.

The formalism described above combines the LB formula 
with DFT and it is widely used to describe steady-state transport 
through nanoscale systems with atomistic detail. 
This level of description  is essential
in the field of {\em molecular electronics} \cite{me-book1,me-book2} 
whose central tenet is to use single molecules as active electronic 
devices.

\subsection{The derivative discontinuity of the DFT exchange-correlation potential} 
\label{discontinuity}

In typical applications of DFT one aims to calculate the ground state energy 
and/or the electronic structure of a molecule or solid for a given, fixed 
number of electrons. However, in the transport problem a nanostructure such as 
a quantum dot or a molecule is connected to leads and the number of electrons 
in the nanostructure fluctuates. In other words, instead of dealing with a 
closed system at fixed particle number, in transport we are studying open 
systems connected to particle reservoirs. Therefore one may expect that the 
dependence of density functional approximations on the particle number becomes 
important. This will be the concern of the present section.

In a seminal paper~\cite{PerdewParrLevyBalduz:82}, Perdew and 
coworkers pointed out 
that the exact xc potential of DFT at zero temperature exhibits discontinuous 
steps as the particle number crosses an integer. To show this, they 
constructed an ensemble DFT based on ensembles of 
states with different (integer) electron numbers. The ensemble expectation 
value of the electron number operator then can yield non-integer values, or, 
in other words, the (ensemble) density integrates to a non-integer
\be
\Id{r}\; n(\vr) = N + \eta 
\ee
with $N\in \mathbbm{N}$ and $0\leq\eta < 1$. Perdew and coworkers 
proved that the ground ensemble 
energy is a piecewise linear function of the (fractional) electron number, 
i.e., 
\be
E_{N+\eta} = (1-\eta) E_N + \eta E_{N+1}
\ee
where $E_N$ is the ground state energy at integer particle number $N$.
Therefore, 
the ground ensemble energy as function of electron number is given by a 
series of straight lines connecting the ground state energies with 
integer numbers of particles. The slope of these straight line segments may be expressed 
in terms of physical quantities: the ionization potential $I(N)$ of the 
$N$-electron system is 
\be 
I(N) = E_{N-1} - E_N,
\label{ioniz}
\ee
while the electron affinity $A(N)$ is given by
\be 
A(N) = E_N - E_{N+1} \;.
\label{affinity}
\ee
As a consequence, the discontinuous jump of the derivative of the ground 
ensemble energy at integer electron number can be expressed as
\bea
\Delta(N) &=& E_{N+1} - 2 E_N + E_{N-1} \nn \\
&=& \lim_{\eta \to 0^+} \left( 
\frac{\delta E[n]}{\delta n(\vr)}\bigg\vert_{N+\eta} - 
\frac{\delta E[n]}{\delta n(\vr)}\bigg\vert_{N-\eta}  \right) \;.
\label{toten_deriv}
\eea
Decomposing the total energy functional into its components, there are only two 
terms which are discontinuous as the particle number crosses an 
integer. The first term comes from the non-interacting kinetic energy 
$T_s[n]$, i.e., 
\bea
\Delta_s(N) &=& \lim_{\eta \to 0^+} \left( 
\frac{\delta T_s[n]}{\delta n(\vr)}\bigg\vert_{N+\eta} - 
\frac{\delta T_s[n]}{\delta n(\vr)}\bigg\vert_{N-\eta}  \right) \nn\\ 
&=& \varepsilon_{N+1}(N) - \varepsilon_N(N)
\label{ks_gap}
\eea
where $\varepsilon_k(N)$ is the $k$-th lowest KS energy eigenvalue of an 
$N$ electron calculation. The second terms 
is the famous derivative discontinuity of the xc potential 
\be
\Delta_{\rm xc}(N) = \lim_{\eta \to 0^+} \left( 
\frac{\delta E_{\rm xc}[n]}{\delta n(\vr)}\bigg\vert_{N+\eta} - 
\frac{\delta E_{\rm xc}[n]}{\delta n(\vr)}\bigg\vert_{N-\eta}  \right) \;.
\label{xc_discont}
\ee
The total discontinuity is the sum of the two, i.e.,
\be
\Delta(N) = \Delta_s(N) + \Delta_{\rm xc}(N).
\label{fund_gap}
\ee

The xc discontinuity not only gives an important contribution to the 
fundamental gap of semiconductors and insulators,
but it is also a highly relevant property of the exact xc energy functional 
in other situations. For instance, it is exactly the xc discontinuity which 
ensures that heteronuclear molecules dissociate into fragments with 
integer electron numbers
\cite{Perdew:85-2,RuzsinszkyPerdewCsonkaVydrovScuseria:06,FuksMaitra:14}. The development 
of the xc discontinuity from solvable systems with fractional electron number 
as the fraction $\eta$ approaches zero has been studied in 
Refs.~\cite{SagvoldenPerdew:08,GoriGiorgiSavin:09}. 
The importance of the derivative discontinuity has emerged in other contexts too. For 
instance, the exact solution of the one-dimensional Hubbard model 
shows that at zero temperature the (uniform) xc potential is 
discontinuous as the number of particles per site crosses unity  
(half-filling)~\cite{LimaSilvaOliveiraCapelle:03,KVOC.2011,CapelleCampo:13}. 
In more than one-dimension it has been shown that the discontinuity 
occurs only for sufficiently strong interactions and that the critical 
value of the interaction strength is the same as that of the 
Mott-Hubbard transition~\cite{KarlssonPriviteiraVerdozzi:11,KKPV.2013}.

Popular density functional approximations such as LDA or GGA's exhibit a 
vanishing derivative discontinuity, at least in the way they are commonly 
used. It has been pointed out only recently, however, that even for these 
functionals one can construct derivative discontinuities 
\cite{KraislerKronik:13}: to do so one interprets the corresponding energy 
functionals as orbital functionals in an ensemble DFT framework and 
consequently calculates the xc potential within the optimized effective 
potential framework. There are also density functional approximations which do exhibit a 
derivative discontinuity, most prominently perhaps explicitly orbital 
dependent functionals such as the exact-exchange functional. 
Other density functional approximations based on best-fitting are available for the xc 
potential of the one-dimensional Hubbard 
model~\cite{LimaSilvaOliveiraCapelle:03,CapelleCampo:13}. In this 
case one can also show  how the discontinuous steps emerge in the low temperature 
limit~\cite{xctk.2012}. 

In the next Section we consider a different 
class of interacting models showing a discontinuity in the xc 
potential. Then, in Sec.~\ref{lb_dft_strong}, we will use the 
LB+DFT approach to study this class 
of models in a quantum transport setup.

\subsection{The derivative discontinuity in a few illustrative examples}

The importance of the derivative discontinuity appears in the description of 
systems whose electron number can fluctuate. This is precisely the 
situation of junctions (the central region $C$ of Sec.~\ref{lb_dft})
connected to leads. In the limit of vanishing contacts the 
equilibrium properties of the junction 
are given by the grand canonical partition function. 
Below we will use the grand canonical  generalization of DFT by Mermin~\cite{m.1965} 
in two paradigmatic model systems and  
provide a somewhat different perspective 
on the derivative discontinuity.

\subsubsection{Single Site Hubbard Model}
\label{ssm_sec}

Our first model for a quantum dot consists of a single level with 
on-site energy $v$ which can hold up to two electrons \cite{sk.2011}. It is 
described by the Hamiltonian
\be
\hat{H}^{\rm dot} = v \hat{n} + U \hat{n}_{\uparrow} \hat{n}_{\downarrow}
\label{single_dot}
\ee
where $U$ is the charging energy, $\hat{n}_{\sigma} = \hat{d}^{\dagger}_{\sigma} 
\hat{d}_{\sigma}$  is the number operator for electrons of spin $\sigma$ on 
the dot, $\hat{n}=\hat{n}_{\uparrow}+\hat{n}_{\downarrow}$, and 
$\hat{d}^{\dagger}_{\sigma}$ and $\hat{d}_{\sigma}$ are the 
corresponding electron creation and annihilation operators, respectively.

The KS Hamiltonian of such a system is 
\be
\hat{H}^{\rm dot}_{s} = v_{s} \hat{n} \;. 
\label{single_dot_nonint}
\ee
According to Mermin's finite-temperature version of DFT \cite{m.1965} 
there exists a unique potential $v_{s}$ for which this KS Hamiltonian yields, 
for a given temperature and chemical potential, the same density as the interacting 
Hamiltonian.
We write  $v_s = v +v_{\rm Hxc}[n]$ 
where the Hxc potential $v_{\rm Hxc}[n]$ of the single site model (SSM) 
is a function of the dot density $n$. Let us derive the 
exact functional form of the SSM Hxc potential. 

\begin{figure}[t]
\centerline{
\includegraphics[width=0.47\textwidth]{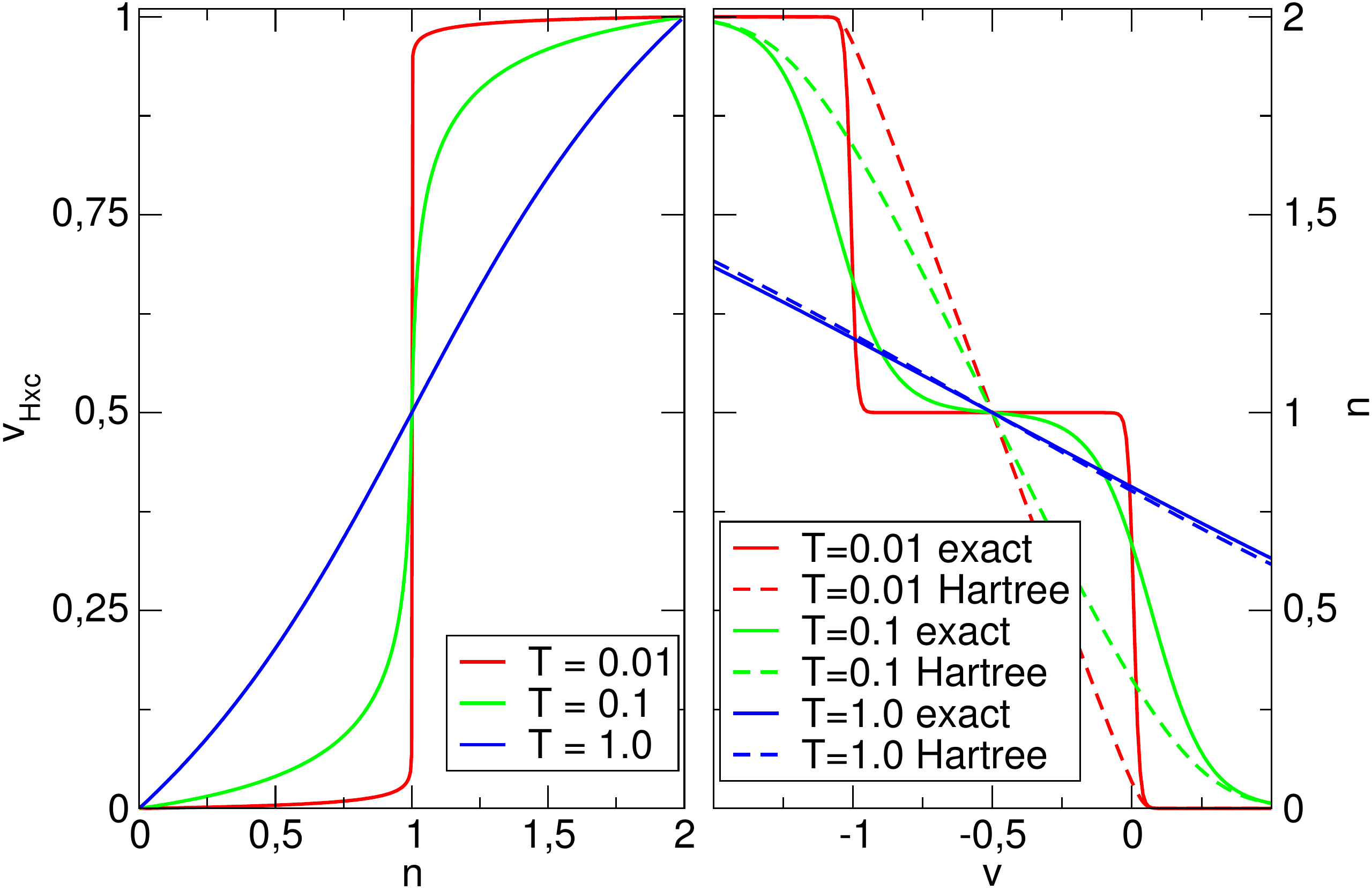}}
\caption{Exact Hxc potential (left panel) and 
density (right panel)  of the single 
site model for different temperatures. 
The density is calculated using the exact Hxc potential (solid lines) 
and, for comparison, the Hartree potential (dashed lines).
Energies are given in units of $U$.}
\label{vhxc_ssm}
\end{figure}

We start by observing that for both the interacting and KS 
Hamiltonian, see Eqs.~(\ref{single_dot}) and (\ref{single_dot_nonint}), the 
eigenstates for electron occupation zero, one, or two are: 
$|0\rangle$, $|\uparrow \rangle$, $|\downarrow \rangle$, and 
$|\uparrow \downarrow \rangle$. The eigenvalues of $\hat{H}^{\rm dot}$ 
corresponding to these states are $0$, $v$, $v$, and 
$2 v+U$, respectively, while for
$\hat{H}^{\rm dot}_{s}$ the eigenvalues are $0$, $v_{s}$, $v_{s}$, 
and $2 v_{s}$. We now consider the single site at thermal equilibrium 
in contact with a bath at inverse temperature $\beta=1/T$ and chemical 
potential $\mu$. The total density on the interacting dot reads
\bea 
n(\tilde{v}) &=& \frac{1}{Z}\tr{\exp(-\beta (\hat{H}^{\rm dot} - \mu \hat{n}) 
\hat{n})} \nn\\
&=& \frac{2}{Z}\left[ \exp(-\beta \tilde{v}) + \exp(-\beta(2 \tilde{v} + U)) 
\right],
\label{dens_pot_ssm}
\eea
where we have defined $\tilde{v}=v-\mu$ and the partition function 
\be
Z = 1 + 2 \exp(-\beta \tilde{v}) + \exp(-\beta(2 \tilde{v} + U)) \;.
\ee
For the KS dot the result simply is
\be
n_{s}(\tilde{v}_{s}) = 2 f(v_{s}) 
\label{dens_pot_ssm_nonint}
\ee
with $\tilde{v}_{s}=v_{s}-\mu$. Both density-potential relations 
(\ref{dens_pot_ssm}) and (\ref{dens_pot_ssm_nonint}) can be inverted 
analytically which allows us to write the exact Hxc 
potential of the single-site model as
\be
v_{\rm Hxc}[n] = \tilde{v}_{s}[n] - \tilde{v}[n] =
\frac{U}{2} +g_{U}(n-1),
\label{hxc_pot_ssm}
\ee
where
\be
g_{U}(x) = \frac{U}{2} + \frac{1}{\beta}
\ln \left( \frac{x + \sqrt{x^2 + e^{-\beta U}(1-x^2)}}{1+x}
\right).
\label{gu-function}
\ee
This is an odd function of its argument, $g_{U}(-x)=-g_{U}(x)$, and therefore
$v_{\rm Hxc}[n=1]=\frac{U}{2}$ for all temperatures. In the left panel of 
Fig.~\ref{vhxc_ssm} we show the Hxc potential (\ref{hxc_pot_ssm}) 
for various temperature. The most prominent feature is the 
(smoothened) step at half-filling ($n=1$) and low temperatures. In the zero 
temperature limit, the Hxc potential approaches a discontinuous step 
function $v_{\rm Hxc}(n) \stackrel{T\rightarrow 0}
{\longrightarrow} U \theta(n-1)$ where $\theta(x)$ is the Heaviside step 
function. This step is nothing but the derivative discontinuity at zero 
temperature discussed above. Hence, the discontinuity 
naturally emerges in the zero-temperature limit of Mermin's formulation of 
DFT. At finite temperature, 
on the other hand, the discontinuity is thermally broadened. Thermal 
broadening is not the only mechanism to smoothen the step in the xc potential. 
As we will discuss below, particularly relevant for transport is the broadening 
due to the presence of contacts.

As a pedagogical illustration of the consequences of the step feature in the xc 
potential, we calculate the density of the single site as function of the 
on-site potential $v$ from solving the exact KS self-consistency condition 
\be
n = 2 f(v+v_{\rm Hxc}[n]) 
\label{ssm_dens_sc}
\ee
as well as the KS self-consistency condition at the Hartree level
\be
n = 2 f(v+v_{\rm H}[n]) 
\label{ssm_dens_sc_hartree}
\ee
where the SSM Hartree potential
\be
v_{\rm H} [n]= \frac{1}{2} U n \;
\label{hartree_pot_ssm}
\ee
is a linear (and hence continuous) function of $n$. 
The right panel of Fig.~\ref{vhxc_ssm} shows the results (for convenience we 
choose $\mu=0$). The most striking difference is the plateau feature of 
the exact density in the low-temperature regime for gate potentials in the 
range $-U \lesssim v \lesssim 0$. This is 
completely missing in the Hartree approximation (or any LDA or GGA for that 
matter) since it is a direct consequence of the step in $v_{\rm Hxc}$.
 For increasing temperatures, 
on the other hand, the Hartree density approaches the exact one 
as the step feature is washed out.

\subsubsection{Constant Interaction Model}
\label{cim_sec}

Our second model is slightly more complicated and is known as the 
Constant Interaction Model (CIM). The CIM Hamiltonian reads
\bea
\hat{H}^{\rm CIM} = \hat{H_0}+\hat{H}_{\rm int} &=&
\sum_{i=1}^M \sum_{\sigma= \uparrow,\downarrow} \varepsilon_{i} 
\hat{n}_{i \sigma} \nn\\
&& + \frac{1}{2} U \sum_{i \sigma \neq j \sigma'} 
\hat{n}_{i \sigma} \hat{n}_{j \sigma'} \;.
\label{cim}
\eea
where $\e_{i}$ are the energies of the single-particle levels and $U$ 
is the repulsive energy between two electrons.
For $M=1$ and $\varepsilon_{1}=v$, the CIM Hamiltonian reduces to the 
SSM Hamiltonian of Eq.~(\ref{single_dot}). Notice that the 
interacting part of the CIM Hamiltonian can be equivalently written 
as 
\be 
\hat{H}_{\rm int} = \frac{1}{2} U \hat{N} (\hat{N}-1),
\label{cim_int}
\ee
where $\hat{N}=\sum_{i\s} \hat{n}_{i\s}$ is the operator of the total number of 
particles.

The KS Hamiltonian of the CIM is 
\be
\hat{H}_s^{\rm CIM} = \hat{H_0} + \Id{r} \; v_{\rm Hxc}[n](\vr) \hat{n}(\vr) 
\label{kshamil_cim}
\ee
with the Hxc potential $v_{\rm Hxc}$ and the density operator 
$\hat{n}(\vr)$ which is related to $\hat{N}$ by 
$\hat{N} = \Id{r} \; \hat{n}(\vr)$. It 
can be shown \cite{StefanucciKurth:13}  that in the limit of zero temperature, 
the exact Hxc potential of the CIM is independent of position and depends only 
on the total number $N$ of electrons in the system. In other words
\be
v_{\rm Hxc}[n](\vr) = v_{\rm Hxc}[N] \;.
\label{xcpot_cim_zero_temp}
\ee
Let us prove this statement.
We denote by $| \Psi^N_{l} \rangle$ the 
linearly independent ground states with $N$ electrons, 
where the index $l=1,\ldots,d_{N}$ and
$d_{N}$ is the degeneracy of the ground-state multiplet of energy $E(N)$.
In the zero-temperature limit of the grand canonical ensemble at chemical 
potential $\mu$, the number of electrons $N$ is the largest integer 
for which the addition energy fulfills  
\be
A(N) = E(N) - E(N-1)= \varepsilon_N + U (N-1) < \mu \;.
\label{cim_affinity}
\ee
The corresponding ground state density reads
\be
n(\vr) = \frac{1}{d_{N}} \sum_{l=1}^{d_{N}} \langle \Psi^N_{l} | 
\hat{n}(\vr) | \Psi^N_{l} \rangle.
\ee
Due to the particular form of the 
interaction $\hat{H}_{\rm int}$, the eigenstates of the CIM 
Hamiltonian are the same as the eigenstates of $\hat{H}_{0}$; hence 
they are many-body Slater determinants 
with every level occupied by either zero or one electron of spin $\s$. This 
implies that
\be
\langle \Psi^N_{l} | 
\hat{n}(\vr) | \Psi^N_{l} \rangle = \sum_{i\s} n_{i\s}(N,l) | 
\q_{i\s}(\vr)|^2
\ee
where $n_{i\s}(N,l)=1$ if $|\Psi^N_{l}\ket$ contains an electron of 
spin $\s$  on 
level $i$ and  zero otherwise. 
We turn now to the  KS Hamiltonian 
in Eq.~(\ref{kshamil_cim}). If the Hxc potential 
is independent of position then the CIM Hamiltonian and the KS 
Hamiltonian have the same eigenstates with the same degeneracies. 
Therefore the ground state densities in the $N$-electron sector are the same 
both in the interacting and in the KS system. Consequently, for given chemical 
potential $\mu$, the only role of the Hxc potential $v_{\rm Hxc}$ is to ensure 
that at zero temperature both the interacting and the KS system have 
the same number of electrons. This is achieved if $N$ is the largest integer 
such that 
\be 
\varepsilon_N + v_{\rm Hxc}[N] < \mu \;.
\ee
Thus, for any real $N$, the explicit form of the CIM Hxc potential at 
zero temperature is 
\be
v_{\rm Hxc}[N] = A(\bar{N}) - \varepsilon_{\bar{N}} = U (\bar{N}-1)
\label{hxcpot_cim_zerotemp}
\ee
where $\bar{N}={\rm Int}[N]$ is the integer part of $N$ and we have used 
Eq.~(\ref{cim_affinity}) in the last step. We conclude that the CIM Hxc 
potential is piecewise constant with discontinuities of height $U$ 
whenever $N$ crosses an integer.

At finite temperature, the CIM Hxc potential in general is more complicated. 
However, if all the single-particle levels $\varepsilon_i$ are degenerate, it 
is again possible to write it as a position-independent constant depending 
only on the total number $N$ in the system. Note that now 
$N=\tr{\exp(-\beta (\hat{H}^{\rm CIM}-\mu))\hat{N}}/
\tr{\exp(-\beta (\hat{H}^{\rm CIM}-\mu))}$ is  to be understood as the 
thermal average of the particle number operator $\hat{N}$. 

\begin{figure}[t]
\includegraphics[width=0.47\textwidth]{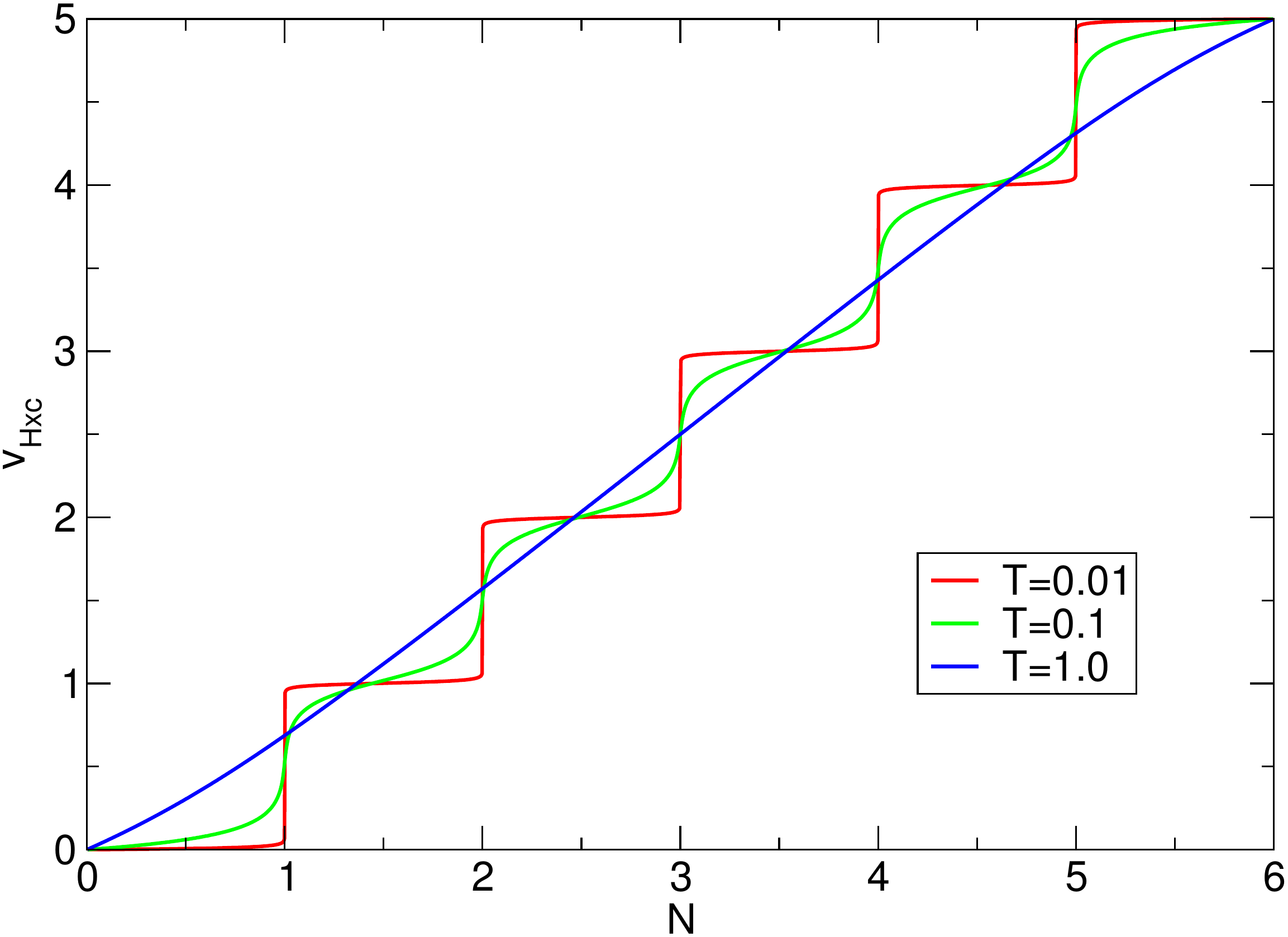}
\caption{Hxc potential of the CIM  with three 
degenerate  levels for different temperatures 
$T=1/\beta$. Energies in units of $U$.}
\label{vhxc_cim_3lev}
\end{figure}

We have constructed the finite temperature Hxc potential of a 6-level CIM with 
degenerate single-particle levels by numerical reverse engineering where we 
have used the scheme described in \ref{cim_uncontacted_exact} to 
calculate the equilibrium occupation of the CIM. The resulting Hxc potential 
of this construction is shown in Fig.~\ref{vhxc_cim_3lev}. The step structure 
at low temperature is apparent and in the zero-temperature limit approaches 
our result of Eq.~(\ref{hxcpot_cim_zerotemp}), as it should be. As in the case 
of the isolated site, at high temperatures the steps are washed out and the 
Hxc potential approaches a linear function of $N$.

\subsection{Successes and limitations of the LB+DFT formalism for strongly correlated transport}  
\label{lb_dft_strong}

\subsubsection{The single impurity Anderson model}
\label{siam-sec} 

In Sec. \ref{ssm_sec} we have considered an 
uncontacted single-level quantum dot. In order to study electronic 
transport through this 
dot we connect it to a left ($L$) and right ($R$) lead described as 
one-dimensional, semi-infinite tight binding chains. The resulting model is 
the celebrated single-impurity Anderson model (SIAM) \cite{Anderson:61} for 
which the Hamiltonian reads
\be
\hat{H}^{\rm SIAM} = \hat{H}^{\rm dot} + \sum_{\alpha = L,R} \hat{H}_{\alpha} + 
\hat{H}_{\rm T} \;. 
\label{siam_hamil}
\ee
Here, the tight-binding lead $\alpha=L,R$ is described by 
\be
\hat{H}_{\alpha} = \sum_{\sigma} \sum_{i=1}^{\infty} \left( V 
\hat{c}^{\dagger}_{i+1 \alpha, \sigma} \hat{c}_{i \alpha, \sigma} + {\rm H.c.} \right)
\ee
and the tunneling Hamiltonian connecting the dot to the leads is 
\be
\hat{H}_{\rm T} = \sum_{\alpha=L,R} \sum_{\sigma} \left( V_{\rm link} 
\hat{c}^{\dagger}_{1 \alpha, \sigma} \hat{d}_{\sigma} + {\rm H.c.} \right) 
\ee
where we have assumed symmetric coupling to left and right leads. The 
embedding self energy for one-dimensional tight-binding leads is known 
analytically. However, here we concentrate on half-filled leads in the 
parameter regime $V_{\rm link} \ll V$, the so-called wide-band limit (WBL). In 
this limit the embedding self energy for lead $\alpha$ becomes a purely 
imaginary, energy independent constant $\gamma_{\alpha} = 2 V_{\rm link}^2/V$  
and the only relevant energy scale for electron tunneling is 
$\gamma = \gamma_L + \gamma_R$. 

For the SIAM the single-particle density matrix $\bgr$ is a $1\times 
1$ matrix whose only entry is the value of the density on the dot. 
Therefore, the self-consistency condition of Eq.~(\ref{dens_matrix_cc_noneq}) 
becomes a nonlinear equation for the only unknown  $\bgr=n$. Taking 
advantage of the WBL nature of the leads it is easy to show that 
in thermal equilibrium (no bias) Eq.~(\ref{neqdensity}) becomes
\be
n = \frac{1}{\pi} \int_{-\infty}^{\infty} {\rm d}\w \; f(\w) 
\frac{\gamma}{(\w - v - v_{\rm Hxc}[n])^2 + \left( \frac{\gamma^2}{4} \right) } 
\; .
\label{siam_dens_sc}
\ee
This equation correctly reduces to Eq.~(\ref{ssm_dens_sc}) in the 
limit of vanishing contacts, i.e., $\g\to 0$.  While for the isolated dot 
the analytic form of the xc potential is known exactly, see 
Eq.~(\ref{hxc_pot_ssm}), this is not the case for the contacted dot. 
However, one may still use $v_{\rm Hxc}[n]$ of 
Eq.~(\ref{hxc_pot_ssm}) as an approximation to the exact xc potential 
if $\g\ll T$. 
With the self-consistent density $n$ solving Eq.~(\ref{siam_dens_sc}) 
we can 
easily compute the KS zero-bias conductance via Eq.~(\ref{lb_conductance})
\be
\frac{G_{s}}{G_0} = - \int_{-\infty}^{\infty} {\rm d} \w\; 
\frac{\partial f(\w)}{\partial \w} 
\frac{\frac{\gamma^2}{4}}{(\w - v - v_{\rm Hxc}[n])^2 + 
\frac{\gamma^2}{4}}.
\label{kscond_siam_wbl}
\ee
where $G_{0}=1/\p$ is the quantum of conductance.

\begin{figure}[t]
\includegraphics[width=0.47\textwidth]{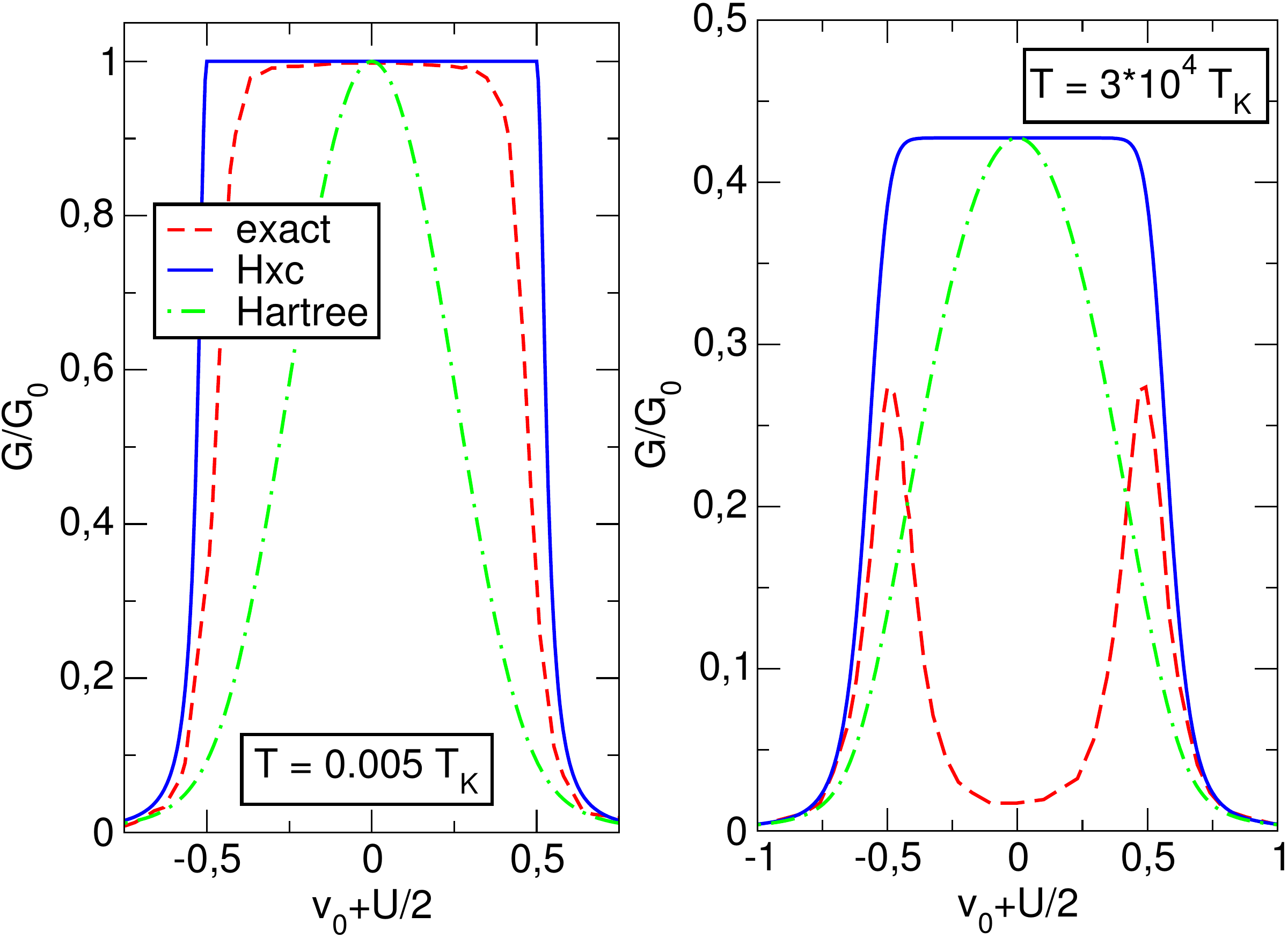}
\caption{KS zero-bias conductances for the Hartree and 
SSM approximations (Eqs.(\ref{hartree_pot_ssm}) and (\ref{hxc_pot_ssm}), 
respectively) as function of the on-site gate potential in comparison 
with NRG results \cite{IzumidaSakai:05} at two different temperatures. 
Reprinted (adapted) with permission from  
Ref.~\protect\onlinecite{sk.2011}. Copyright (2011) American Physical Society. 
Notice that the small difference between this figure and the one in 
Ref.~\protect\onlinecite{sk.2011} is due to a slightly different definition 
of the $T_{\rm K}$ (Eq.~(\protect\ref{tkondo})) used in the present work. 
}
\label{kscond_siam_ssm}
\end{figure} 

In Fig.~\ref{kscond_siam_ssm} we show $G_{s}$ for two different temperatures 
and compare it to the SIAM conductance $G$ as obtained from numerical 
renormalization group (NRG) 
calculations~\cite{IzumidaSakaiSuzuki:01,IzumidaSakai:05}. In the left 
panel, the temperature is much smaller than 
 the Kondo temperature~\cite{JakobsPletyukhovSchoeller:10}
\be
T_{\rm K} = \frac{4 \sqrt{\gamma U}}{\pi} \exp\left( - \frac{\pi}{4} \left( 
\frac{U}{\gamma} - \frac{\gamma}{U} \right) \right),
\label{tkondo}
\ee
whereas in the right panel $T\gg T_{\rm K}$.
Below the Kondo temperature  the most 
prominent feature of $G_{s}$ and $G$  
is the  plateau at one quantum of conductance $G_0$ in the 
region of gate potentials $-U \lesssim v \lesssim 0$. 
Physically
this plateau is due to the screening of the single electron spin at the 
impurity by a spin cloud of opposite-spin electrons at the interface 
of both leads, a phenomenon known as Kondo effect.
In the context of DFT, the plateau 
originates from the pinning of the the KS level to the Fermi 
energy which is a direct consequence of the use of an Hxc potential with a 
step at half filling. This is confirmed by the absence of the conductance 
plateau in DFT calculations using the Hartree potential 
of Eq.~(\ref{hartree_pot_ssm}). Furthermore, we observe that in the DFT calculation 
using $v_{\rm Hxc}$, the conductance plateau extends further and 
terminates more abruptly than in the reference NRG calculation. 

For $T \gg T_{\rm K}$, see right panel of Fig.~\ref{kscond_siam_ssm}, DFT even 
qualitatively disagrees with the  NRG result. While the NRG 
conductance now clearly shows two peaks at $v\approx-U$ and $v 
\approx 0$ due to 
Coulomb blockade, the DFT calculation still exhibits a conductance plateau, 
although at a value smaller than $G_0$. Below we discuss in some detail 
the reasons for the success of the LB+DFT approach at low temperatures and 
its failure at high ones.

At first sight it may come as a surprise that features of complicated 
many-body physics such as the Kondo effect 
can qualitatively be captured with 
a simple DFT model and an explanation is called for. A first explanation 
can be gleaned from the Meir-Wingreen formula of the zero bias 
conductance~\cite{MeirWingreen:92}
\be
G=-\frac{\g}{2}\int \frac{{\rm d} \omega}{2 \pi} \;
f'(\omega) \, A(\omega)
\label{mwgfiniteT}
\ee
with $A(\w)$ the interacting spectral function.
At $T=0$ the Meir-Wingreen formula gives 
\be
\frac{G}{G_0} = \frac{\gamma}{2} |\blG(\mu)|^2 \left( \frac{\gamma}{2} - 
\Im \bgS(\mu) \right)
\label{meir_wingreen_cond}
\ee
where $\blG^{-1}(\w) = [\w - v - \Sigma(\w) + i \frac{\gamma}{2}]$ is 
the $1\times 1$ interacting Green's function and $\Sigma(\w)$ is the many-body 
self-energy. At 
the Fermi energy, quasiparticles have an infinite lifetime, i.e., we have 
$\Im \Sigma(\mu)=0$. Therefore one can see from Eq.~(\ref{meir_wingreen_cond}) 
that it is indeed possible to reproduce the exact conductance from a 
KS calculation if the KS potential at the impurity is 
$v_{s} = v + \Re \Sigma(\mu)$. A second explanation that LB+DFT can give the 
exact zero-bias conductance can be found in the Friedel sum rule 
\cite{Langreth:66,mera-1,mera-2} which states that at zero temperature the 
zero-bias conductance of the SIAM is fully determined by the ground 
state density at the impurity. Since exact DFT by construction gives the exact 
ground state density, it therefore also must yield the exact zero-bias 
conductance, including the conductance plateau due to the Kondo effect. 
The argument of the Friedel sum rule in the context of the LB+DFT conductance 
for the SIAM has independently been discussed in 
Refs.~\cite{sk.2011,blbs.2012,tse.2012}. 

\begin{figure}[t]
\includegraphics[width=0.47\textwidth]{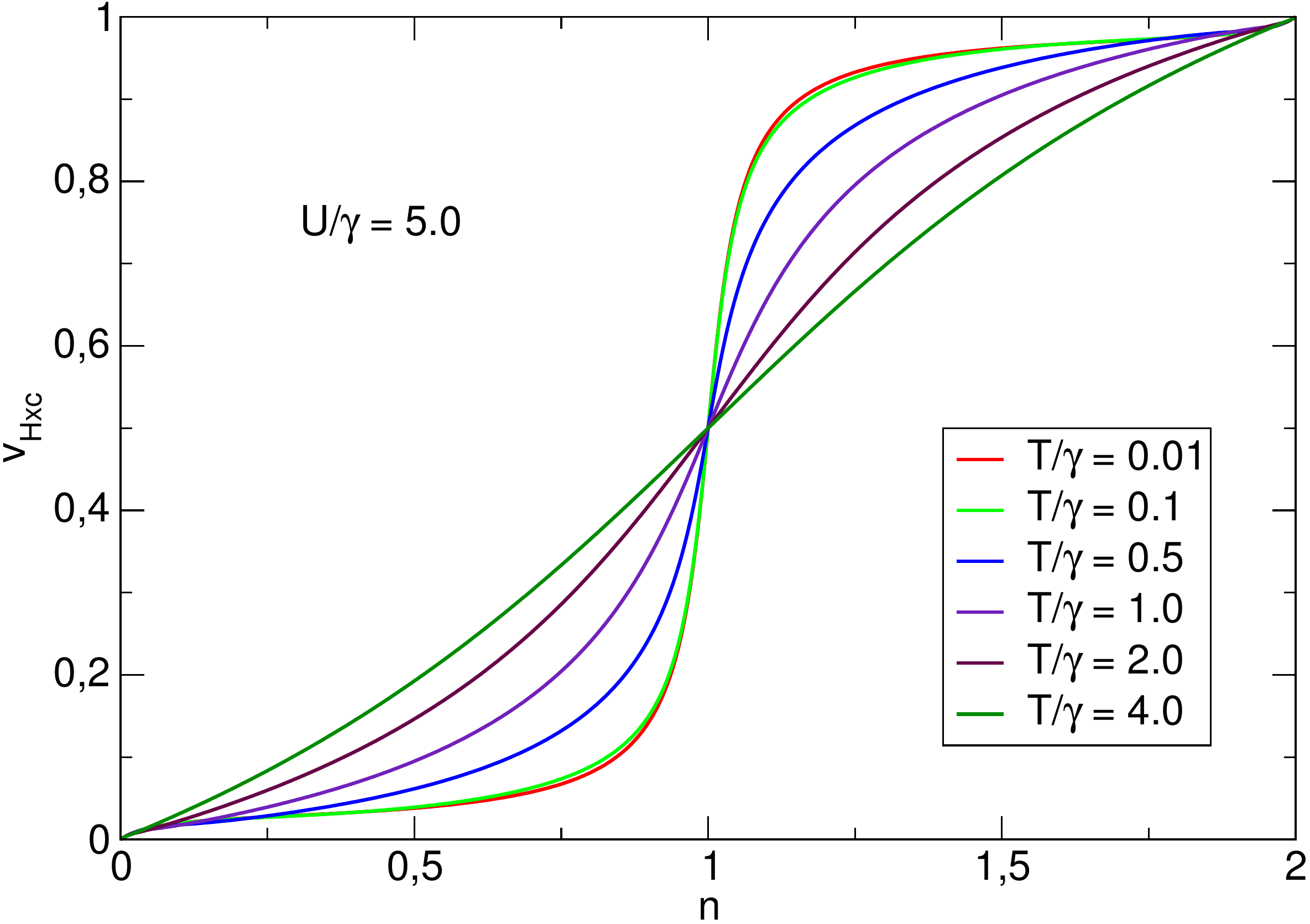}
\caption{Hxc potential (in units of $U$) from Eq.~(\ref{xc_mod_smooth}) for $U/\g=5$ 
and different values of the ratio $T/\g$. }
\label{xcpot_ssm_sm_U5p0g}
\end{figure}

The above arguments also make clear where the small difference between 
the NRG and KS conductances at $T \ll T_{\rm K}$ come from.
The Hxc potential used  in our calculation 
does not contain any information related to the contact to the leads. In 
other words the $v_{\rm Hxc}$ of Eq.~(\ref{hxc_pot_ssm}) is the exact SIAM Hxc potential 
only for $\g\to 0$. 
However, the contacts are responsible for a broadening of the step 
structure and hence $v_{\rm Hxc}$ should exhibit a smeared step 
for finite $\g$ even at $T=0$. To include the broadening due to the 
contacts in the Hxc potential we proceed as follows. The exact 
spectral function of the isolated dot reads
\be
A(\omega) = \left( 1 - \frac{n}{2} \right) 
\delta\left( \omega - v \right) 
+ \frac{n}{2}\delta\left( \omega - v - U \right) \;.
\label{spectral_ssm}
\ee
Broadening  the delta-function $\d(\w)$ into a Lorentzian
$\ell_{\g}(\w)=\g/(\w^{2}+\g^{2}/4)$
of width $\gamma$ we obtain the model many-body (MB) spectral 
function
\be
A^{\rm mod}(\omega) = \left( 1 - \frac{n}{2} \right) 
\ell_{\g}(\w-v)
+ \frac{n}{2}\ell_{\g}(\w-v-U)
\;.
\label{spectral_model}
\ee
With this model spectral function the interacting density as function of 
$\tilde{v}=v-\mu$ is 
\be
n(\tilde{v}) = 2 \int \frac{{\rm d} \omega}{2 \pi} f(\omega) A^{\rm 
mod}(\omega) ,
\label{siam_mod_dens}
\ee
which can be inverted numerically to give $\tilde{v}[n]$. Similarly, for a 
non-interacting impurity with on-site potential $v_s$, through 
Eq.~(\ref{siam_mod_dens}) for $U=0$ the density becomes a function of 
$\tilde{v}_s=v_s-\mu$ which again can be inverted numerically to give 
$\tilde{v}_s[n]$. From these results we then obtain a numerical model 
xc potential for the coupled impurity through 
\be
v_{\rm Hxc}[n] = \tilde{v}_s[n] - \tilde{v}[n] \;. 
\label{xc_mod_smooth}
\ee
This Hxc potential reduces to the $v_{\rm Hxc}$ of Eq.~(\ref{hxc_pot_ssm}) for $\g\to 0$ 
and it has the nice feature of smearing the step at half filling with 
a width $\approx \g$ for $T\ll\g$ and a width $\approx T$ for 
$T\gg\g$. The quantitative effects of thermal broadening versus contact broadening 
are illustrated in Fig.~\ref{xcpot_ssm_sm_U5p0g}. We wish to emphasize that independently 
of the nature of the broadening the qualitative behavior of the 
density does not change: in both cases $n$ exhibits a plateau of 
height $1$ (half-filling) as 
function of $v$ since $v_{\rm Hxc}$ pins the KS level to the chemical 
potential.

At zero temperature a reasonable fit to the model Hxc potential
of Eq.~(\ref{xc_mod_smooth}) is given by 
the  expression 
\be
v_{\rm Hxc}[n] = \frac{U}{2}\left( 1 +  \frac{2}{\pi} 
\arctan\left(\frac{n-1}{W}\right) \right)
\label{xc_mod_smooth_fit}
\ee
where the fitting parameter $W$ is defined as 
\be
W = 0.16 \frac{\gamma}{U} \;.
\label{broadening}
\ee
 We will use this form at various places throughout this review. 
The (numerically) exact Hxc potential for the Anderson model at zero 
temperature has been constructed by a reverse engineering procedure using 
DMRG methods in Refs.~\onlinecite{blbs.2012,LiuBergfieldBurkeStafford:12} 
where the authors also give accurate parametrizations of their results. 

\begin{figure}[t]
\includegraphics[width=0.49\textwidth]{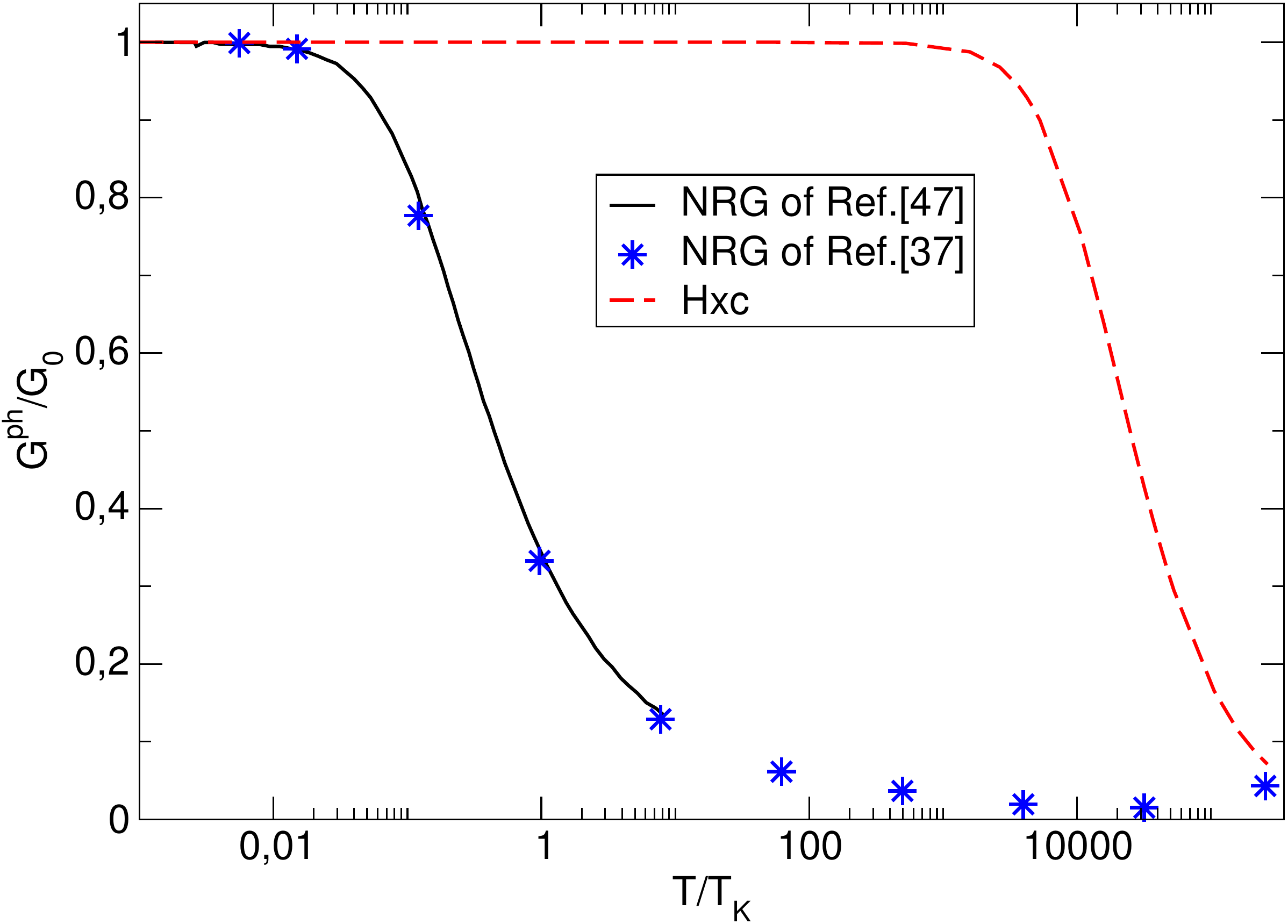}
\caption{NRG conductances from Ref.
\cite{IzumidaSakai:05} [stars(blue)] and Ref.
\onlinecite{Costi:00} [solid(black)] against the exact Hxc
conductance [dashed(red)] at the ph symmetric point versus
temperature. Reprinted (adapted) with permission from  
Ref.~\protect\onlinecite{sk.2011}. Copyright (2011) American Physical Society. 
Notice that the small difference between this figure and the one in 
Ref.~\protect\onlinecite{sk.2011} is due to a slightly different definition 
of the $T_{\rm K}$ (Eq.~(\protect\ref{tkondo})) used in the present work.}
\label{cond_ph}
\end{figure}

We now turn to the discussion of our results for temperatures $T \gg T_{\rm K}$. 
From the right panel of Fig.~\ref{kscond_siam_ssm} we see that 
the KS conductance even qualitativeley 
disagrees with the correct NRG result: instead of the Coulomb blockade 
peaks the DFT results still show a plateau, although at values smaller than 
$G_0$. In fact, both arguments given to explain the correct DFT description 
at zero temperature fail at finite temperature. When calculating the 
conductance from the Meir-Wingreen formula at finite temperature, we cannot 
restrict the discussion of the many-body Green's function to the Fermi 
energy alone and thus the argument given above does not apply. Similarly, 
the validity of the Friedel sum rule is restricted to zero temperature. 
Therefore it is not surprising that the DFT conductance does not give 
the correct physics. This can explicitly be confirmed by considering a 
special value for the gate potential, the so-called particle-hole (ph) 
symmetric point at $v=-U/2$. At this value of the gate, the impurity is 
at half-filling ($n=1$) for all temperatures. In the DFT framework this means 
that the exact Hxc potential must have the value  
$v_{\rm Hxc}[n=1]=\frac{U}{2}$ such that the total KS potential vanishes. 
Our model Hxc potential in Eq.~(\ref{xc_mod_smooth})
correctly satisfies this condition, i.e., the 
value of $v_{\rm Hxc}$ for $n=1$ is {\em exact} for all 
temperatures and interaction strengths $U$. At the 
ph symmetric point, the conductance $G^{\rm ph}$ of the interacting Anderson 
impurity is a universal function of $T/T_{\rm K}$ which is known numerically
\cite{Costi:00,AleinerBrouwerGlazman:02}. Therefore at the ph symmetric point 
we can compare the exact KS conductance $G^{\rm ph}_{s}$ with the 
exact $G^{\rm ph}$. The result is shown in Fig.~\ref{cond_ph}. We see that 
while the exact KS conductance is correct at zero (and very low) 
temperature, it is widely off the mark over a wide temperature range. 

\subsubsection{Constant Interaction Model coupled to leads}
\label{cim_trans} 

In the previous Section we have seen what the LB+DFT formalism can and cannot 
describe for the Anderson model. In the present one we move to multi-level 
systems, in particular systems described by the CIM of Sec.~\ref{cim_sec}. 
There we have  seen that the Hxc potential at zero 
temperature is independent of position and depends only on the total number 
$N$ of electrons on the (multi-level) quantum dot. Thus, in the basis 
of the single-particle eigenstates, the CIM KS Hamiltonian of 
Eq.~(\ref{kshamil_cim}) can be written as 
\be
\hat{H}_s^{\rm CIM} = \sum_{i=1}^M \left( \varepsilon_i + v_{\rm Hxc}[N] \right) 
\hat{n}_i
\ee
where $\hat{n}_i = \sum_{\sigma=\uparrow,\downarrow}\hat{n}_{i\sigma}$
and $M$ is the number of levels.
In order to study the conductance properties of  this system 
we have to connect it to leads. The 
non-interacting leads are described by the Hamiltonian
\be
\hat{H}_{\rm lead} = \sum_{k \sigma} \sum_{\alpha=L,R} 
\varepsilon_{k \alpha} \hat{c}^{\dagger}_{k \sigma \alpha} \hat{c}_{k \sigma \alpha}
\ee
where the $\hat{c}_{k \sigma \alpha}$ ($\hat{c}^{\dagger}_{k \sigma \alpha}$) create 
(annihilate) an electron with energy $\varepsilon_{k \alpha} $ and spin 
$\sigma$ in lead $\alpha$. The tunneling 
Hamiltonian takes the usual form, i.e., 
\be
\hat{H}_{\rm T} = \sum_{k \sigma \alpha} \sum_{i=1}^M \left( T_{k \alpha,i} 
\hat{c}^{\dagger}_{k \sigma \alpha} \hat{c}_{i \sigma} + H.c. \right) 
\ee
which, from Eq.~(\ref{broadmatx}), leads to the broadening matrix 
$\Gamma_{ii'} = 2 \pi \sum_{k \sigma \alpha} T_{k \alpha,i}^* T_{k \alpha,i'} 
\delta(\omega - \varepsilon_{k \alpha})$.
We will again work in the WBL where the $\Gamma_{ii'}$ become independent of 
frequency. We further consider the $\Gamma_{ii'}$ much smaller than 
the level spacings so that we can approximate 
\be
\Gamma_{ii'}=\g_{i}\d_{ii'}.
\ee
in the dot. This assumption is certainly justified at low enough 
temperatures. At low temperatures  we also expect that the 
essential qualitative feature of $v_{\rm Hxc}$  is a 
series of smeared steps at integer $N$. On this ground,
the approximation we propose 
consists in summing the SIAM Hxc potential over all possible charged 
states of the CIM~~\cite{PS.2012}. Taking into account
Eq.~(\ref{xc_mod_smooth_fit}) the explicit form of the 
approximation reads
\be
v_{\rm Hxc}[N] = \sum_{J=1}^{2M-1} \left( \frac{U_J}{2} + \frac{U_J}{\pi} 
\arctan\left( \frac{N-J}{W_J}\right) \right)
\label{HxcfitCIM}
\ee
where, for later use, we already allowed for charging energies $U_J$ and 
widths $W_J$ which depend on the charging state $J$. In the present 
Section we  consider $\g_{i}=\g$ independent of $i$ 
and use both charging energies $U$ and level broadenings 
$W=0.16\; \gamma/U$ (see Eq.~(\ref{broadening})) independent of the charging 
state $J$. 

We study the zero-bias conductance through 
the multi-level CIM. It is therefore sufficient to solve the KS 
problem in equilibrium. Taking into account that the Hxc potential depends only on 
the total $N$, there is only one self-consistency equation 
\be
N = 2 \int\frac{d\w}{2\p} f(\w) \Tr\left[ \blG(\w)\bgG\blG^{\dag}(\w) \right]
\label{CIM-scN}
\ee
with $\bgG = \bgG_L + \bgG_R$ and the trace is over all single-particle states 
of the quantum dot, 
see Eqs.~(\ref{neqN}).
According to Eq.~(\ref{gcceq}) the KS Green's function is given by 
\be
\blG(\w) = \frac{1}{\w - \blH + i \bgG/2}
\label{KSgreenfunction}
\ee
where the KS single-particle Hamiltonian has matrix elements $[\blH]_{jj'} = 
\delta_{jj'} (\varepsilon_j + v_{\rm Hxc}[N]) =:\delta_{jj'} \varepsilon_{s,j}$ 
with the KS single-particle energies $\varepsilon_{s,j}$. At 
self-consistency we can calculate the KS zero-bias conductance using 
Eq.~(\ref{lb_conductance}).

As a first example of transport through multiple correlated levels described 
by the CIM we consider two spin-degenerate single-particle levels, the HOMO 
and the LUMO, coupled to two wide-band leads. As functions of gate 
potential $v$, the KS HOMO and LUMO leves are given by
\be
\varepsilon_s^{\rm H} = -\frac{\Delta \varepsilon}{2} + v + v_{\rm Hxc}[N]
\ee
and
\be
\varepsilon_s^{\rm L} = \frac{\Delta \varepsilon}{2} + v + v_{\rm Hxc}[N] \;. 
\ee

\begin{figure}[t]
\centerline{
\includegraphics[width=0.47\textwidth]{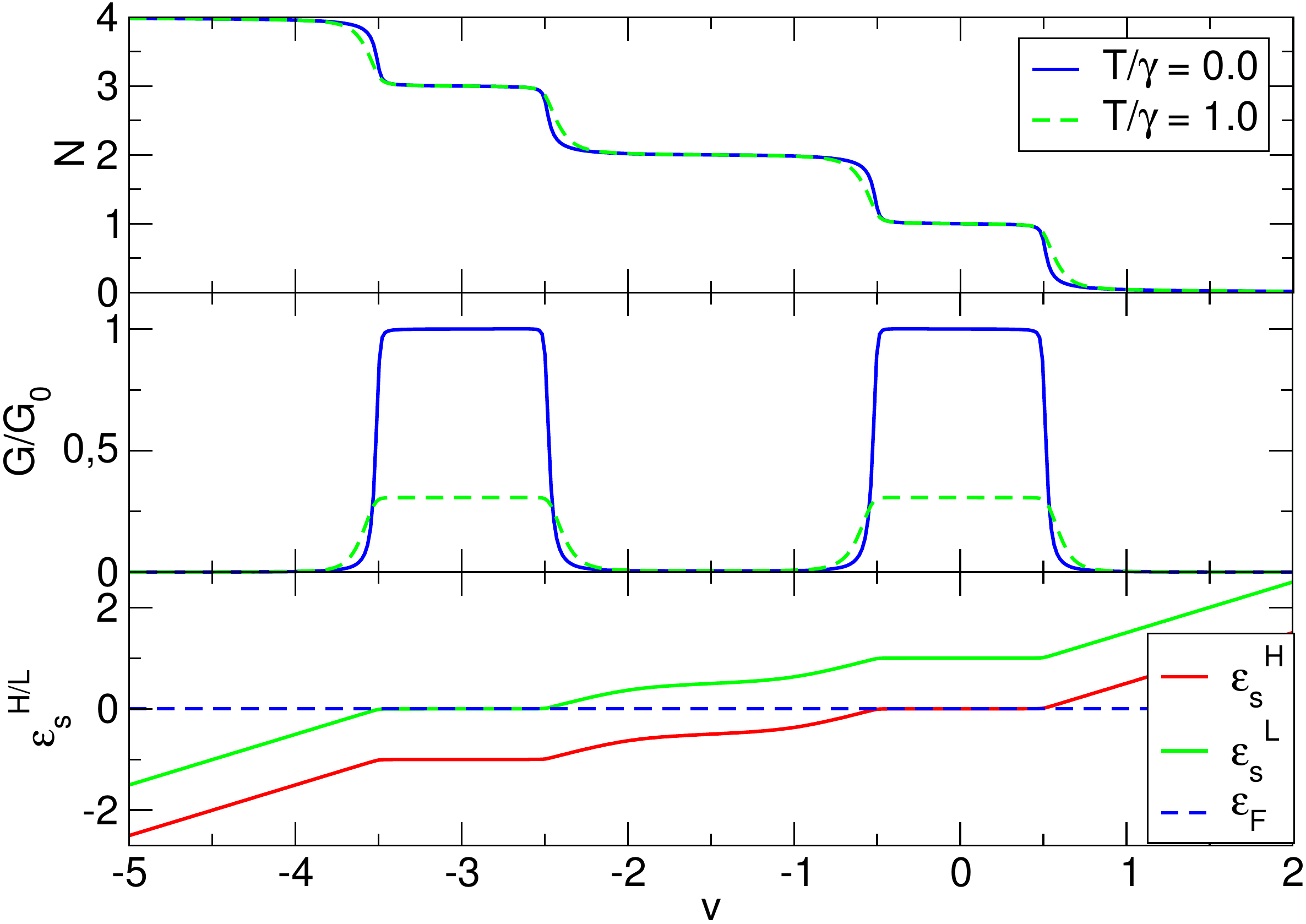}}
\caption{Self-consistent DFT results for the HOMO-LUMO model of two 
spin-degenerate single particle levels coupled to 
two wide-band leads as function of gate potential $v$. The splitting between 
HOMO and LUMO level is $\Delta \varepsilon = 0.5$, the coupling parameter 
to the leads is $\gamma = 0.05$. Energies in units of $U$. 
Upper panel: total occupation for two different temperatures. Middle panel: 
KS conductance for two different temperatures. Lower panel: 
KS HOMO and LUMO eigenvalues, $\varepsilon_s^{\rm H}$ and 
$\varepsilon_s^{\rm L}$, at $T=0$ as well as the Fermi energy 
$\varepsilon_{\rm F}$.}
\label{homo_lumo}
\end{figure}

In Fig.~\ref{homo_lumo} we show zero temperature results for density, 
KS conductance, and KS single-particle levels for the following parameters: 
HOMO-LUMO splitting $\Delta \varepsilon = 0.5$ and  
$\gamma = 0.05$ (all energies are given in units of $U$). At large positive gate potentials, the 
quantum dot is unoccupied (see upper panel). As the gate is lowered, the 
first electron enters the dot when the (KS) HOMO level becomes lower than 
the Fermi energy $\varepsilon_{\rm F}$. Due to the interaction, the second 
electron is blocked from entering the dot at the same gate and only can enter 
when the gate is lowered by $U$. In order for the third electron to enter the 
dot, we have to lower the gate by an additional energy 
$\Delta \varepsilon + U$, i.e., by the single-particle energy difference of the 
levels plus one additional charging energy. Finally, the fourth electron can 
only enter once the gate is lowered by another charging energy $U$. The KS 
conductance (middle panel) at zero temperature shows two regions of gate 
potentials with one 
quantum of conductance $G=G_0$. These regions correspond to those values of 
$v$ with an {\it odd} number of electrons occupying the quantum dot while for 
an even number of electrons the conductance is blocked, exactly the correct 
behavior for the appearance of the Kondo effect in multi-level quantum dots 
\cite{BruusFlensberg:04}. The way how this is achieved within the LB+DFT 
formalism can be deduced by looking at the KS energy levels (lower panel). 
We see that for those gate potentials for which the KS conductance equals 
$G_0$, one of the two KS levels is pinned to the Fermi energy 
$\varepsilon_{\rm F}$. Just like in the SIAM, this pinning leads to one of 
the conductance channels being open, i.e., $G=G_0$, and, of course, is a direct 
consequence of the step feature in the Hxc potential at integer occupation. 
At occupation with even number of electrons, none of the KS levels pins to 
$\varepsilon_{\rm F}$ and the conductance is blocked, despite the fact that 
$v_{\rm Hxc}$ also has steps at even integers of the occupation. As expected from 
our experience with the SIAM, at finite temperature the KS conductance 
qualitatively keeps its $T=0$ shape but with the plateau values now lowered. It 
completely fails to describe the transition from the Kondo to the Coulomb 
blockade regime.

\begin{figure}[t]
\centerline{
\includegraphics[width=0.47\textwidth]{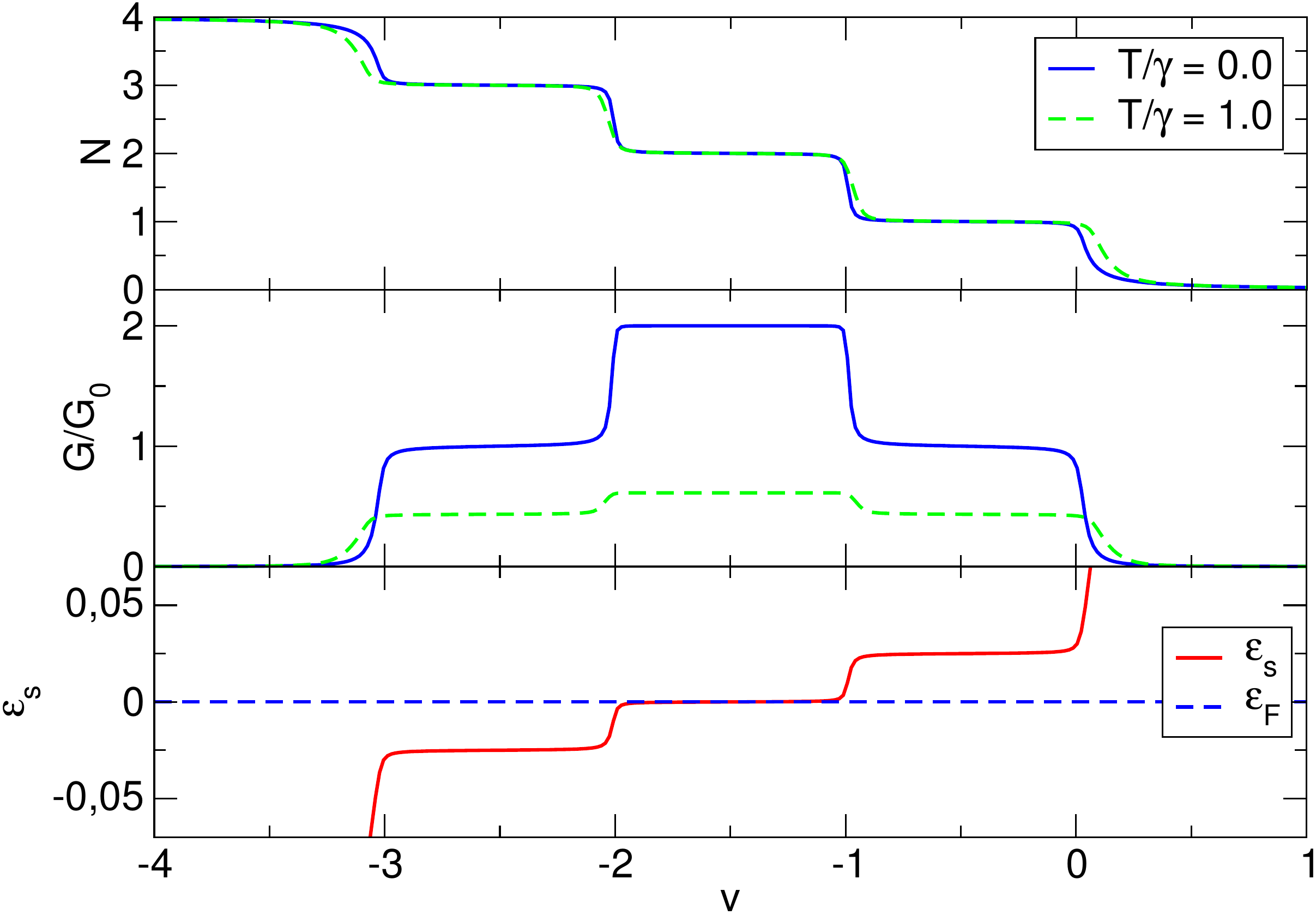}}
\caption{Self-consistent DFT results for the degenerate HOMO-LUMO model 
($\Delta \varepsilon = 0.0$) coupled to two wide-band leads as function of 
gate potential $v$. The coupling parameter to the leads is $\gamma = 0.05$. 
Energies in units of $U$. 
Upper panel: total occupation for two temperatures. Middle panel: KS 
conductance for two temperatures. Lower panel: 
KS eigenvalue $\varepsilon_s$ at $T=0$ and Fermi energy $\varepsilon_{\rm F}$.}
\label{homo_lumo_deg}
\end{figure}

As our second example we again consider the HOMO-LUMO model contacted to two 
leads but now for the case of degenerate HOMO and LUMO levels  
($\Delta \varepsilon=0$) and all other parameters as before. The results are 
shown in Fig.~\ref{homo_lumo_deg}. Again, the occupation exhibits plateaus 
(as function of gate) with integer occupation of the dot. The LB+DFT 
conductance (middle panel), however, now has a different structure. While for 
odd numbers of electrons on the dot, the conductance is still $G_0$, for 
occupation of two we have $G=2 G_0$. This can be easily understood since in 
this case the whole dot is half filled, i.e., both (degenerate) 
single-particle levels are pinned to $\varepsilon_{\rm F}$ (see lower panel) 
and thus two conductance channels are open simultaneously. For those values 
of $v$ where $G=G_0$, the KS level doesn't pin to $\varepsilon_{\rm F}$ but 
to $\pm \gamma/2$ (note the scale on the $y$-axis of the lower panel). 

The examples discussed in this Sec. demonstrate that the LB+DFT  
approach correctly predicts  the zero-temperature Kondo plateaus in $G$ as 
well as the height of the plateaus for degenerate levels. Of course, 
the success of LB+DFT  strongly relies on the use of accurate Hxc 
potentials whose most important feature is the occurrence of steps as the number of 
particles crosses an integer.
However, it is also clear that, just as 
in the SIAM, by increasing the temperature the transition from the Kondo to 
the Coulomb blockade regime  is beyond the 
capabilities of the LB+DFT approach~\cite{StefanucciKurth:13}. 
How to go beyond this approach will be the 
topic of the next two sections.
Before closing this Section we wish to observe that in 
Figs.~\ref{homo_lumo} and \ref{homo_lumo_deg} we used the Hxc 
potential of Eq.~(\ref{HxcfitCIM}) even for $T>\g$.
As we discussed below Eq.~(\ref{xc_mod_smooth}), thermal 
broadening dominates for $T>\g$ and it would be more appropriate to 
use the SSM Hxc potential of Eq.~(\ref{hxc_pot_ssm}). This is precisely what we did in 
the SIAM and, as we can see from Fig.~\ref{kscond_siam_ssm}, the only difference is that 
the width of the flanks of the Kondo plateau is proportional to $T$ 
instead of $\g$. Thus, the qualitative behavior of the total $N$ and 
of the KS conductance is independent of the nature (thermal or due to 
the contacts) of the broadening.

\section{Quantum Transport with Time-Dependent Density Functional Theory}
\label{tddft}

The LB+DFT approach combines the noninteracting steady-state formulation of 
Landauer and B\"uttiker with the ground-state DFT formulation of 
interacting systems. This {\em empirical} combination  
suffers from a  conceptual problem, i.e., the use of an equilibrium
Hxc potential in a nonequilibrium steady-state situation. One way to solve this 
conceptual problem while still remaining in a one-particle framework is 
to reformulate the theory of quantum transport using the Time 
Dependent (TD) version of DFT.  TDDFT allows for 
studying the TD current and density generated by an arbitrary TD bias 
and hence, as a special case, the
real-time evolution of the 
current and density after the switch-on of an external DC bias.
Steady-state quantities can simply be obtained as the 
long-time limit of the TDDFT results.

We consider again the geometry of the previous Section with a central 
region connected to left and right electrodes. The whole
system is initially, say at time $t\leq 0$, in equilibrium 
at a certain temperature and chemical potential. The charge density of the electrodes is 
perfectly balanced and no current flows through the junction.
The system is driven out of equilibrium by an external
electric field,
like the one generated by a battery.
The formation of dipole layers screens the potential
drop along the electrodes and the total potential turns out 
to be uniform in the left and right bulks. Accordingly, the potential 
drop, or bias, is confined in the neighborhood of the junction.

Let $\q_{k}(\blr)$ be a KS eigenstate of the equilibrium system 
with KS energy $\ve_{k}$. 
The time evolution of this state is governed by the time-dependent 
KS equation
\bea
i\frac{\de}{\de t}\q_{k}(\blr,t) =
\left[-\frac{\nabla^{2}}{2}+v_{\rm ext}(\blr,t)+
v_{\rm Hxc}(\blr,t)\right]\q_{k}(\blr,t) \nn\\
\label{tdse}
\eea
where $v_{\rm ext}$ is the external potential, i.e., the sum of the 
nuclear potential and the time-dependent potential of the battery, 
and $v_{\rm Hxc}=v_{\rm H}+v_{\rm xc}$ is the Hartree-xc potential of 
TDDFT. The exact Hxc potential $v_{\rm Hxc}(\blr,t)$
in point $\blr$ at time $t$ depends on the density 
\be
n(\blr',t')=\sum_{k}f(\ve_{k})|\q_{k}(\blr',t')|^{2}
\label{tdn}
\ee
in all points $\blr'$ and at all times $t'<t$. Hence Eq.~(\ref{tdse}) for 
all single-particle indices $k$ together with Eq.~(\ref{tdn}) 
form a nonlinear system of 
coupled differential equations. 

The solution of the differential equations is, in general, a 
difficult task since (i) the 
system is infinitely large (due to the electrodes) and  
spatially non-periodic (due to the junction) and (ii) the 
KS wavefunctions at time $t=0$ are delocalized all over the system. 
Nevertheless, it is 
still  
possible to develop efficient numerical algorithms 
if we take advantage of the uniformity of the time-dependent part of 
the total potential in the left and right bulks. The property of 
uniformity means that
\bea
\lim_{\blr\to\a} \left[v_{\rm ext}(\blr,t)+v_{\rm H}(\blr,t)\right] = 
v_{\rm ext}(\blr,0)+v_{\rm H}(\blr,0)+V_{\a}(t) \nn \\
\label{asymvC}
\eea
where $V_{\a}(t)$ is the (experimentally measured) bias and
\be
\lim_{\blr\to\a} v_{\rm xc}(\blr,t)=
v_{\rm xc}(\blr,0)+V_{\a,\rm xc}(t),
\label{asymvxc}
\ee
where $V_{\a,\rm xc}(t)$ is the xc bias correction predicted by 
TDDFT. In Eqs.~(\ref{asymvC}) and (\ref{asymvxc}) 
the limit $\blr\to\a$ signifies that $\blr$ should be taken deep inside 
lead $\a$. Thus, the time-dependent part of the KS Hamiltonian in 
Eq.~(\ref{tdse}) is spatially constant in the leads. 

In the last 
decade several algorithms have been proposed to obtain the 
one-particle density matrix $\r(\blr,\blr',t)=\sum_{k}f(\ve_{k})
\q^{\ast}_{k}(\blr,t)\q_{k}(\blr',t)$ from which to extract the 
density $n(\blr,t)=\r(\blr,\blr,t)$, see Eq.~(\ref{tdn}), and the 
longitudinal current $I(t)=\int d\blr_{\perp} \,
\Im[(\grad-\grad')\r(\blr,\blr',t)]_{\blr=\blr'}$ (the integral 
is over a surface perpendicular to the current flow). 
Among the methods based on the solution of 
Eq.~(\ref{tdse}) we mention the wavefunction approach with transparent 
boundaries~\cite{ksarg.2005,vsa.2006,spc.2008,spc.2010,kskvg.2010}
and with absorbing 
boundaries~\cite{bsin.2004,v.2011,gwshgw.2014,ww.2016,sk.2016},
the microcanonical approach~\cite{bsdv.2005,cevv.2006,psc.2008,psc.2009},  
the supercell approach~\cite{qlly.2006} and the 
stroboscopic approach~\cite{bcg.2008,kb.2014}.
Other methods are based on the solution of the
Dyson equation for the Green's 
function~\cite{mwg.2006,zwymc.2007,yzcwfn.2011,mgm.2007,pd.2008,cs.2009,PopescuCroy:16,zetal.2010,zcw.2013,tpsvl.2014,cuansing:17},
on the propagation of the Green's function through the
Kadanoff-Baym 
equations~\cite{mssvl.2008,mssvl.2009,pva.2009,pva.2010}, 
on the calculation of Bohm trajectories~\cite{o.2007} and on 
different types of master equations
for the density 
matrix~\cite{sssbht.2006,ly.2007,xetal.2012,zkh.2014,chf.2014}.

In Refs.~\cite{sa-1.2004,sa-2.2004} it has been shown that 
if the bias $V_{\a}$ is constant for large times and {\em if} the KS system attains a 
steady state in the long time limit then 
the steady current flowing through the junction is given by
\bea
I&=&2\int\frac{d\w}{2\p}
\left[f(\w-V_{L,s})-f(\w-V_{R,s})\right] \nn \\
&& \times \Tr\left[ \blG(\w)\bgG_{L}(\w)\blG^{\dag}(\w)\bgG_{R}(\w)
\right].
\label{I-tddft}
\eea
The main difference between Eq.~(\ref{I-tddft}) and the LB+DFT 
formula in Eq.~(\ref{lb_current}) is the appearence of the KS bias $V_{\a,s}$ in the Fermi 
functions. The KS bias is defined according to
\be
V_{\a,s}=\lim_{t\to\iif}\left(V_{\a}(t)+V_{\a,\rm 
xc}(t)\right)=V_{\a}+V_{\a,\rm xc}
\ee
and differs from the physical bias by the addition of an xc 
correction. Another important difference between Eq.~(\ref{I-tddft}) 
and the LB+DFT formula in Eq.~(\ref{lb_current}) lies in the calculation 
of the Green's function $\blG$ and the 
broadening matrix $\bgG_{\a}$. They are calculated as discussed in the previous 
section but the external bias is replaced by the KS bias and the 
Hxc gate is the long-time limit of the 
Hxc potential in the central region. We remind the reader that in the 
LB+DFT scheme the Hxc gate results from a self-consistent calculation of the 
density in the central region and there is no {\em a priori} reason for 
this Hxc gate to be the same as the long-time limit of the 
TDDFT Hxc gate.

It is worth emphasizing that the xc gate and bias in Eq.~(\ref{I-tddft}) are 
functionals of the density {\em everywhere} and {\em at all previous 
times}.

\subsection{Linear response}
\label{lr-tddft.sec}

In this section we work out and discuss the formula for the 
conductance resulting from the TDDFT formulation. To first order in 
the bias the current of Eq.~(\ref{I-tddft}) reads
\bea
\d I&=&-G_{0}(\d V_{L}-\d V_{R}+\d V_{L,\rm xc}-\d V_{R,\rm xc})
\nn\\ &\times&
\int d\w 
f'(\w){\cal T}(\w),
\label{exI}
\eea
where 
\be
{\cal T}(\w)=\Tr\left[ \blG(\w)\bgG_{L}(\w)\blG^{\dag}(\w)\bgG_{R}(\w)
\right]
\ee
is the transmission function calculated with equilibrium (zero bias) DFT 
Green's function $\blG$ and broadening matrix $\bgG_{\a}$. The conductance 
\be
G=\frac{\d I}{(\d V_{L}-\d V_{R})}
\ee
coincides with the KS conductance of the LB+DFT approach, see 
Eq.~(\ref{lb_conductance}), only provided that 
$V_{\a,\rm xc}=0$. In the previous section we showed that the exact KS 
conductance and the exact conductance of the Anderson model 
are different in the Coulomb 
blockade regime. We therefore have to conclude that the xc bias is 
non zero in this case.

From linear-response TDDFT~\cite{ewk.2004,skgr.2007} 
\be
\d V_{\a,\rm xc}=\int 
dt'd{\bf r}' \lim_{t\ra\inf}\lim_{\blr\ra \a} 
f_{\rm xc}({\bf r},{\bf r}';t-t')\d n({\bf r}',t')
\label{dv=fxcdn}
\ee
where $f_{\rm xc}$ is the TDDFT kernel and $\d n({\bf r},t)$ is the 
density variation. 
The assumption of a steady state implies that 
the kernel $f_{\rm xc}\ra 0$ for $|t-t'|\ra\inf$ and 
that $\d n({\bf r},t\ra\inf)=\d n_{\a}$ for ${\bf r}$ deep inside lead $\a$. In 
Eq.~(\ref{dv=fxcdn}) the contribution of the molecular region 
to the spatial integral  is negligible in the thermodynamic 
limit. Hence, it is convenient to define the quantity
\be
f_{\rm xc}^{\a\b}=\int dt'
\int_{\rm lead\;\b} d{\bf r}'_{\parallel}\lim_{\blr\ra \a}
f_{\rm xc}({\bf r},{\bf r}';t')
\ee
and rewrite Eq.~(\ref{dv=fxcdn}) as
\be
\d V^{\a}_{\rm xc}=\sum_{\b=L,R}f_{\rm xc}^{\a\b}\callS_{\b}\d n_{\b},
\ee
where $\callS_{\b}=\int_{\rm lead\;\b} d{\bf r}'_{\perp}$ is the area of 
the transverse section of lead $\b$. Notice that due to charge conservation we 
have
\be
\callS_{L}\d n_{L}=-\callS_{R}\d n_{R}.
\ee

It could be tempting to gain some insight in the behavior of $f_{\rm 
xc}^{\a\b}$ by performing 
equilibrium DFT calculations on leads of finite length and different 
densities. In doing so, however, we would get the equilibrium DFT kernel 
which corresponds to taking the limit $t\ra \inf$ before the limit 
$\blr\to \a$.  In fact, the equilibrium DFT kernel is the response of the
equilibrium xc potential
to a density variation and, deep inside the leads, is
determined by the condition of charge neutrality alone. 
As, in general, the limit $t\ra \inf$ and $\blr\to \a$ do not commute 
we cannot model $f_{\rm xc}^{\a\b}$ using leads of finite length.

Inserting the expression for $\d V^{\a}_{\rm xc}$ into Eq. (\ref{exI}) we find
\be
\d I=(\d V_{L}-\d V_{R})G_{s}-\F G_{s}\callS_{L}\d n_{L}
\label{dI2}
\ee
where
\be
\F\equiv f_{\rm xc}^{RL}+f_{\rm xc}^{LR}
-f_{\rm xc}^{RR}-f_{\rm xc}^{LL}.
\ee
The expression for $\d I$ in Eq.~(\ref{dI2}) is correctly gauge 
invariant. Under a gauge transformation the 
kernel $f_{\rm xc}(\blr,\blr')$ changes by the addition of an arbitrary function 
$q({\bf r})+q({\bf r}')$ \cite{hg.2012} and $\F$ 
is invariant under this transformation.
In conclusion
\be
G = \frac{G_{s}}{1+\F G_{s}/v}.
\label{excond3}
\ee
The quantity $v \equiv  \d I/(\callS_{L}\d n_{L})$ 
is the speed of the charge wavefront propagating in the leads after 
the sudden switch-on of the external bias 
\cite{UimonenKhosraviStanStefanucciKurthLeeuwenGross:11}
and it is of the order of the Fermi velocity.
In the following we refer to $\F G_{s}/v$ as the 
{\em dynamical} xc correction since $\F$ is expressed in terms 
of the TDDFT kernel. An equation similar to Eq.~(\ref{excond3}) can also be
obtained within the framework of time-dependent current 
density functional
theory as has been shown in Ref.~\onlinecite{kbe.2006}.

\subsection{Anderson Model in the Coulomb Blockade regime}
\label{AIM-CB-sec}

We go back to the Anderson model at temperatures higher than the 
Kondo temperature but smaller than the charging energy $U$. This is 
the so called Coulomb blockade (CB) regime where the Abrikosov-Suhl 
resonance (or Kondo peak) in the spectral function has disappeared.
However, as we already pointed out in Sec.~\ref{siam-sec}, 
the CB peaks present in the exact interacting conductance are completely absent 
in the KS conductance. We would like to emphasize that the physical 
situation discussed here is distinct from the one of 
Ref.~\onlinecite{tfsb.2005}. In Ref.~\onlinecite{tfsb.2005} the 
discontinuity is responsible for 
keeping the HOMO doubly 
occupied and the LUMO empty as the gate potential becomes more 
attractive
(closed shell). In this case the discontinuity correctly suppresses 
$G_{s}$ at even $N$. Instead, at odd $N$ the discontinuity
has the opposite effect since it pins the KS gate to the Fermi 
energy, thereby favouring the tunneling of electrons.
Open-shell molecules in the CB regime 
represent a striking example of the inadequacy of 
standard DFT transport calculations. We now show that this is due 
to the lack of the dynamical 
xc correction discussed in Sec.~\ref{lr-tddft.sec}.

To gain some insight into the density dependence of 
$\F/v$ we reason as follows. Away from half-filling the MB and KS 
system behave similarly and 
consequently $G\simeq G_{s}$. On the 
other hand at half-filling, i.e., $N=1$, the exact conductance is 
strongly suppressed whereas the KS conductance is of the 
order of the quantum of conductance 
 $G_{0}$.
Therefore the dynamical 
xc correction has to be small 
for $N\neq 1$ and large for $N=1$. 
Interestingly, this is the same behavior of the {\em derivative} of 
the Hxc potential $\de  v_{\rm Hxc}/\de N$. In fact, we can demonstrate 
that the two quantities are intimately related.

We consider the Anderson model in equilibrium and calculate 
the compressibility $\kappa=\de N/\de \m$ using 
\be
N = 2 \int \frac{{\rm d} \w}{2 \pi} f(\w) A(\w) 
\label{mbdens}
\ee
where $A$ is the interacting spectral function. It is a 
matter of simple algebra to show that
\bea
\kappa&=&\frac{4}{\g}G+
2\ID{\omega} f(\w)\frac{\de A(\w)}{\de \m}\nn\\
&=&\frac{4}{\g}\frac{G}{1+R}
\label{MBcompr}
\eea
where $G$ is the interacting conductance of Eq.~(\ref{mwgfiniteT}) 
and in the last equality 
we defined
\be
R\equiv-2\ID{\omega} f(\w)\frac{\de A(\w)}{\de N}.
\label{defR}
\ee
The interacting and DFT compressibilities are the same by construction since 
the Hxc potential is such that the interacting and DFT densities are identical.
Therefore we can also write
\be
\kappa=
\frac{4}{\g}G_{s}+2\ID{\omega} f(\w)
\frac{\de A_{s}(\w)}{\de \m}
\label{compks}
\ee
where $G_{s}$ is the KS conductance and 
\be
A_{s}(\w)=\ell_{\g}(\w-v-v_{\rm Hxc}[N])
\label{kssf}
\ee
is the KS spectral function.
The latter depends on $\m$  
through  $N$ and the dependence on $N$ is all 
contained in $v_{\rm Hxc}$. 
Taking into account that
$\frac{\de A_s}{\de v_{\rm Hxc}} = - \frac{\de A_{s}}{\de\w}$
we have 
\be
\frac{\de A_{s}}{\de\m}=-\frac{\de A_{s}}{\de\w}\frac{\de v_{\rm Hxc}}{\de N}\frac{\de 
N}{\de\m}. 
\ee
Inserting this result into Eq.~(\ref{compks}), solving for 
$\kappa$ and equating the interacting and DFT expressions we find
\be
\frac{G}{G_{s}}=\frac{1+R}{1+\frac{4}{\g}G_{s}\frac{\de 
v_{\rm Hxc}}{\de N}}.
\label{GT}
\ee
We observe that this relation is valid for any temperature; no 
approximations have been made so far.

We are interested in modelling 
the dependence of $R$ on $N$ for temperatures in the CB regime. In this 
case the many-body (MB) spectral function $A\simeq A^{\rm mod}$, see 
Eq.~(\ref{spectral_model}), and therefore
$R(v)=I(v)-I(v+U)$ where $I(E)\equiv \int f(\w)\ell_{\g}(\w-E)$.
The relation between $v$ and $N$ stems from Eq.~(\ref{mbdens}) and 
reads
\be
N=\frac{2 I(v)}{1+I(v)-I(v+U)},
\ee
from which it follows that
\be
1+R=2I(v)/N.
\ee
For $v<\mu$, or equivalently for $N<1$, we have $I(v+U)\ll 1$.  
Thus for $N<1$ we can write 
$N\simeq 2I(v)/(1+I(v))$ and solving for $I(v)$ we get
$I(v)\simeq N/(2-N)$. The expression of $I(v)$ for $N>1$ can be 
inferred using the ph symmetry and the final result is
\be
1+R= \frac{2}{1+|\d N|},
\label{1+r}
\ee
where $\d N=N-1$. 
Inserting Eq.~(\ref{1+r})  into Eq. (\ref{GT}) we obtain the following DFT 
result for the conductance
\be
\frac{G}{G_{s}}= \frac{2}{1+|\d N|}
\frac{1}{1+\frac{4}{\g}G_{s}\frac{\de v_{\rm Hxc}}{\de N}}.
\label{GT2}
\ee

\begin{figure}[tbp]
    \begin{center}
\includegraphics[width=0.47\textwidth]{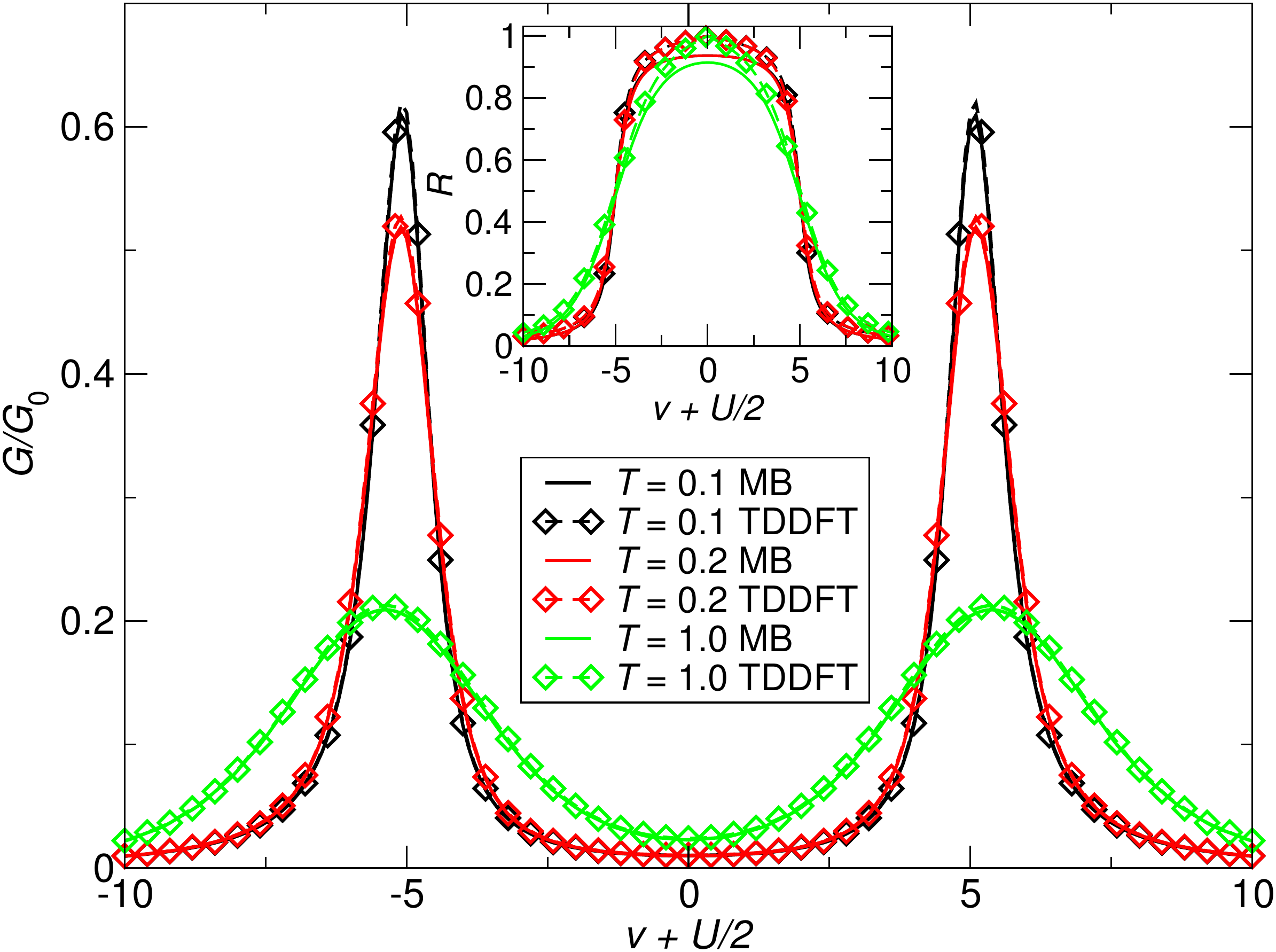}	
\caption{Linear conductance from Eq.~(\ref{mwgfiniteT}) 
using the spectral function $A^{\rm mod}$ of Eq.~(\ref{spectral_model}) (MB, solid) and from 
Eq.~(\ref{GT2}) (TDDFT, dashed).
The inset shows a comparison between the exact and the 
approximate  $R$.
The parameters are (in units of $\g$) $U=10$, $\m=0$. 
Reprinted with permission from Ref.~\protect\onlinecite{KS.2013}. 
Copyright (2013) American Physical Society.}
\label{Gvv}
\end{center}
\end{figure}

This relation is of great utility since it allows to estimate the 
dynamical xc correction from equilibrium DFT.
In fact, the dynamical xc correction of Eq.~(\ref{excond3}) is entirely
expressed in terms of {\em equilibrium} DFT quantities. Moreover, whereas 
$\F$ involves the TDDFT kernel with coordinates in the {\em leads} the 
correction in Eq.~(\ref{GT2}) involves only the DFT $v_{\rm Hxc}$ 
in the {\em molecular junction}. 
The accuracy of Eq. (\ref{GT2}) is examined in Fig.~\ref{Gvv}, 
and benchmarked 
against the interacting conductance of Eq.~(\ref{meir_wingreen_cond}).
Even though the approximate $R$ is 
not on top of the exact one, see inset, the agreement between the
two conductances is extremely good. Most importantly the plateau of $G_{s}$, 
see Fig.~\ref{kscond_siam_ssm}, is completely gone.
 
\subsubsection{Seebeck coefficient of the Anderson model}

The idea to construct the dynamical xc correction to the conductance 
is rather general and can be used to calculate other response 
quantities using equilibrium DFT. As has been shown in 
Ref.~\onlinecite{YPKSD.2016}, 
one such quantity is the Seebeck coefficient that, according to a recently
proposed  DFT  framework for thermal 
transport~\cite{EichVentraVignale:14,EichPrincipiVentraVignale:14},
does also 
contain dynamical xc corrections.

The Seebeck coefficient $S$ is defined as the ratio $S=(\d
V/\d
T)_{I=0}$, where $\d V$ is the voltage that must be applied to
cancel the
current $\d I$ generated by a small temperature difference $\d T$
between the
left and right leads. This definition corresponds to the 
phenomenological Seebeck coefficient of 
Refs.~\onlinecite{Mahan:06,Apertet:15}.
For the Anderson model 
the Seebeck coefficient takes the
form~\cite{Dong:02} 
\begin{equation}
S = -\frac{1}{T} \frac{\ID{\omega} \omega f'(\omega) A(\omega)}{\ID{\omega} 
f'(\omega) \, A(\omega)}, 
\label{seebeck_siam}
\end{equation}
with, according to our notation, $f' \equiv d f/d \omega$ and $A$ the interacting spectral 
function. 
To obtain an expression for $S$ in terms of equilibrium DFT 
quantities we calculate $d N/dT$ from 
Eq.~(\ref{mbdens}). In the CB regime $A\simeq A^{\rm mod}$ 
depends on $T$ and $\m$  
exclusively through $N$.  Using 
$d A/d T = (d A/d N)(d N/d T)$ it is easy to show that
\begin{equation}
\frac{d N}{dT} = - \frac{2}{T} \frac{\ID{\omega} \omega f'(\omega)  
A(\omega)}{1+R}\;,
\label{dn_dtemp}
\end{equation}
where $R$ is defined as in Eq.~(\ref{defR}). Therefore, the numerator of the 
Seebeck coefficient in Eq.~(\ref{seebeck_siam}) is related to the 
temperature derivative of $N$. On the other hand, the denominator in 
Eq.~(\ref{seebeck_siam}) is related to the compressibility $\kappa$ of 
Eq.~(\ref{MBcompr}) since $G=-\frac{\g}{2}\ID{\omega} 
f'(\omega) \, A(\omega)$, see Eq.~(\ref{mwgfiniteT}). Therefore, we can write the 
Seebeck coefficient for the Anderson model in the CB regime as 
\begin{equation}
S = - \frac{d N/dT}{d N/d\mu}.
\label{seebeck_densderiv}
\end{equation}
This is the expression we were looking for as both derivatives $d N/dT$ 
and $d N/d\mu$ can be calculated from equilibrium DFT. In the KS 
system $N=2\ID{\omega}
f(\omega) A_{s}(\omega)$ where $A_{s}$ is the KS spectral function of 
Eq.~(\ref{kssf}). Since the Hxc potential $v_{\rm Hxc}$
depends on $N$ and $T$, at self-consistency 
$A_{s}$ depends implicitly 
(through $N$) on $\m$ and both implicitly (through $N$) and explicitly on $T$.
By calculating the
required density derivatives and using $\frac{d
v_{\rm
Hxc}}{d T}=\left(\frac{\partial v_{\rm Hxc}}{\partial 
N}\right)_{T}\frac{d N}{d
T}+\left(\frac{\partial v_{\rm Hxc}}{\partial T}\right)_{N}$, we
obtain the relation
\begin{equation}
S = S_s + \left(\frac{\partial v_{\rm Hxc}}{\partial T}\right)_{N} \;.
\label{seebeck_corr}
\end{equation}
In this result the KS Seebeck coefficient $S_s$ is defined as in 
Eq.~(\ref{seebeck_siam}) but
with spectral funcion $A_s(\omega)$ in place of $A(\omega)$ 
[notice that the only requirement for the derivation of Eq.~(\ref{seebeck_corr}) 
is that $A_{s}$ is a function of $(\omega-v-v_{\rm Hxc})$].
The KS Seebeck coefficient is precisely 
the coefficient predicted by the LB+DFT approach which
lacks the dynamical xc correction 
$\left(\frac{\partial v_{\rm Hxc}}{\partial T}\right)_{N}$.
Although $v_{\rm Hxc}$ depends 
very weakly on temperature, it turns out that this weak 
dependence is still {\em crucial} to reproduce the MB Seebeck coefficient. 
Let us illustrate this point in more detail.

\begin{figure}[t]
\includegraphics[width=0.47\textwidth]{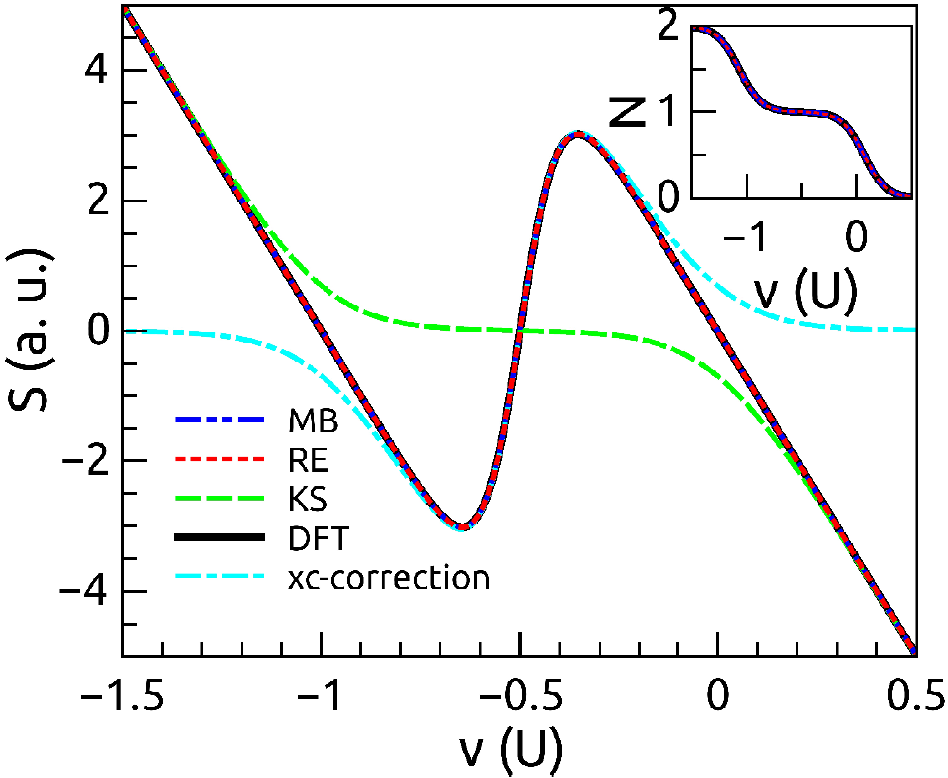}
\caption{Seebeck coefficient $S$ and density $N$
(inset) 
versus gate $v$ for our corrected DFT (black), MB 
(blue) and RE (red). The $S_{s}$ (KS, green) and the xc correction
$\partial v_{\rm Hxc}/\partial T$ (cyan) are also displayed. The 
parameters are $T=0.1$ and  $\gamma=0.01$ (energies in units of $U$).
Reprinted with permission from 
\protect\onlinecite{YPKSD.2016}. Copyright (2016) American Physical Society.}
\label{Siamfig}
\end{figure}

In the CB regime $\gamma$ is the smallest energy scale. 
We consider the limit of very weak contacts $\g\ll T,U$ and 
approximate $v_{\rm Hxc}$ by the exact  Hxc potential of the
isolated ($\gamma=0$) impurity~\cite{sk.2011},
see Eq.~(\ref{hxc_pot_ssm}). Having 
 an analytic expression 
for $v_{\rm Hxc}$  we can evaluate both terms on the r.h.s. of 
Eq.~(\ref{seebeck_corr}). In Fig.~\ref{Siamfig} 
we show $S$ calculated from Eq.~(\ref{seebeck_siam})  (black) versus
the gate $v$ and compare it with 
$S$ calculated from Eq.~(\ref{seebeck_siam}) using the 
MB spectral function of Eq.~(\ref{spectral_model})
(blue) 
as well as with $S$ calculated using the Rate Equation (RE) approach of 
Ref.~\onlinecite{BeenakkerStaring:92} (red), which is exact in the 
limit $\gamma \to 0$.
All three approaches give the same Seebeck coefficient 
and densities (see inset). 

It is instructive to analyze how 
the two terms in Eq.~(\ref{seebeck_corr}) contribute separately to 
yield to correct Seebeck coefficient.
We see in Fig.~\ref{Siamfig} that the KS Seebeck coefficient 
$S_{s}$ (green) accounts for the correct 
linear behavior (with slope proportional to $T^{-1}$) at large 
values of $|v|$. This can easily be understood since as $\gamma\to 0$ 
the KS spectral function $A_s(\omega) \to 2 \pi 
\delta(\omega -v -v_{\rm Hxc})$
 and consequently $S_s = -(v+v_{\rm
Hxc})/T$. 
Less obvious, instead, is  the 
plateau of $S_{s}$ for $v\in (-U,0)$. The plateau is a direct 
consequence of the step in $v_{\rm Hxc}[N]$ at $N=1$,
responsible for blocking electrons 
with energy below $v+U$ from entering the impurity site (see inset).
The CB opens a gap in the 
noninteracting straight line $-v/T$, shifting it leftward by $U$ for 
$v<-U$ and generating the correct behavior at large negative values 
of $v$. 
We may say that the  plateau is a manifestation of the CB,
an equilibrium property of the impurity occupation. 
Although the  KS Seebeck coefficient correctly captures the 
aforementioned gap it entirely misses the oscillation of $S$ for $N\approx 1$.
The dramatic consequence of this fact is that $S_{s}$
severely underestimates the interacting Seebeck coefficient.
It is remarkable that 
this problem is perfectly cured by 
the dynamical xc correction $\partial v_{\rm Hxc}/\partial T$, see  
Fig.~\ref{Siamfig} (cyan).
Thus, the explicit temperature dependence of $v_{\rm Hxc}$
is the key ingredient for the Seebeck coefficient not to vanish in 
the CB
regime~\cite{BeenakkerStaring:92,StaringMolenkamp:93,Dzurak:93,Dzurak:97}.

One last remark before closing this section. The use of any 
temperature-dependent LDA potential \cite{GuptaRajagopal:82,ksdt.2014}
in a LB+DFT calculation would not only miss the oscillation induced by 
$\partial v_{\rm Hxc}/\partial T$ but also
the plateau in $S_{s}$ due to the lack of the step in $v_{\rm Hxc}$ 
at $N\approx 1$.

\subsection{Constant Interaction Model in the Coulomb Blockade regime}
\label{CIM+TDDFTsec}

In this section we extend the analysis on the Anderson model to the 
Constant Interaction Model (CIM) introduced in Sec. \ref{cim_sec}. For 
simplicity we assume 
that the broadening matrix $\bgG_{\a,mn}=(\g/2)\d_{mn}$ is diagonal and 
proportional to the identiy matrix.
In Sec. \ref{cim_sec} we proved that the Hxc potential of the 
zero-temperature isolated CIM is 
a uniform shift depending on the total number of particles only, 
i. e., $v_{{\rm Hxc}}[n](\blr)=v_{\rm Hxc}[N]$, see Eq.~(\ref{xcpot_cim_zero_temp}). 
For a broadening $\g$ and temperature $T$
much smaller than both the level spacings and the charging 
energies, the inhomogeneity of the Hxc potential as well as the 
dependence of the Hxc potential on the local occupations can be safely discarded.
Then we can go through the 
same steps of the single-level derivation of 
Sec.~\ref{AIM-CB-sec} and find again Eq.~(\ref{GT2}). 
The only difference  is that 
$\d N$ is given by the deviation of  $(N-{\rm Int}[N])$ from 
unity.

To illustrate the importance of the dynamical xc correction to the 
conductance we approximate $v_{\rm Hxc}$ as in Eq.~(\ref{HxcfitCIM}). 
We recall that the charging energies $U_{J}$ are given by the xc part of the 
derivative discontinuity of the 
CIM with $J$ electrons~\cite{PerdewParrLevyBalduz:82}. 
For the widths we take $W_{J}= 0.16\, \g/U_{J}$ which is consistent with 
Ref.~\onlinecite{ES.2011}. We mention that in the CB regime the 
temperature  $T>\g$ and hence the smeared steps of $v_{\rm Hxc}$ 
should be broadened by $T$ and not by $\g$ as in Eq.~(\ref{HxcfitCIM}). 
Nevertheless, as we already discussed, 
this quantitative feature has no effect in the qualitative behavior 
of the number of particles $N$ and KS conductance $G_{s}$ as 
functions of the gate $v$.

Possible physical realizations of the CIM are quantum dots made 
from metallic single-wall nanotubes
(SWNT)~\cite{lbp.2002,BBNIS.2002,SJKDKvdZ.2005}. 
This has been shown by Oreg 
{\em et al.}~\cite{OBH.2000} who were able to reproduce 
the observed fourfold periodicity in the electron addition energy of 
a SWNT of finite length using the following CIM Hamiltonian
\bea
H &=& \sum_{l\n\s}\e_{l\n}n_{l\n\s}+\frac{1}{2}E_{C}\left(N-N_{0}\right)^{2} \nn\\
&& +\d U\sum_{l\n}n_{l\n\ua}n_{l\n\da} + J_{\rm x} N_{\ua}N_{\da}.
\label{oregham}
\eea
Here $\s$ is the spin index, $\n=0,1$ is the band index and $l$ is the 
integer of the quantized quasi-momentum of the electrons. The entire 
Hamiltonian is expressed solely in terms 
of the occupation numbers $n_{l\n\s}$ since 
$N_{\s}\equiv \sum_{l\n}n_{l\n\s}$ (total number of electrons 
with spin $\s$) and $N=N_{\ua}+N_{\da}$ (total number of 
electrons). The finite length of the SWNT causes a finite subband 
mismatch $\d$ so that the  single-particle energies are
\be
\e_{l\n}=\left\{
\begin{array}{ll} l\D-\d &\quad{\rm for}\quad \n=0 \\
l\D&\quad{\rm for}\quad \n=1
\end{array}
\right.
\ee
with $\D$ the average level spacing. In Eq.~(\ref{oregham}) 
$E_{C}$ is the charging energy ($N_{0}$ is the number of 
electrons of the charge neutral SWNT quantum dot), $\d U$ is the 
extra charging energy for two electrons in the same energy level and 
$J$ is the exchange energy between electrons of opposite spin. With 
the parameters of Ref.~\onlinecite{OBH.2000}, when an extra electron enters the 
nanotube it occupies the lowest available single-particle energy level.
Thus the {\em aufbau} is the same as that of the noninteracting Hamiltonian.
Accordingly the spin of the ground state 
is $0$, $1/2$, $0$, $1/2$, \ldots for 0, 1, 2, 3,  \ldots extra 
electrons. Let $E(N)$ be the ground state energy of the SWNT with $N$ extra 
electrons. Using Eq.~(\ref{oregham}) it is straightforward to obtain
\bea
E(0)&=&0,
\nn\\
E(1)&=&\D-\d+\frac{1}{2}E_{C},
\nn\\
E(2)&=&2\D-2\d+\frac{1}{2}4E_{C}+\d U +J_{\rm x},
\nn\\
E(3)&=&3\D-2\d+\frac{1}{2}9E_{C}+\d U+2J_{\rm x},
\eea
and so on.
The function $E(N)$ with $N$ a real 
continuous variable has a discontinuous derivative at integers $N$ 
and the size of this 
discontinuity  is given by 
$\D(N)=E(N+1)-2E(N)+E(N-1)$, see Ref.~\onlinecite{PerdewParrLevyBalduz:82}. 
One finds
\bea
\D(1)&=&E_{C}+\d U +J_{\rm x},
\nn\\
\D(2)&=&\d+E_{C}-\d U ,
\label{delta}
\eea
and $\D(N)=\D(J)$ if $N-J=0$ mod2 with $J=1,2$.

The Hxc potential $v_{\rm Hxc}[N]$ has the property that for a given 
chemical potential the ground state occupations $n_{l\n\s}$ of the KS 
system are the same as those of the CIM.  From Eqs.~(\ref{delta}) we 
find that the charging energies $U_{J}$ are given by
\bea
U_{1}&=&E_{C}+\d U +J_{\rm x},
\nn\\
U_{2}&=&E_{C}-\d U ,
\label{u}
\eea
and $U_{J}=U_{K}$ if $J-K=0$ mod2 with $K=1,2$. 
The average values of these parameters 
can be found in Ref.~\onlinecite{OBH.2000}. In order to match the 
position of the conductance peaks of the SWNT of 
length $\simeq 100$ nm we use (all energies are in meV): $\D=9.2$, $\d=2.27$,  
charging energy $E_{C}=2.485$, exchange energy $J_{\rm x}=0.7$, extra charging 
energy for doubly occupied levels $\d U=0.37$~\cite{KS.2013}. 
\begin{figure}[tbp]
    \begin{center}
\includegraphics[width=0.47\textwidth]{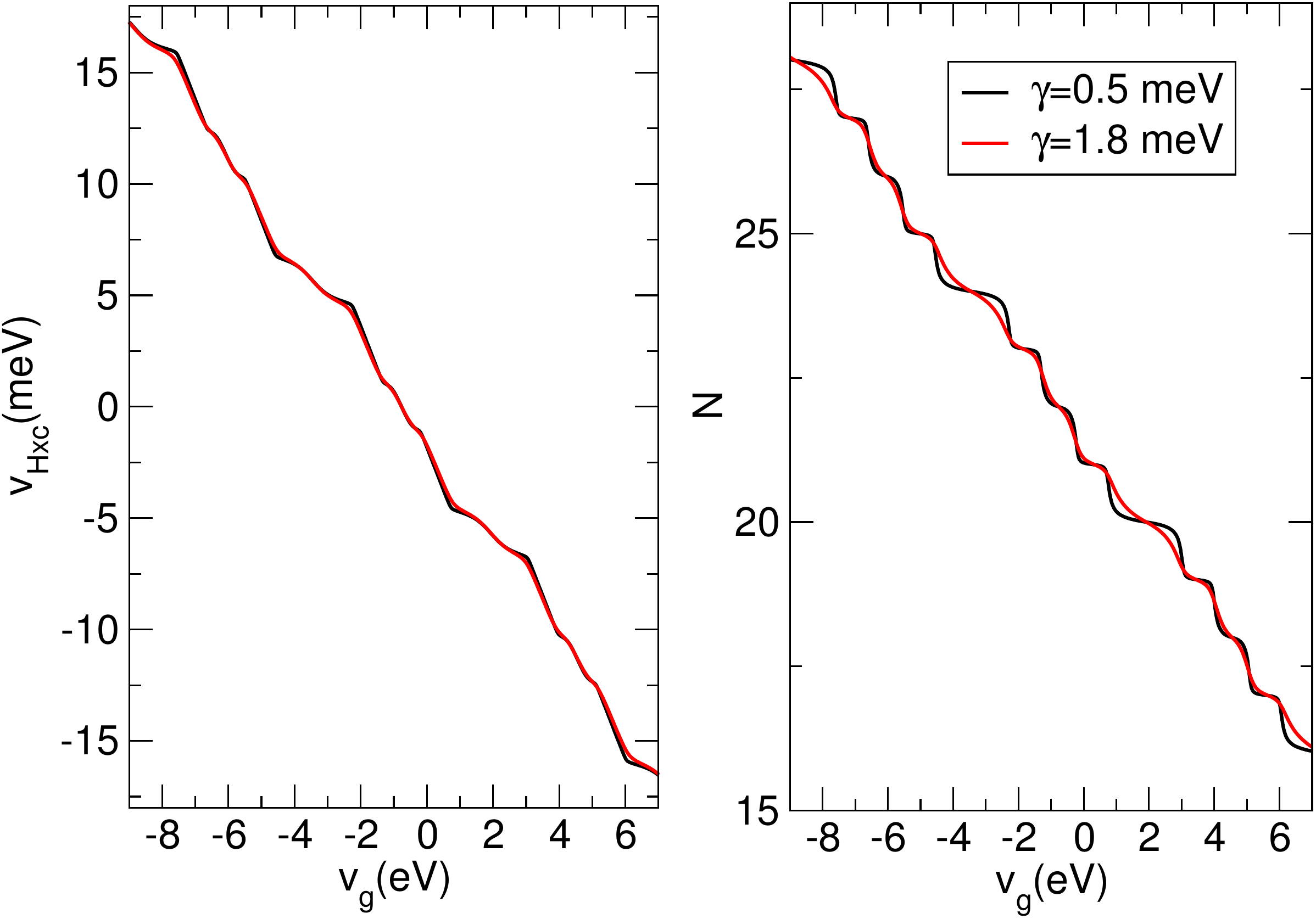}	
\caption{KS Hxc potential of Eq.~(\ref{HxcfitCIM}) (left panel) and number of 
electrons on the SWNT quantum dot (right panel) for different values of the 
coupling $\gamma$ to the leads.
Reprinted with permission from Supplemental Material to 
Ref.~\protect\onlinecite{KS.2013}. Copyright (2013) American Physical Society.}
\label{suppl_vhxc_n}
\end{center}
\end{figure}

The total number of particles is obtained from the self-consistent 
solution of the KS equation (\ref{neqN}) which in equilibrium can be 
written as
\be
N=2\ID{\omega} f(\w)\Tr[\blA_{s}(\w)],
\ee
where $\blA_{s}(\w)=\blA_{L,s}(\w)+\blA_{R,s}(\w)$ is the total KS spectral function and 
$\blG$ is the Green's function of Eq.~(\ref{KSgreenfunction}).
Taking into account that the broadening matrices are diagonal and 
that $v_{\rm Hxc}$ is uniform then $\blA_{s}$ is diagonal. For the 
SWNT with Hamiltonian in Eq.~(\ref{oregham}) the trace of the KS 
spectral function reads
\be
\Tr[\blA_{s}(\w)]=\sum_{l\n}\frac{\g}{(\w-\e_{l\n}-v_{\rm 
Hxc}[N]-v)+\g^{2}/4}.
\ee
In Fig.~\ref{suppl_vhxc_n} we plot $v_{\rm Hxc}[N]$  as 
well as $N$ at self-consistency  as a function of 
$v_{g}=v_{0}+\a v$ for two different values of $\gamma$. 
The potential energy $v_{0}$ is determined by requiring that our reference 
energy is the same as  in 
Ref.~\onlinecite{lbp.2002}. The parameter $\a=C/C_{g}$ is the ratio 
between the total capacitance and the gate capacitance. For the 
experiment in Ref.~\onlinecite{lbp.2002} this ratio is about 250. 
\begin{figure}[tbp]
    \begin{center}
\includegraphics[width=0.47\textwidth]{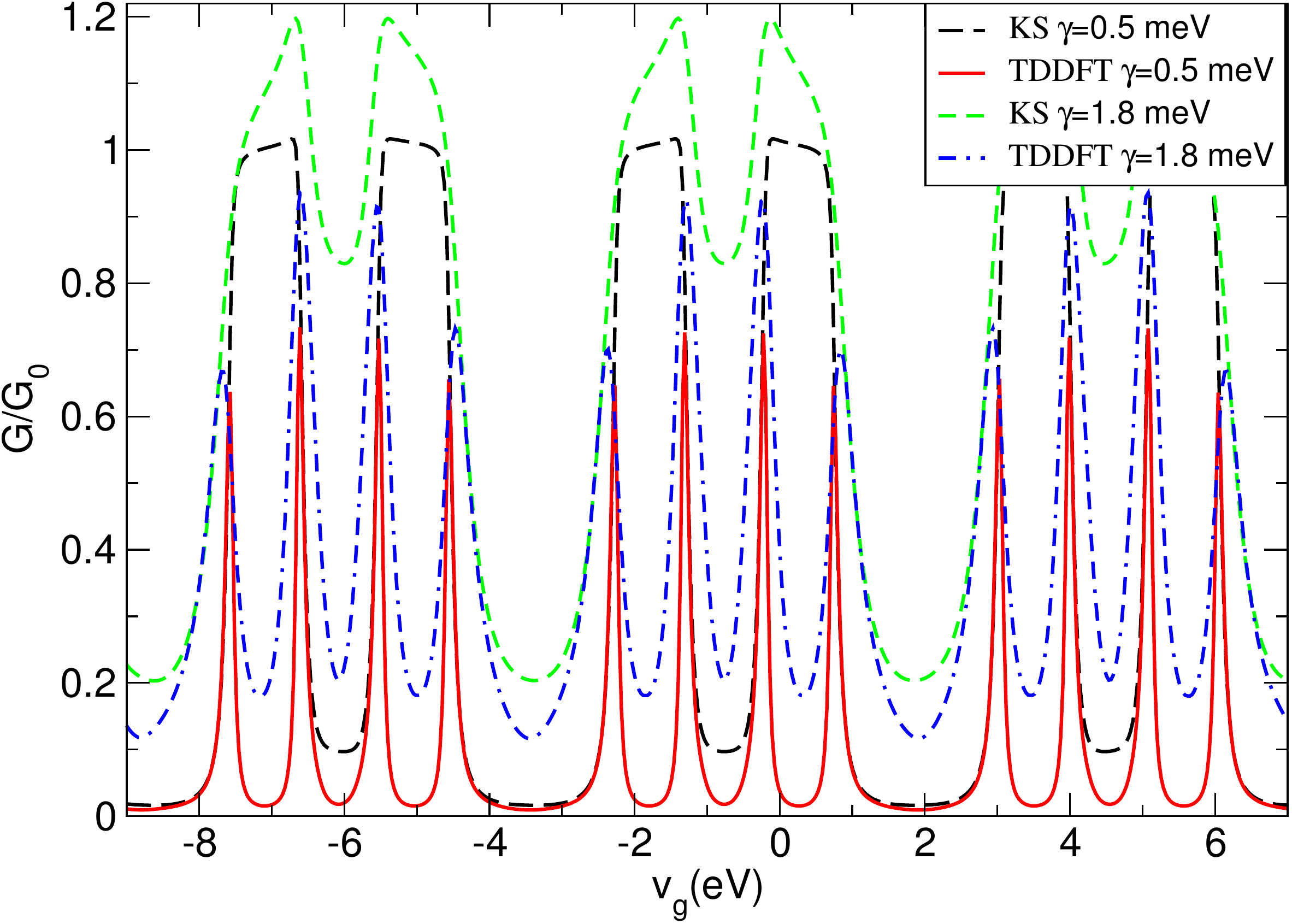}	
\caption{KS and TDDFT corrected conductances for the SWNT quantum dot 
for different values of the broadening $\gamma$.
Reprinted with permission from Supplemental Material to 
Ref.~\protect\onlinecite{KS.2013}. Copyright (2013) American Physical Society.}
\label{suppl_cond}
\end{center}
\end{figure}

To obtain the KS conductance at a certain value of $v$ we evaluate 
the KS spectral function with $N=N[v]$
and then calculate the KS conductance in accordance with 
Eq.~(\ref{lb_conductance}), i.e.,
\be
G_{s}=-\frac{\g}{2}\int\frac{d\w}{2\p}f'(\w)\Tr[\blA_{s}(\w)].
\ee
Subsequently we correct $G_{s}$ according to 
\be
\frac{G}{G_{s}}= \frac{2}{1+|\d N|}
\frac{1}{1+\frac{4}{\g}G_{s}\frac{\de v_{\rm Hxc}}{\de N}},
\label{GTSWNT}
\ee
see Eq.~(\ref{GT2}). The KS conductance $G_{s}$ as well as the conductance with dynamical xc 
corrections $G$ are shown in Fig.~\ref{suppl_cond} 
for different values of the broadening 
parameter $\g$. 
As expected, for small $\g$ the KS conductance  behaves like in 
Fig.~\ref{homo_lumo}, i. e., it exhibits a Kondo plateau whenever the number of 
particles $N$ is close to an odd 
integer. The dynamical xc correction 
suppresses this plateau and yields 
the correct CB pattern.

\begin{figure}[tbp]
    \begin{center}
\includegraphics[width=0.49\textwidth]{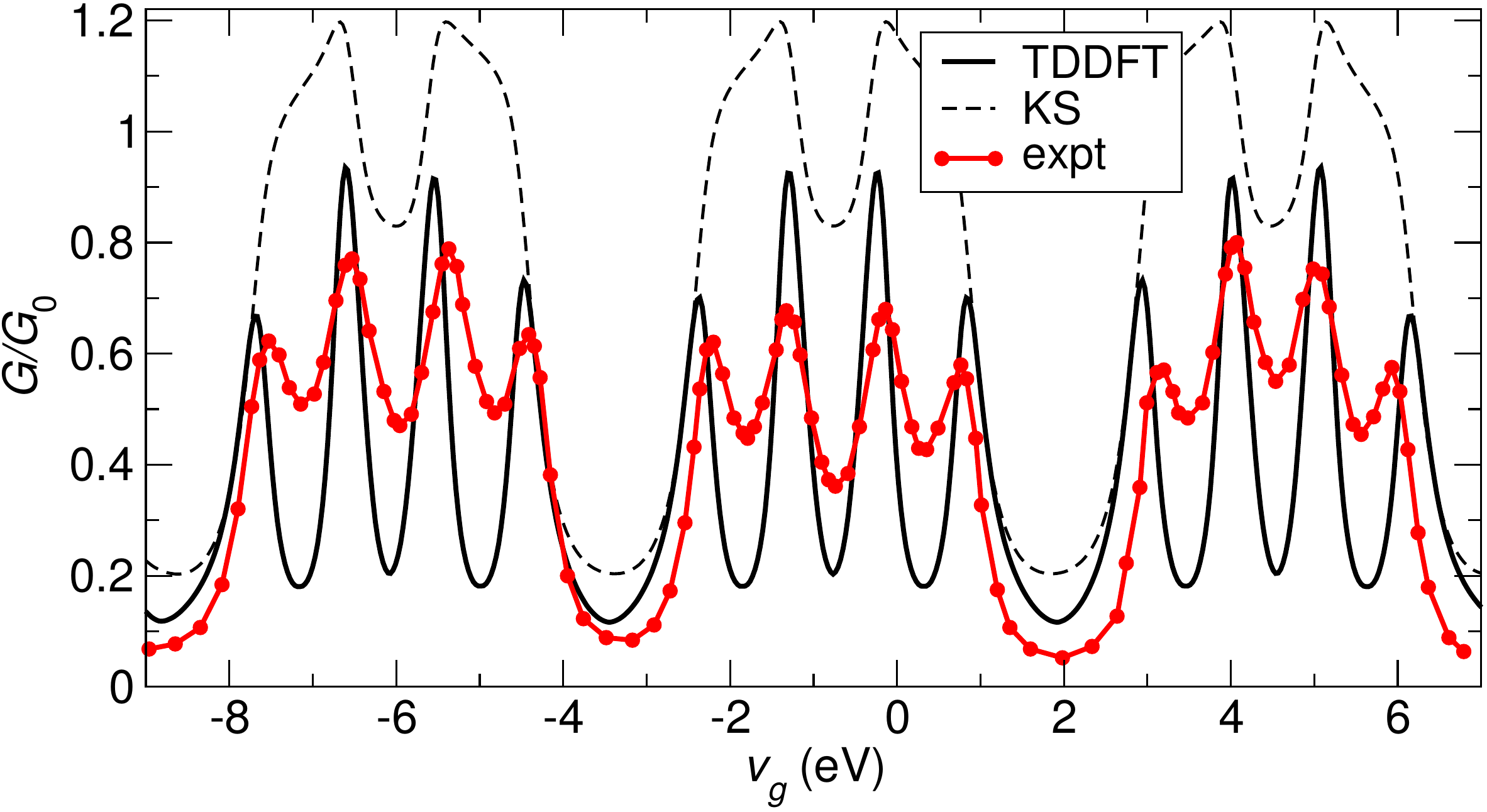}	
\caption{Linear KS and TDDFT conductance (Eq.~(\ref{GT2})) 
for a SWNT quantum dot in comparison to 
experimental conductance from Ref. \onlinecite{lbp.2002}, 
as function of gate voltage.
Reprinted with permission from 
\protect\onlinecite{KS.2013}. Copyright (2013) American Physical Society.}
\label{SWNT-comparison}
\end{center}
\end{figure}

In Fig. \ref{SWNT-comparison} we compare $G_{s}$ and $G$ with  
the experimental conductance. 
The conductance $G$  represents a considerable improvement 
over $G_{s}$ which, instead, shows two deformed Kondo plateaus per period. 
Notice that the fourfold periodicity is also captured. More details 
can be found in Ref.~\onlinecite{KS.2013}.

\subsubsection{Seebeck coefficient of the CIM}

For temperatures $T\gg\gamma$ the Seebeck coefficient of the CIM exhibits a sawtooth behavior
as a function of gate voltage, with ``jumps'' occurring when the 
number of particles crosses an integer. In addition to these jumps,
a superimposed fine structure of wiggles spaced by $\D\ve$ 
emerges whenever the level spacing $\D\ve$ is larger than the 
temperature~\cite{BeenakkerStaring:92}. The wiggles are associated 
to charged excitations from the ground state with $(N-1)$ particles 
to some excited state with $N$ particles.

For diagonal broadening matrices $\bgG_{\a,mn}=\d_{mn}\g/2$
we can again express the Seebeck coefficient as in 
Eq.~(\ref{seebeck_densderiv}). The derivation is identical provided 
that we replace the interacting spectral function $A$ with its 
trace $\Tr[\blA]$. Since $N$ can be calculated from DFT, 
Eq.~(\ref{seebeck_densderiv}) allows us to express the Seebeck coefficient
in a pure DFT fashion. Approximating the Hxc potential as 
a uniform shift, see discussion
at the beginning of Sec.~\ref{CIM+TDDFTsec}, it  is straightforward to 
show that
\begin{equation}
S=S_{s}+
\left(\frac{\partial v_{{\rm Hxc}}}{\partial T}\right)_N.
\label{mult_level_seebeck}
\end{equation}

Like in Eq.~(\ref{HxcfitCIM}) we construct $v_{\rm Hxc}[N]$ by summing over all 
possible charged states the Hxc potential of the Anderson 
model. However, due to the importance of the temperature dependence 
we use the finite-temperature Hxc potential of the isolated Anderson 
impurity, see Eq.~(\ref{hxc_pot_ssm}). Thus we have
\begin{equation}
v_{\rm Hxc}[N] = \sum_{J=1}^{2M-1} \left[\frac{U_{J}}{2}+
g_{U_{J}}^{\rm ext}(N-J)\right],
\label{vxc_multi_sum}
\end{equation}
where $U_{J}$ is the charging energy
and the extended $g^{\rm ext}_{U}$
function 
is defined according to
\begin{equation}
g_{U}^{\rm ext}(N-1) = \left\{ 
\begin{array}{cl}
-U/2 & \mbox{ $N<0$} \\
g_{U}(N-1) & \mbox{ $0\leq N \leq 2$}~~, \\
U/2 & \mbox{ $N>2$} 
\end{array}
\right.
\end{equation}
with $g_{U}$ given in Eq.~(\ref{gu-function}). Like the Hxc potential in 
Eq.~(\ref{HxcfitCIM}) also this Hxc potential has a staircase 
behavior with smeared steps of 
width $U_{J}$ between two consecutive integers but the smearing is 
governed by $T$ instead of $\g$. 

\begin{figure}[tbp]
        \begin{center}
    \includegraphics[width=0.47\textwidth]{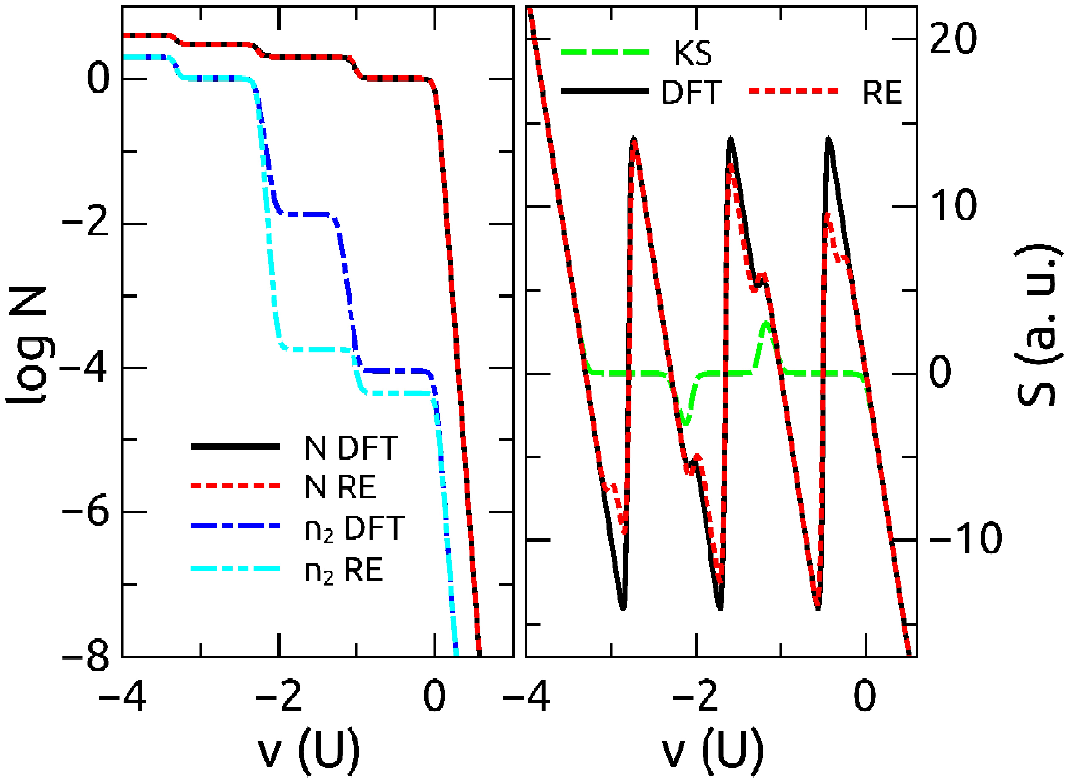}
    \caption{ Density (left) and Seebeck coefficient
(right) of CIM with two spin-degenerate levels computed from RE and DFT using
the approximate functional of Eq.~(\protect\ref{vxc_multi_sum}). The KS
Seebeck coefficient is also shown.
Reprinted with permission from 
\protect\onlinecite{YPKSD.2016}. Copyright (2016) American Physical Society.}
\label{homo_lumo-seebeck}
\end{center}
\end{figure}

It is worth noting that the property of the Hxc potential of being the same for 
all energy levels 
is an exact feature only at zero temperature. 
In fact, our approximate $v_{\rm Hxc}$ can reproduce only the 
occupations corresponding to a thermal mixture of ground states with 
different number of particles. To illustrate what physics is lost in 
this way we first consider a two-level CIM with repulsion energy $U$.
We choose a temperature  $T=0.03$ much larger than $\gamma=0.001$ and 
the energy of the levels $\varepsilon_{i}=\varepsilon_{i}^{0}+v$ with 
$\varepsilon^{0}_{1}=0$
and $\varepsilon_{2}^{0}=0.3$ and $v$ the external gate potential. 
Here all energies are given in units of $U$. The left panel of 
Fig.~\ref{homo_lumo-seebeck} shows the 
total number of particles  $N=n_{1}+n_{2}$ as well as the  
occupation $n_{2}=\sum_{\sigma}n_{2\sigma}$ of the highest level
as obtained using DFT with Hxc potential in 
Eq.~(\ref{vxc_multi_sum}) and the RE approach.
The approximation of a uniform Hxc potential has no effect on $N$, which is 
identical in both approaches, but it introduces exponentially small 
discrepancies in $n_{2}$ (and hence in $n_{1}$). These discrepancies are due to the neglect of 
excited states in the thermal mixture represented by $v_{\rm Hxc}$. 
Therefore, we  expect that some 
of the wiggles in the Seebeck coefficient are not captured by our 
approximation (which accounts only for the addition of electrons 
in the lowest available level). This is confirmed by 
the right panel of Fig.~\ref{homo_lumo-seebeck} where the wiggles associated
to the addition energies of excited states emerge using 
a rate equations approach (red) and are absent using our approximate DFT treatment
(black). Nevertheless, we  emphasize that the 
wiggles stem from $S_{s}$ and are not due to the xc 
correction. The latter is responsible for the large sawtooth 
oscillations and, as Fig.~\ref{homo_lumo-seebeck} clearly 
shows, it is the dominant contribution to $S$. 

Experimental measurements of the Seebeck coefficient for an individual single-wall carbon 
nanotube in the 
CB regime have been reported in Ref.~\onlinecite{SPK.2003}.
In order to show the performance of our DFT scheme 
we extracted both single-particle energies and 
charging energies  from the
experimental results.
We again consider the Hxc potential of Eq.~(\ref{vxc_multi_sum}) but,
in contrast to the model 
calculations described previously, the charging energies $U_J$
depend on the charging state $J$. Details on the parameters can be 
found in Ref.~\onlinecite{YPKSD.2016}

\begin{figure}
    \begin{center}
\includegraphics[width=0.47\textwidth]{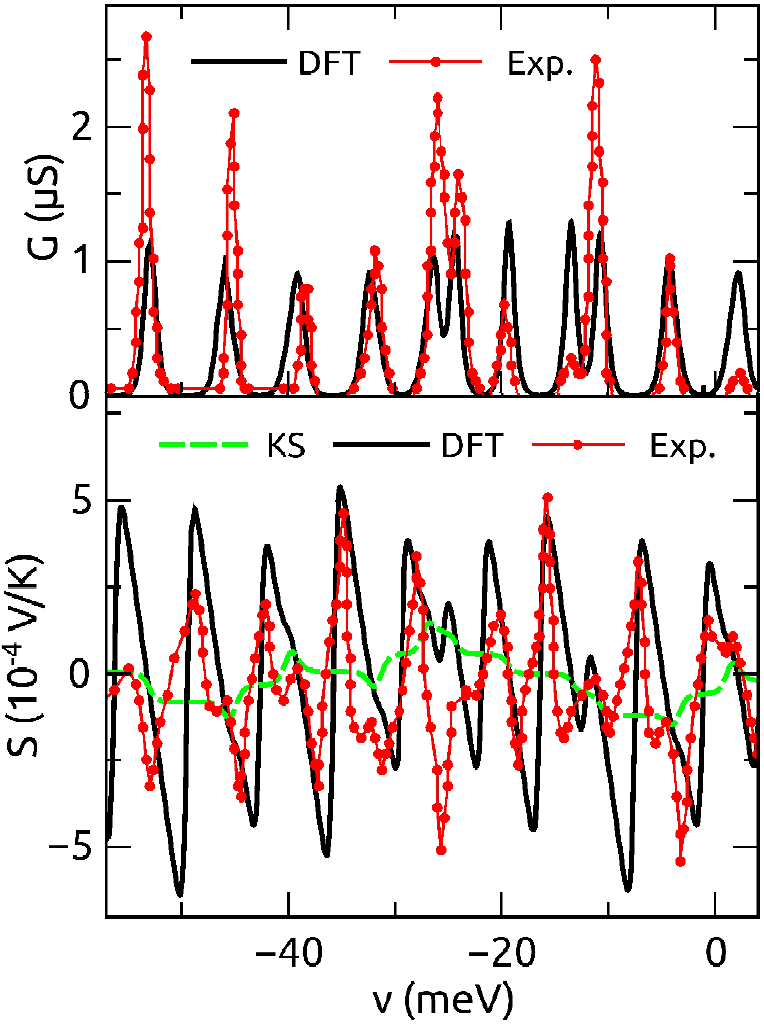}
\caption{Conductance (upper panel) and Seebeck
coefficient 
(lower panel) of a single-wall carbon nanotube from DFT (black) and 
experiment (red, data from Ref.~\onlinecite{SPK.2003}). 
Also shown is the KS Seebeck coefficient (dashed green). 
Reprinted with permission from 
\protect\onlinecite{YPKSD.2016}. Copyright (2016) American Physical Society.}
\label{CNT}
\end{center}
\end{figure}

In Fig.~\ref{CNT} we show the interacting 
conductance $G$ calculated using Eq.~(\ref{GTSWNT}), see also 
Ref.~\onlinecite{KS.2013}, (upper panel)  
and the Seebeck coefficient $S$ 
calculated from 
Eq.~(\ref{mult_level_seebeck}) (lower panel) 
versus the gate voltage $v$ 
for temperature $T=4.5$ 
K and coupling $\gamma=0.02$ meV. For comparison we also show the KS Seebeck 
coefficient $S_{s}$. The latter fails in reproducing the characteristic sawtooth 
behaviour of the experimental results. Instead, the interacting Seebeck 
coefficient calculated from Eq.~(\ref{mult_level_seebeck}) clearly shows the peak
and valley structures observed in experiment, confirming again the
crucial 
role of the xc correction.
Remarkably, the fine 
structure wiggles (kinks in some cases) are correctly captured too.

\section{Steady-State Density Functional Theory for Transport at Finite Bias}
\label{idft}

In Sec.~\ref{lb_dft} we have discussed the standard approach to transport 
within a DFT framework, the LB+DFT formalism which combines static 
(ground state or equilibrium) DFT with the Landauer-B\"uttiker formalism. We 
have pointed out that formally this approach is incomplete as even the knowledge 
of the {\em exact} xc potential of DFT doesn't guarantee that the corresponding 
(steady-state) currents are  exact. In the regime of linear response we 
have shown with explicit examples that LB+DFT transport properties such as 
the zero-bias conductance or the Seebeck coefficient can capture some 
but not all of the correct physics of transport. 

In Sec.~\ref{tddft}, on the other hand, we have discussed TDDFT as a truly 
non-equilibrium density functional approach which in principle can describe 
electronic transport correctly. This is of course true in the time domain if 
one is interested in explicitly time-dependent currents such as transients or 
currents in response to a time-dependent bias. However, it is also true for 
the steady-state regime which is interpreted as the long-time limit of the 
time evolution of a system driven out of equilibrium by a DC bias. In fact, we 
have seen in Eq.~(\ref{I-tddft}) that TDDFT leads to an expression for the 
steady current which is structurally identical to the original LB+DFT 
expression (\ref{lb_current}) but with the crucial difference that 
the applied bias has to be corrected by an xc contribution. 

In principle, the TDDFT xc potential (and thus also the xc correction to the 
bias) is a functional with ``memory'', i.e., it depends not only on the 
instantaneous density but also on its entire history at previous 
times. Moreover, it not only depends on the density in the central device 
region but also on the one deep inside the leads. On the other hand, for 
the models we studied we have derived expressions for the xc correction 
to the bias (at least in the linear regime) which depend only on the density 
of the device. If we are honest, though, while our interpretation of the  
xc bias correction was based on TDDFT ideas, its derivation was not. This 
latter fact maybe shouldn't come as a surprise: the construction of TDDFT 
functionals beyond the adiabatic approximation is a notoriously difficult 
task. 

In the present Section we will present a DFT formulation, which we will call 
i-DFT, for {\em steady-state} transport, i.e., we here aim to describe a 
system in its (non-equilibrium) steady state and are not interested in the 
time evolution towards this state (just as the LB+DFT formalism does). We will 
try to incorporate some of the lessons learned from previous (model) studies 
into the new framework, one of which is the overriding importance of an xc 
correction to the bias, another one the formulation in terms of quantities 
defined in the device region only. At the very fundamental level we will first 
choose a set of basic ``density'' variables for which, under certain 
conditions, one can prove a one-to-one correspondence between this set of 
densities and a set of potentials. Once the basic formalism is established, 
we will again focus on its usefulness in the description of transport through 
strongly correlated systems. First we will show how the new formalism can 
handle Coulomb blockade both at zero and at finite bias. Then we will apply 
the insights gained here again to the problem of the SIAM and will construct 
a functional which is able to correctly describe both the Kondo and the Coulomb 
blockade regime as well as the transition from one to the other. 

\subsection{Foundations of the i-DFT formalism}
\label{idft_formal}

As before, we consider a central region attached to left and right leads. 
In equilibrium, the system is subject to an electrostatic potential 
$v_0(\vr)$, e.g., the potential generated by the nuclei. Out of equilibrium, 
the system is exposed to an external bias potential $v_{b}(\vr)$ 
generated by an external battery. In addition the system may also be subject 
to an additional gate potential $v_{g}(\vr)$ which vanishes deep inside the 
leads. In these regions of space the classical potential 
$v_0(\vr)+v_{b}(\vr)+v_{\rm H}(\vr)$ (where $v_{\rm H}(\vr)$ is the Hartree 
potential), differs by a uniform shift $V/2$ ($-V/2$) from its equilibrium 
value, where $V$ is the potential drop (bias) across the junction. Now we 
split the different components of the potentials according to the different 
regions, e.g., we write $v_0(\vr) = v_{0,L}(\vr) + v_{0,C}(\vr) + v_{0,R}(\vr)$ 
where $v_{0,\alpha}(\vr)=v_0(\vr)$ for $\vr \in \alpha$ and zero otherwise. 
The total potential in region $C$ then is $v_{C}(\vr) = v_{0,C}(\vr) + 
v_{{b},C}(\vr)$. We can also split the density in a similar way, i.e., 
$n(\vr) = n_{L}(\vr) + n_{C}(\vr) + n_{R}(\vr)$. With these definitions we 
will show below that there is a one-to-one correspondence between the 
pair $(v_{C}(\vr), V)$ and the pair $(n_{C}(\vr),I)$ where $I$ is the 
steady-state current through region $C$. The fundamental theorem of 
i-DFT can then be formulated as

{\em Theorem}: For any {\em finite} temperature and for fixed potentials 
$v_{\alpha}(\vr)$ in leads $\alpha \in \{L,R\}$ the map $(v_{C}(\vr), V) 
\longrightarrow (n_{C}(\vr),I)$ is invertible in a finite bias window 
around $V=0$.  

{\em Proof}: In order to prove the theorem we show that the Jacobian
\be
J_{V=0} = \det \left[ 
\begin{array}{cc}
\frac{\delta n_{C}(\vr)}{\delta v_C(\vr')} & 
\frac{\partial n_C(\vr)}{\partial V} \\
\frac{\delta I}{\delta v_C(\vr')} & \frac{\partial I}{\partial V}
\end{array}
\right]_{V=0}
\label{jacobian}
\ee
is non-vanishing. The upper left block $\chi_C(\vr,\vr') = \frac{\delta 
n_{C}(\vr)}{\delta v_C(\vr')}_{V=0}$ is the static, equilibrium density response 
function for the contacted $L-C-R$ system but evaluated for both $\vr$ and $\vr'$ 
in region $C$ and $G \equiv \frac{\partial I}{\partial V}$ is the zero-bias 
conductance. As a first observation we note that the variation
$\frac{\delta I}{\delta v_C(\vr')}\big\vert_{V=0}$ vanishes since at zero bias 
a change 
in the central potential does not induce a steady current. Thus, it remains 
to be shown that both entries on the diagonal of the Jacobian (\ref{jacobian}) 
have a definite sign. We first look at the equilibrium density response 
function $\chi_C(\vr,\vr')$ which can be calculated using leads of finite 
length $L$ at taking the limit $L\to \infty$ at the end. At finite temperature 
$1/\beta$ and at chemical potential $\mu$, the Lehmann representation of 
$\chi_C(\vr,\vr')$ reads
\bea
\chi_C(\vr,\vr') &=& \frac{1}{Z} \sum_{i,j} 
\frac{f_{ij}(\vr) f_{ij}(\vr')}{\Omega_{ij}^2 + \eta^2} \nn \\
&& \Omega_{ij} \left( e^{-\beta E_i} - e^{-\beta E_j} \right) e^{\beta \mu N_i} \;.
\label{densresp_c} 
\eea
Here $Z$ is the partition function and the sum is over a complete set of 
many-body eigenstates $|\Psi_i\rangle$ of the contacted system with energy 
$E_i$ and particle number $N_i$. 
Furthermore we have defined the excitation energies $\Omega_{ij}=E_i-E_j$, the 
excitation amplitudes $f_{ij}(\vr) = \langle \Psi_i | \hat{n}(\vr) | 
\Psi_j \rangle - \delta_{ij} n(\vr)$ (with the density operator 
$\hat{n}(\vr)$) 
and $\eta$ is a positive infinitesimal. In order to prove the invertibility 
of $\chi_C(\vr,\vr')$ we have to show that for an arbitrary test function 
$t(\vr)$ we have
\bea
\lefteqn{
\int_C {\rm d}^3r \; {\rm d}^3r\, t(\vr) \chi_C(\vr,\vr') t(\vr') }\nn\\
&&= \frac{1}{Z} 
\sum_{ij} \frac{|T_{ij}|^2 \Omega_{ij}}{\Omega_{ij}^2 + \eta^2} 
\left( e^{-\beta E_i} - e^{-\beta E_j} \right) e^{\beta \mu N_i} \neq 0
\label{int_densresp_c}
\eea
with $T_{ij} \equiv \int_C {\rm d}^3 r \; f_{ij}(\vr) t(\vr)$. It is easy to see 
that for $E_i \neq E_j$ we have $\Omega_{ij} \left( e^{-\beta E_i} - e^{-\beta E_j} 
\right) <0$ and thus the l.h.s. of Eq.~(\ref{int_densresp_c}) can be zero 
only if $T_{ij}=0$ for any $i,j$ with $E_i \neq E_j$. Obviously, for an 
arbitrary test function $t(\vr)$ this cannot happen~\cite{StefanucciKurth:15} 
and we have to conclude 
that $\chi_C$ is invertible. 

Also for the zero-bias conductance one can write down the Lehmann 
representation \cite{bsw.2006} which reads
\be
G = -\frac{1}{Z} \sum_{ij} \frac{2 \eta |I_{ij}|^2 \Omega_{ij}}
{(\Omega_{ij}^2 + \eta^2)^2} 
\left( e^{-\beta E_i} - e^{-\beta E_j} \right) e^{\beta \mu N_i} 
\label{lincond_lehmann}
\ee
where $I_{ij} \equiv \langle \Psi_i | \hat{I} | \Psi_j \rangle$ with the 
longitudinal current operator $\hat{I}$. From this expression one can see that 
$G>0$. Therefore we find that the Jacobian $J_{V=0} = \det[\chi_C] G < 0$. Since 
$J_V$ is a continuous function of $V$ around $V=0$, there exists a finite 
interval around $V=0$ (whose size depends on $v_C$) for which $J_V<0$. 
Therefore in this domain the map $(v_{C}(\vr), V) 
\longrightarrow (n_{C}(\vr),I)$ is invertible. 

In what follows, we will omit the subscript $C$ again but it is understood 
that all local quantities (potentials, densities) refer to the central 
device region only. Let $(n(\vr),I)$ be the density and steady current induced 
by the potentials $(v(\vr),V)$ in an {\em interacting} junction. We assume that 
the pair $(n(\vr),I)$ is non-interacting $v$-representable, i.e., it is a 
physically realizable pair for a {\em non-interacting} system as well. 
Then our theorem guarantees that the pair of potentials $(v_s(\vr),V_s)$, which 
leads to the same density and current in a non-interacting system, is unique. 
Following the usual KS procedure we can then define the Hxc 
gate potential and the xc bias as
\be
v_{\rm Hxc}[n,I](\vr) = v_s[n,I](\vr) - v[n,I](\vr) \;,
\label{hxc_gate_idft}
\ee
\be
V_{\rm xc}[n,I] = V_s[n,I] - V[n,I].
\label{xc_bias_idft}
\ee
Of course, these are purely formal definitions and in practice $v_{\rm Hxc}[n,I]$ 
and $V_{\rm xc}[n,I]$ have to be approximated. 

The self-consistent KS equations of the i-DFT formalism are then 
given by Eqs.~(\ref{neqdensity}) and (\ref{lb_current}) with 
$V_{\a}\to V_{\a}+V_{\a,\rm xc}$. For a symmetric bias these 
equations read
\bea
n(\vr) = 2\!\!\! \sum_{\alpha=L,R} \int \frac{{\rm d} \omega}{2 \pi}
f\!\left(\!\w + s_{\alpha} \frac{V+V_{xc}}{2}\right) 
\!A_{\alpha,s}(\vr,\w)
\nn\\
\label{idft_dens}
\eea
\bea
I &=& 2\!\!\!  \sum_{\alpha=L,R} \int \frac{{\rm d} \omega}{2 \pi} 
 s_{\alpha} f\!\left(\!\w + s_{\alpha} \frac{V+V_{xc}}{2}\right) \nn \\ 
&& \;\;\;\;\;\;
\times\Tr\left[\blG(\w) \bgG_{L}(\w) \blG^{\dagger}(\w) \bgG_{R}(\w) \right]
\label{idft_curr}
\eea
where $s_{R/L}=\pm 1$. Although 
the i-DFT equations are very similar  
to the LB+DFT equations there are important differences: 
in LB+DFT, the Hxc gate potential is a functional of the density alone and 
therefore there is a self-consistency condition only for the density, see 
Eq.~(\ref{neqdensity}). The current is then evaluated with this self-consistent Hxc 
gate potential from Eq.~(\ref{lb_current}) which has the same structure as 
Eq.~(\ref{idft_curr}) but with vanishing xc bias. In i-DFT, on the other hand, 
we have to take into account the generally non-vanishing xc contribution 
to the bias. Both this xc bias as well as the Hxc gate depend on density and 
current and thus the self-consistency conditions for these two quantities are 
coupled and have to be solved together. 

Also in comparison to TDDFT (for an applied DC bias in the long-time limit), 
the i-DFT equation for the current is structurally identical to 
Eq.~(\ref{I-tddft})~\cite{sa-1.2004,sa-2.2004}. 
However, the TDDFT Hxc gate potential and xc bias are 
functionals of the density {\em everywhere}, i.e., both in the device region 
and in the leads. Furthermore, the TDDFT potentials at given time 
$t$ depends on the full history of the density at all previous times. 
In contrast, the i-DFT functionals are independent of history (which is not 
surprising since i-DFT only deals with steady states) and depend only on 
the density in the device region as well as on the steady current. 
The augmented local character of the i-DFT xc potentials agrees with 
similar findings in time-dependent current density functional theory~\cite{v.1995,nptvc.2007}.

The zero-bias conductance in the i-DFT formalism can be derived by 
linearising Eq.~(\ref{idft_curr}) in the bias leading to the simple but 
exact result \cite{StefanucciKurth:15}
\be
G = \frac{G_s}{1 - G_s \frac{\partial V_{\rm xc}}{\partial I} 
\big\vert_{V=0}} \;. 
\label{idft_zb_cond}
\ee
Compared to the TDDFT result (\ref{excond3}) for the conductance, which 
involves the zero-frequency and zero-momentum limit of the xc kernel, the 
above expression is more transparent. Its simple form will later also give 
a hint on the design of approximate i-DFT functionals.

\subsection{i-DFT functionals for the Coulomb blockade regime}
\label{clb_block_idft}

So far we have presented the formal foundations of the i-DFT framework. 
In order for i-DFT to be applied, however, we need approximate xc 
functionals. In the present Section we will construct such approximations 
for the  model systems we studied before, i.e., the SIAM and the CIM. For 
now, our aim is to construct approximations which capture the essential 
physics of Coulomb blockade both at zero and at finite bias. The 
construction of these approximations will be done by reverse-engineering 
from standard techniques typically used to describe Coulomb blockade. 

We start again with the SIAM and again, for simplicity, we  restrict 
ourselves to the wide-band limit. In order to obtain the i-DFT xc potentials 
we need a model for the density and current of the biased, interacting system. 
Fortunately, such a model is easily constructed using the ingredients 
already introduced in Sec.~\ref{siam-sec}. Using the model spectral function 
$A^{\rm mod}(\w)$ of Eq.~(\ref{spectral_model}), the steady-state density 
$N=n$ and current $I$ of the SIAM can be calculated from
\be
N = \int \frac{{\rm d} \omega}{2 \pi} \left[ f(\omega-V/2) + f(\omega+V/2) 
\right] A^{\rm mod}(\omega)
\label{dens_siam_cb}
\ee
and 
\be
I = \frac{\gamma}{2} \int \frac{{\rm d} \omega}{2 \pi} \left[ f(\omega-V/2) - 
f(\omega+V/2) \right] A^{\rm mod}(\omega) .
\label{curr_siam_cb}
\ee
The resulting densities and currents are in excellent agreement with the 
results of the rate equations \cite{rate-paper1,rate-paper2}, the standard 
technique to describe Coulomb blockade for weakly coupled systems. In the 
reverse-engineering procedure we numerically invert Eqs.~(\ref{dens_siam_cb}) 
and (\ref{curr_siam_cb}) for a given, fixed pair $(N,I)$ both for the 
interacting and the non-interacting system and then extract the i-DFT 
xc potentials according to Eqs.~(\ref{hxc_gate_idft}) and (\ref{xc_bias_idft}). 
One can actually prove \cite{StefanucciKurth:15} that for the SIAM in the WBL 
the map $(v,V) \to (N,I)$ is invertible for any value of the bias $V$ 
(infinite bias window). The domain spanned by $N$ and $I$ is 
$|I|\leq (\gamma/2)N$ for $N\in[0,1]$ and $|I|\leq (\gamma/2)(2-N)$ for 
$N\in[1,2]$. 

\begin{figure}[t]
\includegraphics[width=0.5\textwidth]{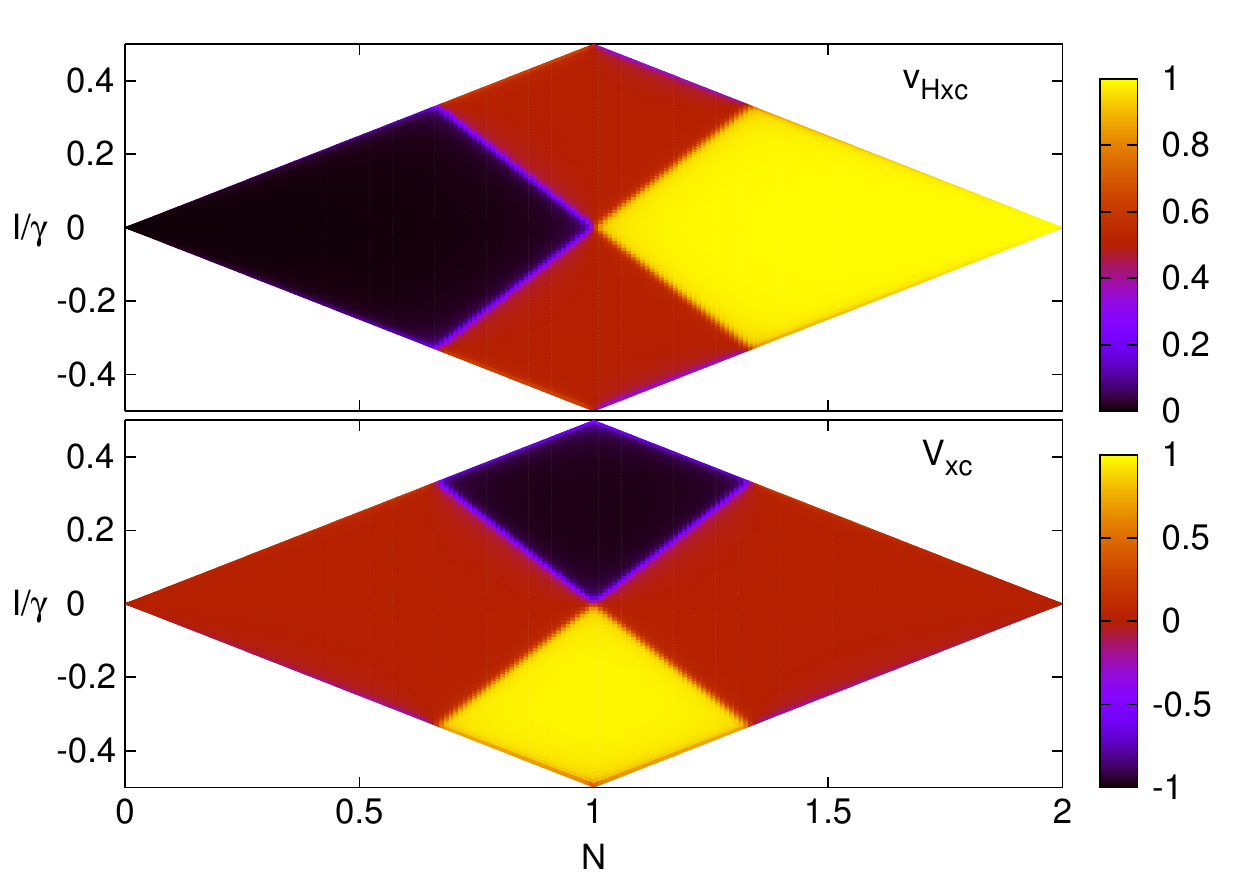}
\caption{Hxc gate (top) and xc bias (bottom) for the SIAM for $U/\gamma=40$. 
Energies in units of $U$. Reprinted with permission from 
\protect\onlinecite{StefanucciKurth:15}. Copyright (2015) American 
Chemical Society.}
\label{siam_xc_gate_bias}
\end{figure}

The Hxc gate $v_{\rm Hxc}[N,I]$ and the xc bias $V_{\rm xc}[N,I]$ resulting from 
the reverse engineering procedure are shown in Fig.~\ref{siam_xc_gate_bias}. 
The most prominent features are smeared steps of height $U/2$ for $v_{\rm Hxc}$ 
and of height $U$ for $V_{\rm xc}$ along the lines $N=1\mp I/\gamma$. The DFT xc 
discontinuity of $v_{\rm Hxc}[N,0]$ bifurcates as current starts flowing. The 
sign of the xc bias is opposite to the current, i.e., the effective KS bias 
$V+V_{\rm xc}$ is lower than the external bias $V$. This is in agreement with 
the model study of Ref.~\onlinecite{sds.2013}. The derivative 
$(\partial V_{\rm xc}/\partial I)_{I=0}<0$ and therefore, according to 
Eq.~(\ref{idft_zb_cond}), $G_s$ becomes the upper limit of the interacting 
zero-bias conductance. 

The i-DFT xc potentials at finite current can be parametrized in the spirit of 
Eq.~(\ref{xc_mod_smooth_fit}) for the zero-current case as 
\bea
v_{\rm Hxc}[N,I] = \frac{U}{2} + \sum_{s=\pm} \frac{U}{2 \pi} 
\arctan\left(\frac{N+(s/\gamma)I -1}{\lambda_1 W}\right)
\label{xcgate_siam_idft}
\nn\\
\eea
and
\be
V_{\rm xc}[N,I] = -\sum_{s=\pm} \frac{sU}{\pi} 
\arctan\left(\frac{N+(s/\gamma)I -1}{\lambda_1 W}\right) 
\label{xcbias_siam_idft}
\ee
where $W$ is defined according to Eq.~(\ref{broadening}) and we have 
introduced for later use an extra parameter which here we set to unity, 
$\lambda_1=1$. Note that 
Eq.~(\ref{xcgate_siam_idft}) reduces to Eq.~(\ref{xc_mod_smooth_fit}) in the 
limit of zero current while Eq.~(\ref{xcbias_siam_idft}) vanishes in this 
limit. We have verified that the self-consistent i-DFT results using the xc 
potentials of Eq.~(\ref{xcgate_siam_idft}) and (\ref{xcbias_siam_idft}) are in 
excellent agreement with the results of the rate equations (as they should be). 

\begin{figure}[t]
\includegraphics[width=0.5\textwidth]{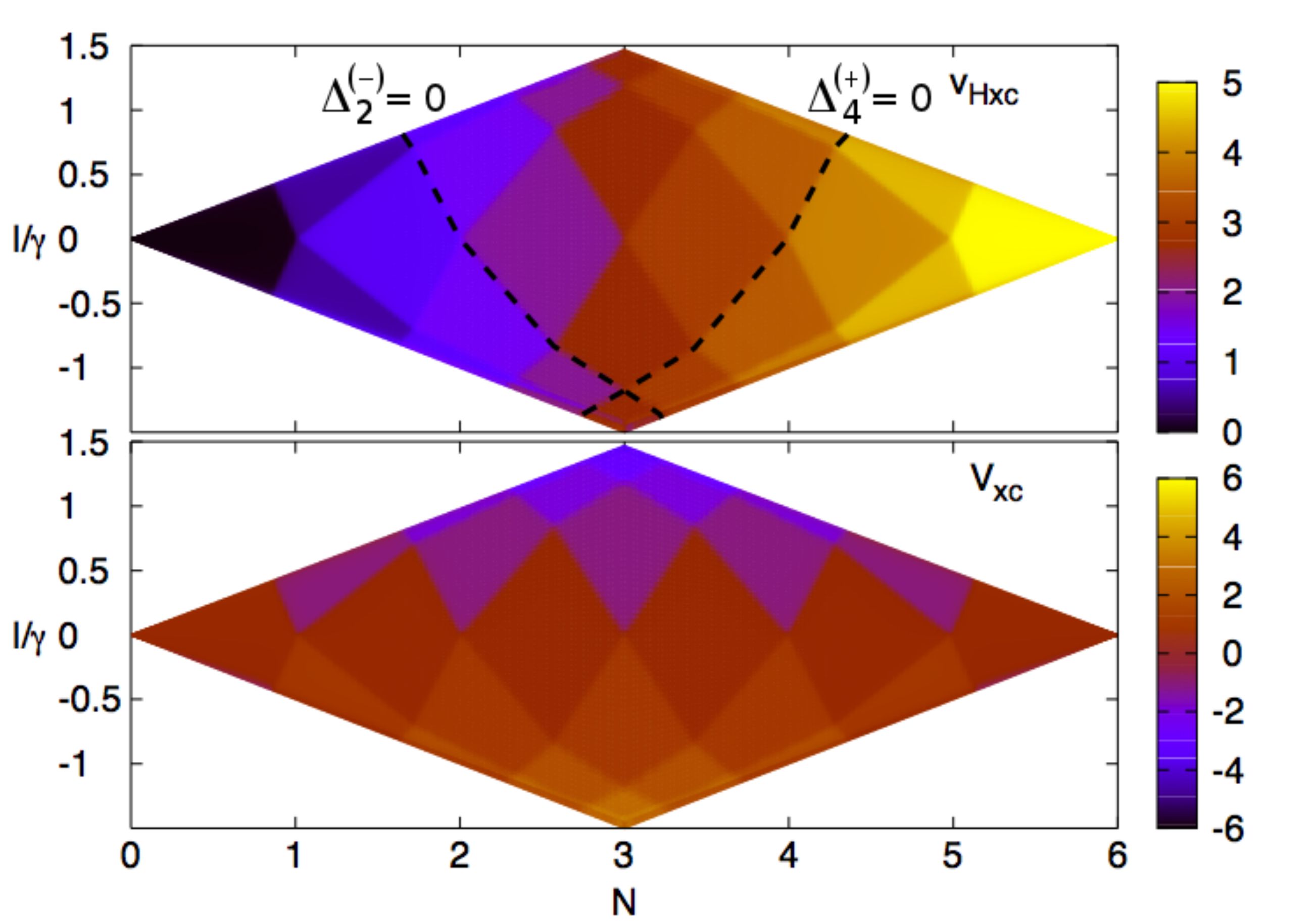}
\caption{Hxc gate (top) and xc bias (bottom) for the CIM with $M=3$ degenerate 
single-particle levels for $U/\gamma=40$. Energies in units 
of $U$. Reprinted with permission from \protect\onlinecite{StefanucciKurth:15}. 
Copyright (2015) American Chemical Society.}
\label{cim_xc_gate_bias_3lev}
\end{figure}

The reverse-engineering procedure can also be applied to the CIM. We consider 
a CIM with $M$ levels described by the Hamiltonian (\ref{cim}) and coupled 
to wide band leads, $\G_{\a,ij}(\w) = \delta_{ij} \gamma/2$. In general, the 
i-DFT xc potentials depend on all level occupations. For simplicity, we 
here  restrict ourselves to the case of $M$ degenerate single-particle 
levels. In this case, by symmetry, the xc potentials become functionals 
only of the total occupation $N=\sum_{i=1}^M \sum_{\sigma} n_{i \sigma}$ and the 
total current $I$. Above the Kondo temperature $T_{\rm K}$, both $N$ and $I$ can be 
obtained by solving the rate equations \cite{rate-paper1}. For given $(N,I)$ 
we numerically invert the map $(v,V)\to(N,I)$ both for the interacting and the 
non-interacting case and then obtain the xc potentials according to 
Eqs.~(\ref{hxc_gate_idft}) and (\ref{xc_bias_idft}). Again, in the wide-band 
limit the map is invertible for all $V$ and the codomain is 
$|I|\leq (\gamma/2) N$ for $N\in[0,M]$ and $|I|\leq (\gamma/2)(2M-N)$ for 
$N\in[M,2M]$. In Fig.~\ref{cim_xc_gate_bias_3lev} we show the xc potentials 
for $M=3$. As in the SIAM, the Hxc gate (xc bias) potential exhibits smeared 
steps of height $U/2$ ($U$) with a rather complex pattern for the edges which 
follow piecewise straight lines in the $(N,I)$-plane. Again, the xc 
discontinuity at integer $N$ and $I=0$ bifurcates as the current starts flowing 
with the edges having different slopes depending on $N$. The edges connect 
``vertices'', i.e., points in the $(N,I)$-plane where two edges meet, and 
typically the edges change slope at the vertices. We denote by 
$\Delta_K^{(s)}(N,I)$ the piecewise linear function of $N$ and $I$ which 
vanishes along the step edge passing through $(K,0)$ with positive ($s=+1$) 
or negative ($s=-1$) slopes (see top panel of 
Fig.~\ref{cim_xc_gate_bias_3lev} for examples). 

In order to model the reverse-engineered xc potentials we have to 
understand the position of the vertices. We realized an interesting 
duality \cite{StefanucciKurth:15}: 
the vertices occur exactly at those points in the $(N,I)$-plane which 
correspond to plateau values of the particle number and the current 
in the gate-bias $(v,V)$ plane. Moreover, these plateau values for 
$N$ and $I$ can be calculated for any degenerate $M$-level CIM using 
simple expressions. From the rate equations, the degenerate CIM with $M$ levels 
leads to $(2M+1)^2$ distinct plateau values. A plateau can be 
uniquely identified by a pair of two integers $(m,n)$ with $m,n = 0,\ldots,2M$. 
In the $(m,n)$ plateau for $n\leq m$, the probability $P(q)$, $q=m,\ldots,n$ 
of finding $q$ particles are all identical and given by $P^{-1}(q) \equiv 
P_{n\leq m}^{-1} = \sum_{j=m}^n \left(\!\!\begin{array}{c} 2\mathcal{M} \\ j 
\end{array}\!\!\right)$, whereas all other probabilities vanish. 
The density and current of the $(m,n)$-plateau are then given by 
\be
N=N_{n\leq m} = P_{n\leq m} \sum_{j=m}^n j 
\left(\!\!\begin{array}{c} 2\mathcal{M} \\ j 
\end{array}\!\!\right)
\ee
and
\be
I=I_{n\leq m} = \frac{\gamma}{2} P_{n\leq m} \sum_{j=m}^n (2M - j) 
\left(\!\!\begin{array}{c} 2\mathcal{M} \\ j 
\end{array}\!\!\right)
\ee

Knowing the plateau values one can define 
the step edges $\Delta_K^{(s)}(N,I)$ and with these one can 
parametrize the Hxc gate and xc bias of the degenerate $M$-level CIM as
\bea
\lefteqn{
v_{\rm Hxc}^{(M)}[N,I] =} \nn\\
&& \;\;\;\;\;\;\;\; \frac{U}{4} \sum_{K=1}^{2 M -1} \sum_{s=\pm} \left[ 
1 + \frac{2}{\pi} \arctan\left( \frac{\Delta_K^{(s)}(N,I)}{\lambda_1 W} 
\right) \right]
\label{xcgate_cim_deg}
\eea
and
\bea
\lefteqn{
V_{\rm xc}^{(M)}[N,I] =} \nn\\
&& \;\;\;\;\;\;\;\; -U \sum_{K=1}^{2 M -1} \sum_{s=\pm} 
\frac{s}{\pi} \arctan\left( \frac{\Delta_K^{(s)}(N,I)}{\lambda_1 W} \right)  
\label{xcbias_cim_deg}
\eea
where, again, $W$ is defined according to Eq.~(\ref{broadening}) and 
$\lambda_1=1$. Again, for 
an $M$-fold degenerate CIM the self-consistent i-DFT results using the 
xc potentials (\ref{xcgate_cim_deg}) and (\ref{xcbias_cim_deg}) are in 
excellent agreement with the rate equation results. 

So far, we have used the rate equations to construct i-DFT xc potentials 
for the degenerate case. For the non-degenerate CIM, according to our i-DFT 
philosophy, the xc potentials are now functionals of the local occupations 
$n_i$ and the current $I$ instead of the total $N$ and $I$. However, from 
what we have learned so far we can still construct useful approximations to the 
xc potentials for the non-degenerate case without having to do 
the full reverse engineering from the rate equations. 

Let $n=\{n_1,\ldots,n_M\}$ be the occupations of the levels ${1,\ldots,M}$ of 
an $M$-level CIM with arbitrary single-particle level structure. Let 
$\mathcal{M}_p[n]$ be the degeneracy of the $p$-th largest occupation and 
$\mathcal{D}[n]$ the number of distinct densities. For instance if $M=5$ and 
$n=\{\frac{1}{3},\frac{1}{2},\frac{1}{2},\frac{1}{3},\frac{1}{3}\}$ then
$\mathcal{M}_{1}=2$, $\mathcal{M}_{2}=3$ and $\mathcal{D}=2$.
We further define
$\mathcal{N}_{p}[n]=2\sum_{q=1}^{p-1}\mathcal{M}_{q}[n]$ as the
maximum number of particles in the first $(p-1)$ levels with degenerate 
occupations ($\mathcal{N}_{1}=0$). The degeneracies $\mathcal{M}_{p}$ are used 
to construct the following i-DFT potentials
\bea
\lefteqn{
v_{\rm Hxc}[n,I]\!\! = \!\!\sum_{p=1}^{\mathcal{D}[n]}v_{\rm Hxc}^{(\mathcal{M}_{p}[n])}
\big[N-\mathcal{N}_{p}[n],I\big] +\frac{U}{4} } \nn \\
&& \times \sum_{p=1}^{\mathcal{D}[n]-1}\sum_{s=\pm}\!\left[1+\frac{2}{\p}\,
\arctan\left(\frac{N+\frac{2s}{\g}I-\mathcal{N}_{p+1}[n]}{\lambda_1 W}\right)
\right]
\nn\\
\label{xcgate_cim_gen}
\eea
\bea
\lefteqn{
V_{\rm xc}[n,I]\!\! = \!\!\sum_{p=1}^{\mathcal{D}[n]}V_{\rm xc}^{(\mathcal{M}_{p}[n])}
\big[N-\mathcal{N}_{p}[n],I\big] } \nn \\
&& \;\; -\sum_{p=1}^{\mathcal{D}[n]-1}\sum_{s=\pm}\frac{s U}{\p}\,
\arctan\left(\frac{N+\frac{2s}{\g}I-\mathcal{N}_{p+1}[n]}{\lambda_1 W} \right) .
\label{xcbias_cim_gen}
\eea
The dependence on the local occupations enters exclusively through the
$\mathcal{M}_{p}$. At the joining points ($N=\mathcal{N}_{p+1}[n]$ and
$I=0$) between two consecutive
$v_{\rm Hxc}^{(\mathcal{M}_{p}[n])}$ we add a discontinuity with
slopes $\pm 2/\g$. In fact, the slope of the lines delimiting the
domain of the i-DFT potentials of a $M$-fold degenerate CIM are 
independent of $M$ (see Figs.~\ref{siam_xc_gate_bias} and 
\ref{cim_xc_gate_bias_3lev}). 

\begin{figure}[t]
\includegraphics[width=0.5\textwidth]{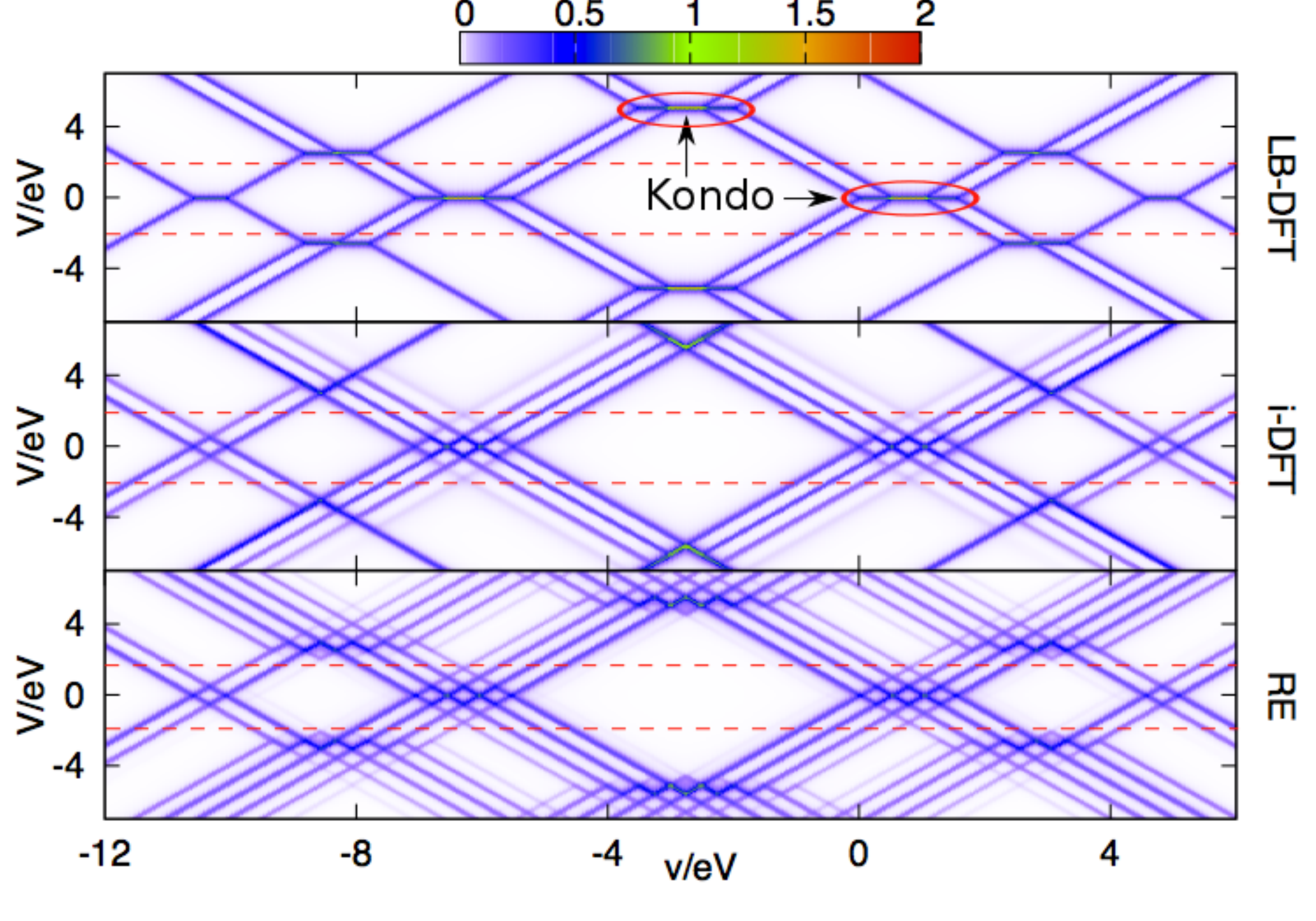}
\caption{Differential conductance (in units of $G_0$) for a six-level CIM 
model of benzene from LB+DFT (top panel), i-DFT (middle), and rate equations 
(bottom). The red lines delimit the low bias region where i-DFT and rate 
equations agree. Reprinted with permission from 
\protect\onlinecite{StefanucciKurth:15}. Copyright (2015) American Chemical 
Society.}
\label{didv_benzene}
\end{figure} 

In order to show the performance of i-DFT we have calculated the finite-bias 
differential conductances of a benzene junction. Here we 
model the benzene molecule by a six-level CIM with $U=0.5$ eV and the 
single-particle energies $\varepsilon_i = \varepsilon_i^0 + v$ with 
$\varepsilon_1^0=-\varepsilon_6^0= 5.08$ eV,  $\varepsilon_2^0=\varepsilon_3^0= 
-\varepsilon_4^0=-\varepsilon_5^0= 2.54$ eV. These parameters are taken from 
a Pariser-Parr-Pople model of benzene \cite{BursillCastletonBarford:98} 
while the coupling to the leads is $\gamma=0.05$ eV. 
We show the differential conductances from LB+DFT (top panel), i-DFT (middle), 
and the rate equations (bottom). Due to the step structure in the 
(current-independent) Hxc gate potential, in LB+DFT we see Kondo plateaus at 
zero bias for odd occupations. Moreover, we also see ``Kondo-like'' 
plateaus at finite bias, a feature which is certainly unphysical. Furthermore, 
many of the lines with finite $dI/dV$ at finite bias which are present in the 
rate equations, in LB+DFT are simply missing. In i-DFT, on the other hand, 
we don't see any Kondo plateaus neither at zero nor at finite bias. At zero 
bias, this is not surprising since we reverse-engineered our (H)xc potentials 
from the rate equations which doesn't describe the Kondo effect. At finite 
bias it is comforting that i-DFT cures the spurious Kondo plateaus of LB+DFT. 
Furthermore, in the low-bias regime (delimited by the dashed horizontal lines) 
i-DFT describes {\em all} lines in the $dI/dV$ correctly as in the rate 
equations. At higher bias, on the other hand, some lines are missing. This, 
however, can be attributed to our particular approximation 
in Eq. (\ref{xcgate_cim_gen})
where the Hxc potential acts as a uniform constant shift for all 
levels. 

\subsection{i-DFT functional for the SIAM: from Kondo to Coulomb blockade regime}
\label{siam_simple}

In Section \ref{lb_dft_strong} we have seen that, at zero temperature, already 
the LB+DFT formalism can correctly describe the Kondo plateau in the zero-bias 
conductance. In Section \ref{tddft}, on the other hand, we have designed 
approximations to linear transport coefficients (zero-bias conductance, 
Seebeck coefficient) which capture the correct physics in the Coulomb blckade 
regime, assigning the corrections over LB+DFT to dynamical 
xc corrections of TDDFT. In the previous Section, in yet another DFT 
framework, we have constructed approximations to the i-DFT xc potentials 
which are designed to work in the Coulomb blockade regime, now not only at 
zero but also at finite bias. The question then arises if it is possible to 
design i-DFT functionals which capture {\em both} the Kondo and the Coulomb 
blockade regimes correctly and also the transition from one regime to the 
other. In the present Section we address this question for the 
SIAM~\cite{KurthStefanucci:16}.

The KS self-consistency conditions in Eqs.~(\ref{idft_dens}) and 
(\ref{idft_curr}) applied to the SIAM simplify to 
\be
N = \sum_{\alpha=L,R} \int \frac{{\rm d} \w}{2 \pi} \; 
f\left(\w + s_{\alpha} \frac{V+V_{\rm xc}}{2}\right) A_s(\w) 
\label{idft_ksdens_siam}
\ee
\be
I = \frac{\gamma}{2} \sum_{\alpha=L,R} \int \frac{{\rm d} \w}{2 \pi} \; 
f\left(\w + s_{\alpha} \frac{V+V_{\rm xc}}{2}\right) s_{\alpha} A_s(\w) 
\label{idft_kscurr_siam}
\ee
where $s_{R/L}=\pm$ and the KS spectral function is given in 
Eq.~(\ref{kssf}).

At first we consider the case of zero temperature, $T=0$. 
We know from Sec.~\ref{siam-sec} that if we use an approximate Hxc gate 
potential at zero current $v_{\rm Hxc}[N,I=0]$ which has a step at half 
occupation $N=1$, the resulting KS zero-bias conductance $G_s$ 
exhibits a Kondo plateau as function of gate voltage. This is 
exactly the case if we use the approximation of Eq.~(\ref{xcgate_siam_idft}) 
(which reduces to Eq.~(\ref{xc_mod_smooth_fit}) for $I=0$). On the other hand, 
we also know that in order to obtain the interacting zero-bias conductance 
$G$, we have to correct $G_s$ according to Eq.~(\ref{idft_zb_cond}). If we use 
Eq.~(\ref{xcbias_siam_idft}) as an approximation to the xc bias, it is easy 
to show that $\partial V_{\rm xc}/\partial I |_{I=0}$ (and thus the 
correction to $G_s$) is non-vanishing. The resulting $G$ will, instead of 
showing the Kondo plateau, exhibit the typical two-peak structure associated 
with Coulomb blockade. It is thus immediately clear what 
we have to do in order to recover the Kondo plateau in the i-DFT framework: 
we have to design the approximation to the xc bias in such a way that at $T=0$ 
the correction to $G_s$ according to Eq.~(\ref{idft_zb_cond}) vanishes. 
As a second requirement on our improved functional we want the Hxc gate at 
zero current to be not just qualitatively correct but also quantitatively as 
accurate as possible. Fortunately, in Ref.~\onlinecite{blbs.2012} an accurate, 
ready-to-use parametrization $v_{\rm Hxc}^{(0)}[n]$ of the Hxc gate 
potential at $T=0$ has been designed. We thus can make the following ansatz for 
our modified SIAM xc potentials \cite{KurthStefanucci:16}
\bea
\lefteqn{
v_{\rm Hxc}^{\rm SIAM}[N,I] =}\nn\\
&&\;\;\;\;\;\; \left( 1- \tilde{a}^{(0)}[I]\right) v_{\rm Hxc}[N,I] + 
\tilde{a}^{(0)}[I] v_{\rm Hxc}^{(0)}[N] 
\label{xcgate_siam_acc}
\eea
and 
\be
V_{\rm xc}^{\rm SIAM}[N,I] = \left( 1- \tilde{a}^{(0)}[I]\right) V_{\rm xc}[N,I] 
\label{xcbias_siam_acc}
\ee
where $v_{\rm Hxc}[N,I]$ and $V_{\rm xc}[N,I]$ are the functionals of 
Eqs.~(\ref{xcgate_siam_idft}) and (\ref{xcbias_siam_idft}), respectively, 
with the parameter $\lambda_1$ introduced there now kept as a fitting 
parameter to be determined. For the choice of the function $\tilde{a}^{(0)}[I]$ 
there are a few restrictions: by symmetry, it should be an even function of 
the current and for the correction $\partial V_{\rm xc}^{\rm SIAM}\partial I 
|_{I=0}$ to vanish, its value at $I=0$ should be unity. Furthermore, we want 
the effect of $\tilde{a}^{(0)}$ to fade out for increasing currents since 
Eqs.~(\ref{xcgate_siam_idft}) and (\ref{xcbias_siam_idft}) already captured the 
high-current Coulomb blockade physics correctly. We thus suggest the following 
form:
\be
\tilde{a}^{(0)}[I] = 1 - \left[ \frac{2}{\pi}\arctan\left(\frac{I}{\gamma W} 
\right) \right]^2 \;. 
\label{a0_func}
\ee
Finally, we fix the value of the parameter $\lambda_1=2$ such that the i-DFT 
$I-V$ characteristic at the particle-hole symmetric point is in good agreement 
with results from the functional renormalization group (fRG) 
\cite{EckelHeidrichJakobsThorwartPletyokhovEgger:10}. 

\begin{figure}[t]
\includegraphics[width=0.5\textwidth]{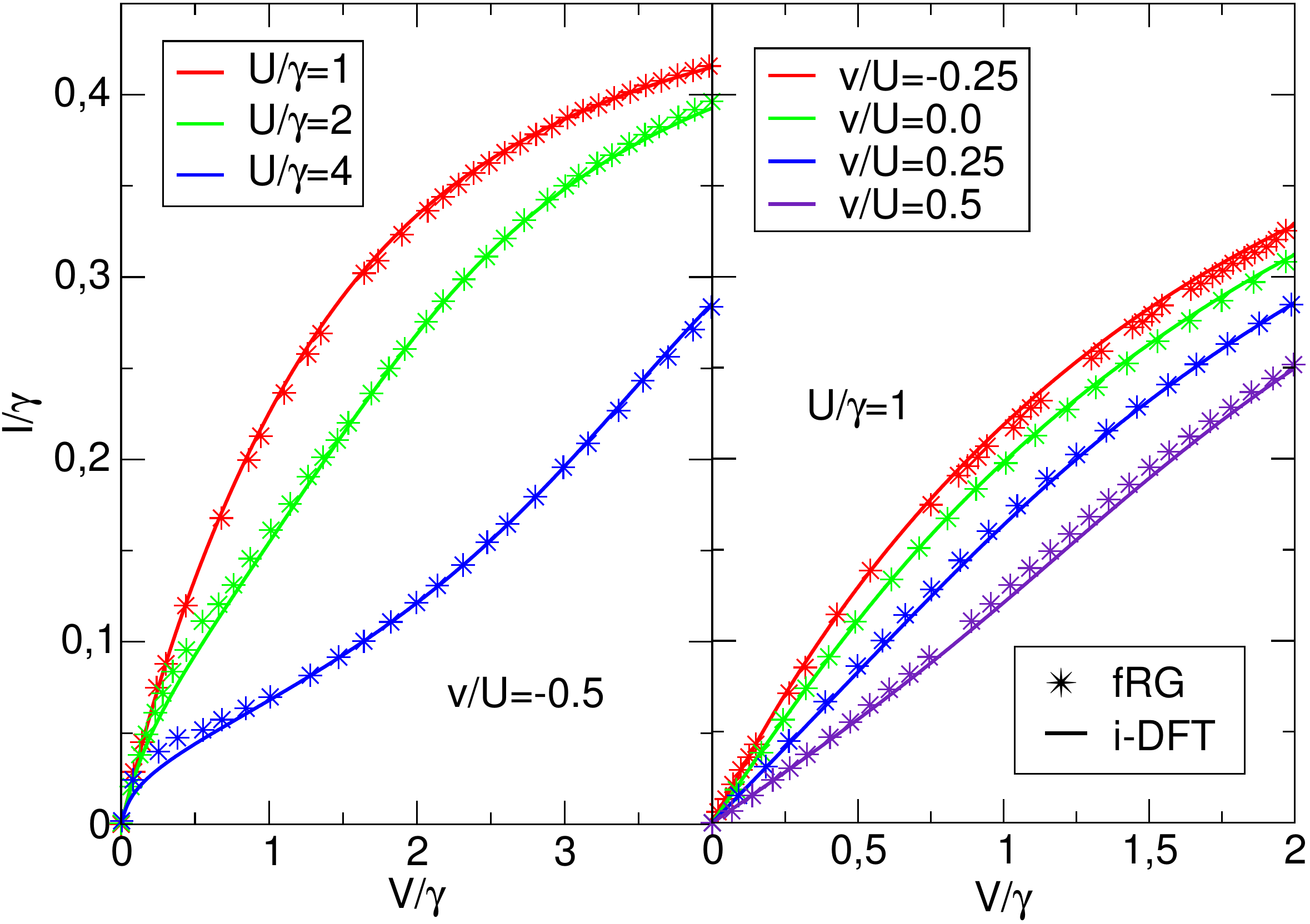}
\caption{Comparison of i-DFT and fRG $I-V$ characteristics at zero 
temperature. Left panel: at the particle-hole symmetric point $v=-U/2$ for 
different $U/\gamma$. Right panel: at fixed $U/\gamma=1$ for different $v$. 
fRG results from 
Ref.~\onlinecite{EckelHeidrichJakobsThorwartPletyokhovEgger:10}.
Reprinted with permission from 
\protect\onlinecite{KurthStefanucci:16}. Copyright (2016) American 
Physical Society.}
\label{IV_fRG_idft}
\end{figure} 

In the left panel we compare $I-V$ characteristics from i-DFT with those of 
fRG for $v=-U/2$. The agreement is excellent with small deviations at small 
biases. These deviations become somewhat more pronounced as $U$ increases. 
In the right panel we compare again $I-V$ characteristics but this time at 
fixed value of $U/\gamma=1$ for different gates $v$. Again, the agreement is 
excellent. Note that the fitting parameter $\lambda_1$ was chosen only to 
optimize the agreement at $v=-U/2$ but we obtain excellent results also for 
other values of the gate. 

After having fixed our parametrization for the SIAM (H)xc potentials at $T=0$, 
we now turn our attention to an extension for finite temperatures. Here we 
do not repeat all the details entering into the finite-$T$ approximations 
(which can be found in Ref.~\onlinecite{KurthStefanucci:16}) but rather 
sketch the main physical ingredients entering its construction. 

\begin{figure}[t]
\includegraphics[width=0.5\textwidth]{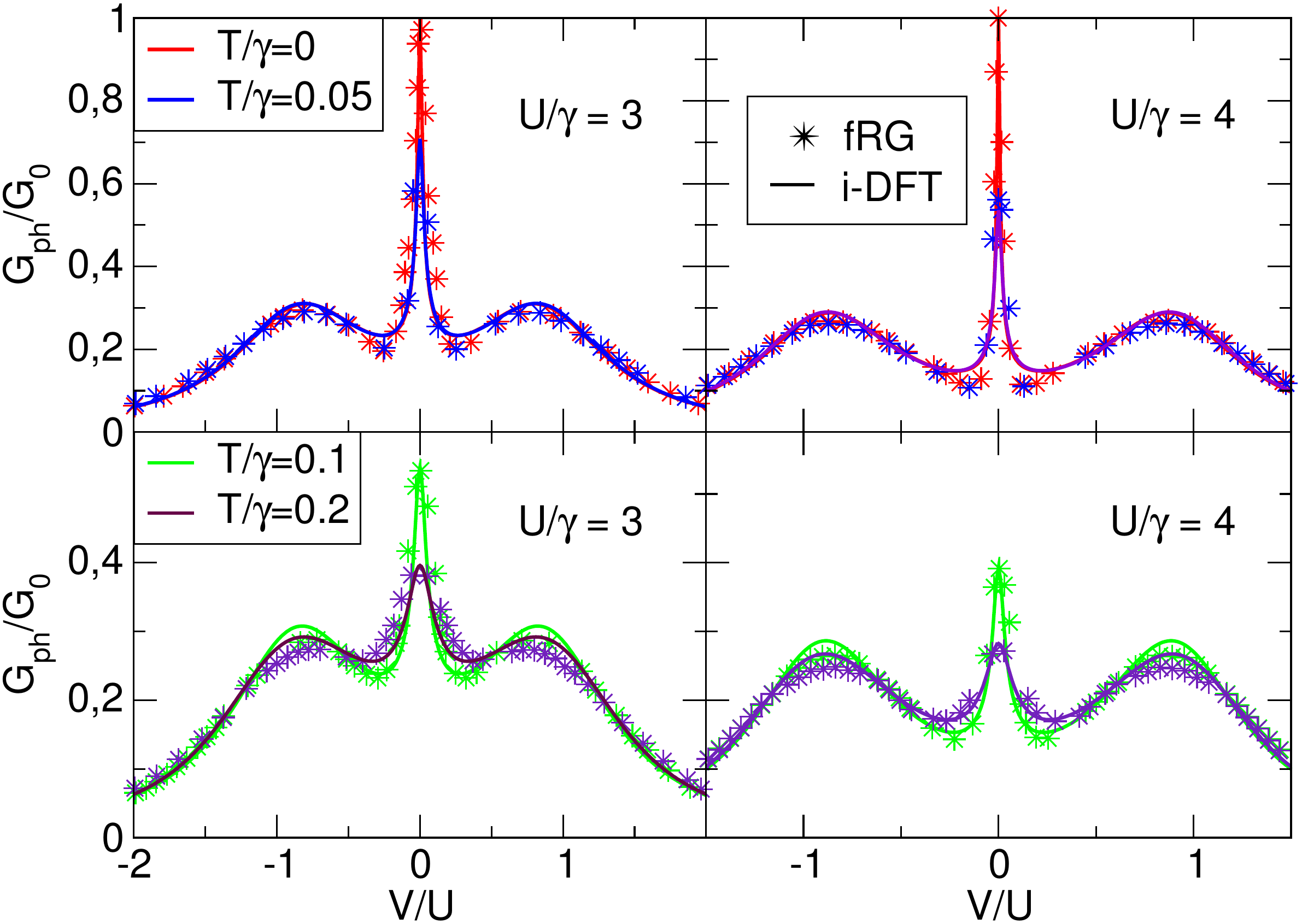}
\caption{Comparison of fRG and i-DFT differential conductances at $v=-U/2$ 
(in units of the quantum of conductance $G_0$) as function of bias for 
different temperatures. Left panels: $U/\gamma=3$. Right panels:  
$U/\gamma=4$. fRG results from Ref.~\onlinecite{JakobsPletyukhovSchoeller:10}.
Reprinted with permission from 
\protect\onlinecite{KurthStefanucci:16}. Copyright (2016) American 
Physical Society.}
\label{dIdV_fRG_idft}
\end{figure}

As we have already discussed in Sec.~\ref{siam-sec}, at the particle-hole 
symmetric point the zero-bias conductance $G^{\rm ph}$ of the SIAM at finite 
temperature is a universal function of $T/T_{\rm K}$ where $T_{\rm K}$ is the 
Kondo temperature defined in 
Eq.~(\ref{tkondo})~\cite{Costi:00,AleinerBrouwerGlazman:02}. 
This universal function is shown in Fig.~\ref{cond_ph}. When constructing 
our finite-$T$ approximations for the i-DFT functionals, we want to recover 
this exact property. This can be achieved by replacing $\tilde{a}^{(0)}[I]$ 
in Eqs.~(\ref{xcgate_siam_acc}) and (\ref{xcbias_siam_acc}) by 
\be
a^{(T)}[N,I] = b^{(T)}[N] \tilde{a}^{(T)}[I]
\ee
where $b^{(T)}[N]$ is chosen such that at $N=1$ the exact $G^{\rm ph}$ is 
recovered. The function $\tilde{a}^{(T)}[I]$ is obtained by just replacing 
the broadening $W$ in $\tilde{a}^{(0)}[I]$, see Eq.~(\ref{a0_func}), by a 
temperature-dependent function $W(T)$ \cite{KurthStefanucci:16} 
defined as
\be
W(T) = W \left[ 1 + 9 \left( \frac{T}{\gamma}\right)^2 \right] \;.
\ee
The same replacement $W\to W(T)$ is also done in the functional forms 
of $v_{\rm Hxc}[N,I]$ and $V_{\rm xc}[N,I]$. Physically, this replacement 
reflects the expectation that at small temperature the broadening in the steps 
of the (H)xc potentials is dominated by $\gamma$ while at large temperatures it 
is dominated by $T$. The particular form for $W(T)$ was chosen such as to 
best reproduce the fRG differential conductances of 
Ref.~\onlinecite{JakobsPletyukhovSchoeller:10}, see 
Fig.~\ref{dIdV_fRG_idft}.
The two physical ingredients for designing a finite-temperature approximations 
(reproduction of the universal $G^{\rm ph}$ as function of $T$, temperature
dependent broadening of the step features) are augmented by some smaller 
tweaks to reproduce well \cite{KurthStefanucci:16} the fRG zero-bias 
conductances of Ref.~\onlinecite{JakobsPletyukhovSchoeller:10}. 

In Fig.~\ref{dIdV_fRG_idft} a comparison of fRG and i-DFT finite-bias 
differential conductances at $v=-U/2$ is shown for two different interaction 
strengths and temperatures. By construction, the proper reduction of the 
Kondo peak at $V=0$ with increasing temperature is correctly reproduced. Also 
the Hubbard sidebands are reproduced with good accuracy. 

\begin{figure}[t]
\includegraphics[width=0.5\textwidth]{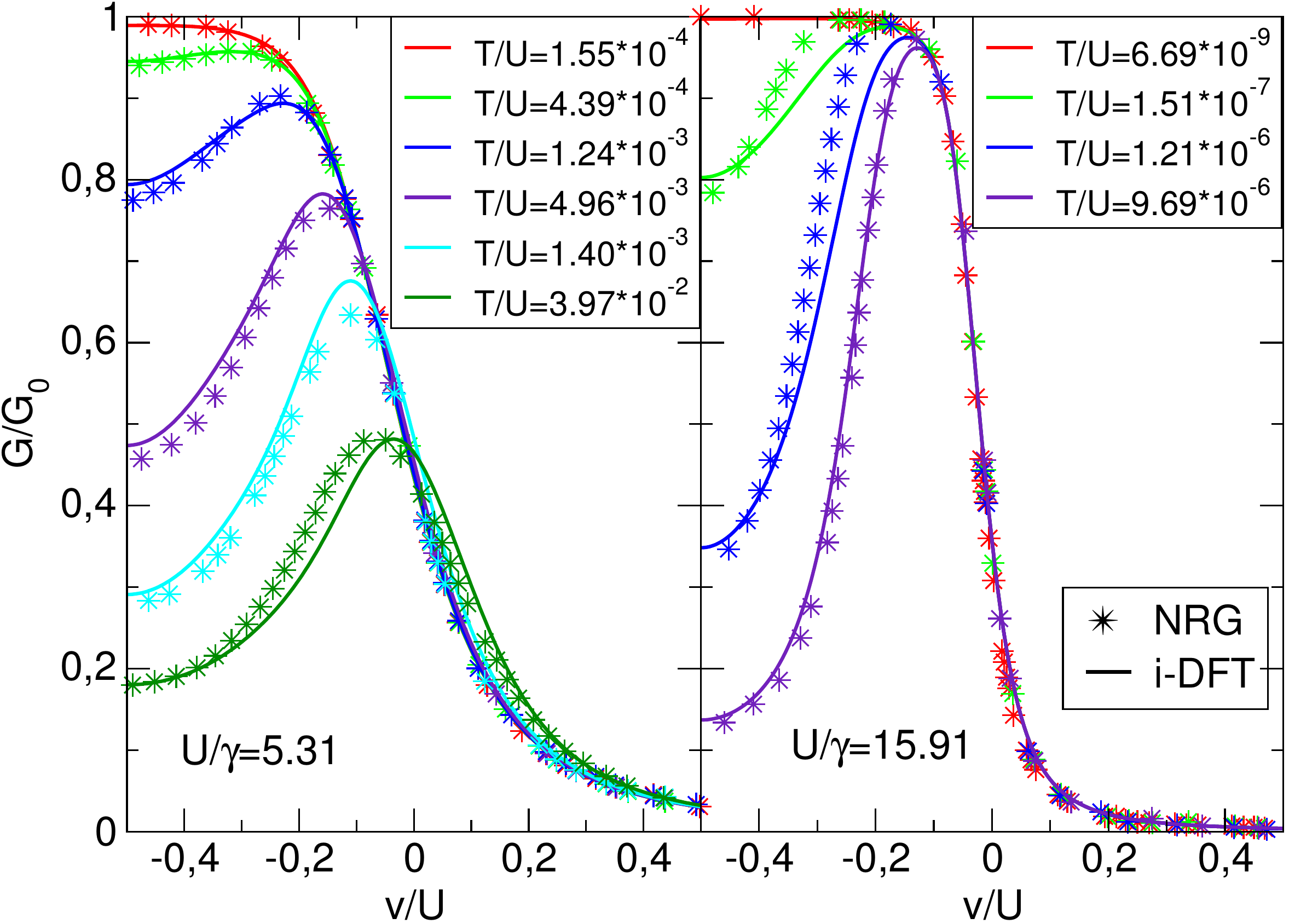}
\caption{Comparison between NRG and i-DFT zero-bias conductances (in units 
of $G_0$) as function of gate for different temperatures. Left panel: 
for $U/\gamma=5.31$. Right panel: for $U/\gamma=15.91$. NRG results from 
Ref.~\onlinecite{IzumidaSakaiSuzuki:01}.
Reprinted with permission from 
\protect\onlinecite{KurthStefanucci:16}. Copyright (2016) American Physical Society.}
\label{conduct_NRG_idft}
\end{figure}

Finally, in Fig.~\ref{conduct_NRG_idft} we compare NRG and i-DFT 
zero-bias conductances as function of gate for two interaction strengths and 
various temperatures. In general, the agreement is very good. Only for high 
$U/\gamma$ (right panel) and low temperatures the shape of the side peaks 
is slightly different. 

In summary we can say that we have indeed been able to design i-DFT 
functionals which capture both Kondo and Coulomb blockade physics in the 
SIAM as well as the transition between the two regimes as temperature 
increases. These functionals allow for the accurate calculation of densities 
and currents of the SIAM in the steady state over a wide range of parameters 
at negligible numerical cost. 

Of course, the design of our functionals relied heavily on the 
availability of accurate solutions of the SIAM obtained from other methods, 
just like in standard DFT the construction of the local density approximation 
(LDA) relies on the availability of accurate xc energies of the uniform 
electron gas from Quantum Monte Carlo calculations. Nevertheless, with the 
design of our functionals we have explicitly demonstrated that it is indeed 
possible to accurately describe transport through strongly 
correlated systems with DFT, thus disproving common wisdom that DFT is not 
suited to deal with strong correlations. 

At this point it is worth to emphasize that the 
construction of our i-DFT xc potentials was guided by a few rather simple 
considerations. The lessons learned here are easily transferrable 
to more complicated systems. For instance, the incorporation of Kondo 
physics in the i-DFT description of the contacted CIM can easily be achieved 
by modifying the CIM functionals of the previous Sec.~\ref{clb_block_idft} 
along similar lines to the ones discussed here for the SIAM: since the 
zero-bias KS conductance $G_s$ already contains the correct Kondo features 
also for multi-level systems (see Sec.~\ref{cim_trans}) 
one only has to ensure that the correction to $G_s$ 
vanishes for $I=0$. For instance, for a degenerate CIM with $M$ levels this 
can be achieved by using the following Hxc potentials (at $T=0$)
\bea
\lefteqn{v_{\rm Hxc}[N,I] = }\nn\\
&& \;\;\;\;\;\; \left( 1 - \tilde{a}^{(0)}[I] \right) v_{\rm Hxc}^{(M)}[N,I] 
+ \tilde{a}^{(0)}[I] \bar{v}_{\rm Hxc}^{(0)}[N]
\label{xcgate_cim_deg_mod}
\eea
\be
V_{\rm Hxc}[N,I] = \left( 1 - \tilde{a}^{(0)}[I] \right) V_{\rm xc}^{(M)}[N,I] 
\label{xcbias_cim_deg_mod}
\ee
where $v_{\rm Hxc}^{(M)}$ and $V_{\rm Hxc}^{(M)}$ are given by 
Eqs.~(\ref{xcgate_cim_deg}) and (\ref{xcbias_cim_deg}), respectively, 
with the only difference that we now use the parameter value $\lambda_1=2$. 
The function $\tilde{a}^{(0)}[I]$ is the same one as defined in 
Eq.~(\ref{a0_func}) and for the equilibrium Hxc potential 
$\bar{v}_{\rm Hxc}^{(0)}$ we use  
\be
\bar{v}_{\rm Hxc}^{(0)}[N] = \sum_{K=1}^{2M-1} v_{\rm Hxc}^{\rm ext}[N-(K-1)] \;. 
\ee
Here we have defined the extended function
\be 
v_{\rm Hxc}^{\rm ext}[N] = 
\left\{ 
\begin{array}{cl}
0 & \mbox{ $N<0$} \\
v_{\rm Hxc}^{(0)}[N] & \mbox{ $0\leq N \leq 2$}~~, \\
U & \mbox{ $N>2$} 
\end{array}
\right.
\ee
with the parametrization $v_{\rm Hxc}^{(0)}[N]$ of the equilibrium SIAM 
Hxc potential of Ref.~\onlinecite{blbs.2012}. Note that for $M=1$, 
Eqs.~(\ref{xcgate_cim_deg_mod}) and (\ref{xcbias_cim_deg_mod}) reduce exactly 
to Eqs.~(\ref{xcgate_siam_acc}) and (\ref{xcbias_siam_acc}), i.e., our 
accurate $T=0$ parametrizations for the i-DFT (H)xc potentials of the SIAM. 

\begin{figure}[tb]
\includegraphics[width=0.48\textwidth]{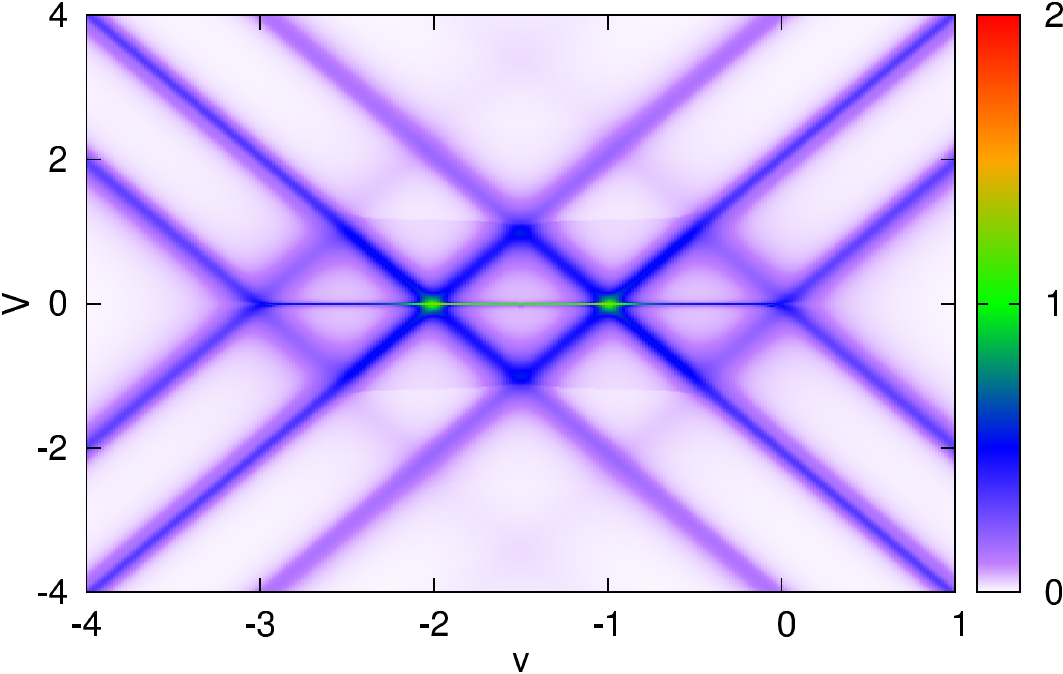}
\caption{Zero-temperature differential conductance in the gate-voltage plane 
of a a degenerate HOMO-LUMO CIM for $U/\g=8$. Energies in units of $U$.}
\label{didv_HOMO_LUMO_deg_cimmod}
\end{figure}

In Fig.~\ref{didv_HOMO_LUMO_deg_cimmod} we show the 
zero-temperature differential conductance in the gate-voltage plane 
of a degenerate HOMO-LUMO CIM for $U/\g=8$ (top 
panel) and $U/\g=4$ (bottom panel). 
One can clearly appreciate the 
Kondo strip at zero voltage for gates $v\in (-3U,0)$ as well as the 
fact that the height of the strip for $v\in (-2U,-U)$ is twice as 
large as the height for $v\in (-3U,-2U)$ and $v\in (-U,0)$.
This simple incorporation of Kondo physics in more complicated 
models seems hard to achieve within other frameworks to deal with 
transport through correlated systems. 

\section{Conclusion and outlook}

The chemical complexity of molecular junctions calls for a 
first-principle description of the molecule attached to leads 
in order to make quantitative predictions  
and/or comparisons with experiments. DFT is a computationally 
efficient theory which, in principle, is well suited to achieve this 
goal. 
However, DFT strongly relies on the quality of the xc functional 
and, at present, most available approximations are inadequate for 
strongly correlated junctions. Furthermore, the applicability of 
(equilibrium) DFT in quantum transport needs to be carefully 
discussed.

In this Topical Review we have revisited the standard LB+DFT approach to 
quantum transport and showed that, even with the {\em exact} xc 
functional this approach is {\em not exact}. We have presented two 
exact frameworks which can cure the 
deficiencies of LB+DFT. The first is based on TDDFT and leads to an xc 
corrections to the applied bias. Unfortunately this xc bias correction 
as well as the Hxc potential in the molecular region are functionals 
of the density {\em everywhere} and {\em at all previous times}. This 
circumstance makes the development of practical approximations very 
difficult, especially if we are not interested in the full time evolution 
but only in steady-state properties.  
The second theory is i-DFT and, like TDDFT, predicts the existence of an 
xc bias 
correction. However, the basic variables of i-DFT are the 
steady-state density in the molecular region and the steady-state 
longitudinal current. It is therefore possible to obtain both 
these quantities by a steady-state self-consistent calculation 
without any need of knowing how the system has attained the steady 
state. i-DFT reduces to the LB+DFT approach if the xc bias correction 
is discarded and if the Hxc potential is the one of standard DFT 
depending only on the density.

Of course, for i-DFT to prove useful for the calculation of the transport 
properties of strongly correlated junctions one needs accurate 
approximations to the 
i-DFT functionals. Unfortunately, we are still not in the position of 
offering (H)xc potentials ready to use in available first-principles 
codes. Yet, we have been able to identify crucial and general properties 
that any good approximation to these functionals should fulfill.  
We have shown that the step of the equilibrium Hxc potential 
as the total number of particles crosses an integer {\em bifurcates} 
at finite current. Furthermore, a bifurcating step structure at integer 
number of particles occurs also for the xc bias as the current starts 
flowing. This latter property is of utmost importance
 since the derivative $\de V_{\rm xc}/\de I$ enters 
explicitly into the i-DFT formula for the zero-bias conductance and it plays a 
crucial role in suppressing  the Kondo plateau in the 
CB regime. We have also identified an interesting duality between the 
values of the current and number of particles inside a CB diamond on 
one side 
and the intersections between discontinuity lines in the (H)xc 
potentials on the other side. These qualitative features should be incorporated in any 
approximation in order to reproduce the CB pattern of the 
differential conductance.

As a proof of concept we have examined in detail the SIAM and 
provided an accurate parametrization of the i-DFT potentials by best 
fitting the numerical results from fRG and NRG. The performance of i-DFT 
has turned out to be a full success. The differential conductance 
at any finite bias calculated by solving the i-DFT equations is accurate for any value 
of the interaction strenght and for temperatures ranging from zero to  
well above the Kondo temperature. Last but not least, owing to the simplicity of the 
i-DFT equations the calculation of an $I-V$ characteristics takes 
less than a CPU second.

An interesting possible extension of i-DFT which we are currently 
exploring consists in considering the Hxc potential and xc bias as 
adiabatic functionals of current and  density for 
time-dependent calculations. This study generalizes previous 
investigations in strongly correlated models where 
the equilibrium xc 
potential of DFT~\cite{LimaOliveiraCapelle:02,LimaSilvaOliveiraCapelle:03,SchoenhammerGunnarssonNoack:95}
was turned, through the adiabatic approximation, into the xc
potential of TDDFT~\cite{Verdozzi:08,kskvg.2010,UimonenKhosraviStanStefanucciKurthLeeuwenGross:11,KhosraviUimonenStanStefanucciKurthLeeuwenGross:12}. 
In Ref.~\onlinecite{UimonenKhosraviStanStefanucciKurthLeeuwenGross:11} it 
was pointed out that the inaccurate value of the TDDFT current in the 
SIAM was due to the neglect of the xc bias. However, as we have shown, 
this is not the whole story. It is crucial that the xc bias is  
also a functional of the current. In fact, the time-local 
dependence on the current in i-DFT translates into a time- and space-nonlocal dependence 
on the density in TDDFT, see Ref.~\onlinecite{nptvc.2007}. The entangled 
space and time nonlocality of the functionals are strongly related by 
conservation laws~\cite{vk.1996}. Thus, i-DFT holds promise for 
an improved description of time-dependent phenomena like, e.g., 
transient processes or AC  responses.

\begin{acknowledgments}
S.K. acknowledges funding by a grant of the "Ministerio de Economia y
Competividad (MINECO)" (FIS2016-79464-P) and by the
``Grupos Consolidados UPV/EHU del Gobierno Vasco'' (IT578-13). 
G.S. acknowledges funding by MIUR FIRB Grant No. RBFR12SW0J and EC funding 
through the RISE Co-ExAN (GA644076).
\end{acknowledgments}

\appendix
\section{Exact equilibrium density of the Constant Interaction Model}
\label{cim_uncontacted_exact}

In this Appendix we describe an algorithm which allows for the calculation of 
the equilibrium density of the (uncontacted) Constant Interaction Model (CIM) 
described by the Hamiltonian $\hat{H}^{\rm CIM}$ of Eq.~(\ref{cim}) at 
arbitrary temperature. Our aim is to calculate the densities 
(occupations) 
\be 
n_{k} = \Tr{ \hat{\rho} \hat{n}_k}
\ee
where 
\be
\hat{\rho} = \frac{\exp(-\beta (\hat{H}^{\rm CIM} - \mu))}
{Z_{K}(\tilde{\varepsilon}_1, \ldots, \tilde{\varepsilon}_K)}
\ee
with the partition function $Z_{K}(\tilde{\varepsilon}_1, \ldots, 
\tilde{\varepsilon}_K)$ of 
the CIM with $K$ single-particle levels with energies 
$\tilde{\varepsilon_k}=\varepsilon_k - \mu$. 
Suppose that we know the partition function 
$Z_{K-1}(\tilde{\varepsilon}_1, \ldots, \tilde{\varepsilon}_{K-1})$ of the 
system with $K-1$ levels. Then the partition function for $K$ levels can 
be calculated by the recursion relation 
\bea
\lefteqn{
Z_{K}(\tilde{\varepsilon}_1, \ldots, \tilde{\varepsilon}_K)
=Z_{K-1}(\tilde{\varepsilon}_1, \ldots, \tilde{\varepsilon}_{K-1})}\nn\\
&&\!\! + \exp(-\beta \tilde{\varepsilon}_K)
Z_{K-1}(\tilde{\varepsilon}_1+U, \ldots, \tilde{\varepsilon}_{K-1}+U)
\label{cim_partfunc_rec}
\eea
Defining the quantity 
\be
R_K(\tilde{\varepsilon}_1, \ldots, \tilde{\varepsilon}_K) := 
\frac{Z_{K}(\tilde{\varepsilon}_1+U, \ldots, \tilde{\varepsilon}_{K}+U)}
{Z_{K}(\tilde{\varepsilon}_1, \ldots, \tilde{\varepsilon}_{K})} 
\ee
from Eq.~(\ref{cim_partfunc_rec}) one can then easily derive the recursive 
relation
\bea
\lefteqn{
R_{K}(\tilde{\varepsilon}_1, \ldots, \tilde{\varepsilon}_K) = 
R_{K-1}(\tilde{\varepsilon}_1, \ldots, \tilde{\varepsilon}_{K-1})}\nn\\
&& \times \frac{1 + \exp(-\beta (\tilde{\varepsilon}_M+U)) 
R_{K-1}(\tilde{\varepsilon}_1+U, \ldots, \tilde{\varepsilon}_{K-1}+U)}
{1 + \exp(-\beta \tilde{\varepsilon}_M) 
R_{K-1}(\tilde{\varepsilon}_1, \ldots, \tilde{\varepsilon}_{K-1})}\nn\\
\eea
where we have also defined
\be
R_1(\tilde{\varepsilon}) = \frac{ 1 + \exp(-\beta (\tilde{\varepsilon} + U))}
{ 1 + \exp(-\beta \tilde{\varepsilon})} \;.
\ee
With these definitions the occupation of level $k$ becomes
\bea
n_k &=& -\frac{1}{\beta} \frac{\partial}{\partial \varepsilon_k} \ln
Z_{K}(\tilde{\varepsilon}_1, \ldots, \tilde{\varepsilon}_K) \nn\\
&=& \frac{\exp(-\beta \tilde{\varepsilon}_k) 
R_{K-1}(\tilde{\varepsilon}_1, \ldots, \tilde{\varepsilon}_{K-1})}
{1 + \exp(-\beta \tilde{\varepsilon}_k ) 
R_{K-1}(\tilde{\varepsilon}_1, \ldots, \tilde{\varepsilon}_{K-1})}
\eea

\section*{References}

\providecommand{\newblock}{}


\end{document}